\pdfoutput=1

\documentclass[twocolumn]{aastex62}

\usepackage{nameref} 
\DeclareUnicodeCharacter{2212}{\ensuremath{-}}
\newcommand{\Msun}{$M_{\odot}$}

\usepackage{xcolor}
\usepackage{graphicx}	
\usepackage{amsmath}	
\usepackage{multirow}
\usepackage{appendix}
\usepackage{chngcntr} 

\usepackage{longtable} 
\usepackage{subfigure}

\def\arcsec{\hbox{$^{\hbox{\rlap{\hbox{\lower4pt\hbox{$\,\prime\prime$}}}\hbox{$\frown$}}}$}}

\graphicspath{{./}{figures/}}

\shorttitle{}
\shortauthors{}

\begin{document}


\title{
Empirical color correction to MIST and PARSEC isochrones on Gaia BR$-$RP and G$-$RP with benchmark open clusters
}

\correspondingauthor{Min Fang}
\email{mfang@pmo.ac.cn}

\author{Fan Wang}
\affiliation{Purple Mountain Observatory, Chinese Academy of Sciences, 10 Yuanhua Road, Nanjing 210023, China}
\affiliation{University of Science and Technology of China, Hefei 230026, China}

\author{Min Fang}
\affiliation{Purple Mountain Observatory, Chinese Academy of Sciences, 10 Yuanhua Road, Nanjing 210023, China}
\affiliation{University of Science and Technology of China, Hefei 230026, China}

\author{Xiaoting Fu}
\affiliation{Purple Mountain Observatory, Chinese Academy of Sciences, 10 Yuanhua Road, Nanjing 210023, China}
\affiliation{University of Science and Technology of China, Hefei 230026, China}

\author{Yang Chen}
\affiliation{School of Physics and Optoelectronic Engineering, Anhui University, 230601, Hefei, China}

\author{Lu Li}
\affiliation{Key Laboratory for Research in Galaxies and Cosmology, Shanghai Astronomical Observatory, Chinese Academy of Sciences, 80 Nandan Road, Shanghai 200030, China}
\affiliation{University of Chinese Academy of Sciences, No. 19A Yuquan Road, Beijing 100049, China}

\author{Xiaoying Pang}
\affiliation{Department of Physics, Xi’an Jiaotong-Liverpool University, 111 Ren’ai Road, Dushu Lake Science and Education Innovation District, Suzhou 215123, Jiangsu Province, China}
\affiliation{Shanghai Key Laboratory for Astrophysics, Shanghai Normal University, 100 Guilin Road, Shanghai 200234, China}

\author{Zhongmu Li}
\affiliation{Institute of Astronomy, Dali University, Dali 671003, China}

\author{Jing Tang}
\affiliation{National Astronomical Observatories, Chinese Academy of Sciences, 100101, Beijing, China}

\author{Wenyuan Cui}
\affiliation{College of Physics, Hebei Normal University, Shijiazhuang 050024, China}

\author{Haijun Tian}
\affiliation{School of Science, Hangzhou Dianzi University, Hangzhou 310018, China} 
\affiliation{Big Data Institute, Hangzhou Dianzi University, Hangzhou 310018, China}

\author{Chao Liu}
\affiliation{National Astronomical Observatories, Chinese Academy of Sciences, 100101, Beijing, China}



\begin{abstract}
Recent literature reports a color deviation between observed Gaia color-magnitude diagrams (CMDs) and theoretical model isochrone predictions, particularly in the very low-mass regime. To assess its impact on cluster age determination via isochrone fitting, we quantified the color deviations for three benchmark clusters, Hyades, Pleiades, and Praesepe, both for the Gaia color (BP$-$RP) and (G$-$RP). In general, the (G$-$RP) color deviations are smaller than the (BP$-$RP) ones. Empirical color correction functions based on these benchmarks are derived  for the currently available MIST and PARSEC 1.2S isochrone models. Applying the correction functions to 31 additional open clusters and 3 moving groups results in a significantly improved alignment between the isochrones and observed CMDs. With our empirical corrections, isochrones provide age estimates consistent with literature values obtained through the spectral Lithium Depletion Boundary method, validating the effectiveness of our approach. The corresponding metallicities with PARSEC 1.2S also show a good agreement with the spectroscopic results. The empirical color correction function we present in this work offers a tool for a consistent age determination within the full mass range of stellar clusters using the isochrone fitting method.
\end{abstract}



\section{Introduction} 

\label{sec:intr}
The isochrone fitting method is commonly employed to derive star cluster parameters like age, metallicity, distance, and reddening \citep[see e.g., ][]{Pietrinferni2004ApJ, Tognelli2011AA, Dotter2016ApJS, Bressan2012MN,Li2022ApJ}. Cluster age, crucial for understanding formation and evolution \citep{Soderblom2009AJ1292S, Galli2023MNRAS}, relies on both precise observational data and robust isochrone models. The Gaia mission\footnote{\url{https://www.cosmos.esa.int/web/gaia}} has revolutionized star cluster studies by providing high-precision photometry and astrometry (proper motion and parallax) for a vast number of stars. With this support, fitting isochrones on  color-magnitude diagram (CMD) to derive the age of nearby stellar populations, including open clusters and moving groups, has become significantly more accurate than in the pre-Gaia era. Isochrones of two stellar evolutionary models are widely used in the community, MIST \citep[MESA Isochrones \& Stellar Tracks,][]{Dotter2016ApJS, Choi2016ApJ} and PARSEC \citep[the PAdova \& TRieste Stellar Evolution Code,][]{Bressan2012MN, Tang2014MN, Chen2014MN, Chen2015MN}.

However, different models used in isochrone fitting introduces  systematic offsets in age determinations. \citet{Bell2015MN} found that fitting the observed photometry of the $\beta$ Pic moving group with various
isochrone models (BHAC15, Dartmouth, PARSEC, Pisa) yielded age estimates ranging from 10\,Myr to 20\,Myr. The issue has  been observed in other nearby moving groups and open clusters \citep[e.g., ][]{Mamajek201469M, Bell2015MN, Kerr2022, Lee20244760L}. The age variations with different isochrones can be attributed  to the stellar models' physics input, treatment of mixing in the stellar interior, and stellar atmosphere models used for the bolometric corrections.

No model is perfect, though. It has been discussed for a long time that, for given photometry systems  or evolutionary stages, there are color discrepancies on the CMD between observations and model predictions. \citep[see e.g., ][]{Castellani2001MNRAS66C, Preibisch2012, Bell2015MN, Kopytova2016AA7K, Siegel2019AJ35S,Li2020ApJ}. This is particularly conspicuous in the Gaia era due to the high precision of Gaia photometry and astrometry. In a series of papers, \citet{Brandner2023AJ108B, Brandner2023b, Brandner2023MN} show that the CMDs of open clusters with Gaia DR3 photometry exhibit deviations from the MIST and PARSEC model isochrones at the low-mass range.

Modelling low mass stars is complex. Discussions in the literature \citep[e.g., ][]{Stauffer1998ApJ199S, Baraffe2002, DaRio2010, Bell2014MN3496B} suggest that the observation and model deviation on CMD at the low-mass part could be a combination of factors, including the stellar interior structure and atmosphere models, especially at lower temperatures $T_{eff}$, as well as other factors affecting  photometric data such as magnetic activity \citep{Bell2014MN3496B}. 

The isochrone color deviation is not unique to Gaia\,DR3 photometry. It is also seen with Gaia\,DR2 data \citep[see isochrone fitting figues in e.g.][]{Bossin2019AA108B} and other passbands \citep[see e.g.][]{Fritzewski2019AA}. The color deviation is also considered a  contributor to age differences between different studies \citep[see e.g.][]{Lee2022}, especially for young moving groups \citep[e.g., ][]{Feiden2016, Naylor2009MN432N, Herczeg2015}. Magnetic inhibition of convection in cool stars, verified by \citet{Feiden2016, Jeffries2017MN1456J}, could be a possible explanation.

Color deviation can affect the determination of age and binary fraction. At the upper main sequence, the model isochrone roughly matches the actual photometric data of the star cluster. However, at the lower main sequence, color deviation between the model isochrone and the actual photometric data leads to inconsistent age determinations based on the upper main sequence end. A bluer color of the model isochrone can also lead to an overestimation of the binary fraction. For older star clusters, the single star and binary sequences are usually distinguishable in CMDs. For example, the ridge line method can be used effectively to determine the binary fraction of the cluster. However, for young clusters, their distributions in CMDs are scattered, which makes it difficult to determine their binary fractions without correct model isochrones. 

This study aims to address this issue by deriving an empirical color correction function for Gaia DR3 photometry (BP$-$RP) and (G$-$RP) using benchmark clusters with minimal extinction and high-confidence age determinations. The goal is to improve agreement between model isochrones and observed clusters for a better age determination. The Hyades, Pleiades, and Praesepe clusters serve as our benchmark targets.

The structure of the paper is outlined as follows. Section~\ref{sec:Data} describes the data and isochrone models we use in this work. Section~\ref{sec:benchmark_isochrones} demonstrates the color deviations between the CMDs of the Hyades, Pleiades, and Praesepe clusters and model isochrones. Section~\ref{sec:process_color_correction} describes the color correction process. Section~\ref{sec:result-disc} verifies the validity of the color correction function using 31 clusters and 3 moving groups. Section 6 is the summary and discussions.

\begin{figure*}
    \centering
    {%
    \includegraphics[width=0.331\textwidth, trim=0.3cm 1.22cm 0.2cm 1.7cm, clip]{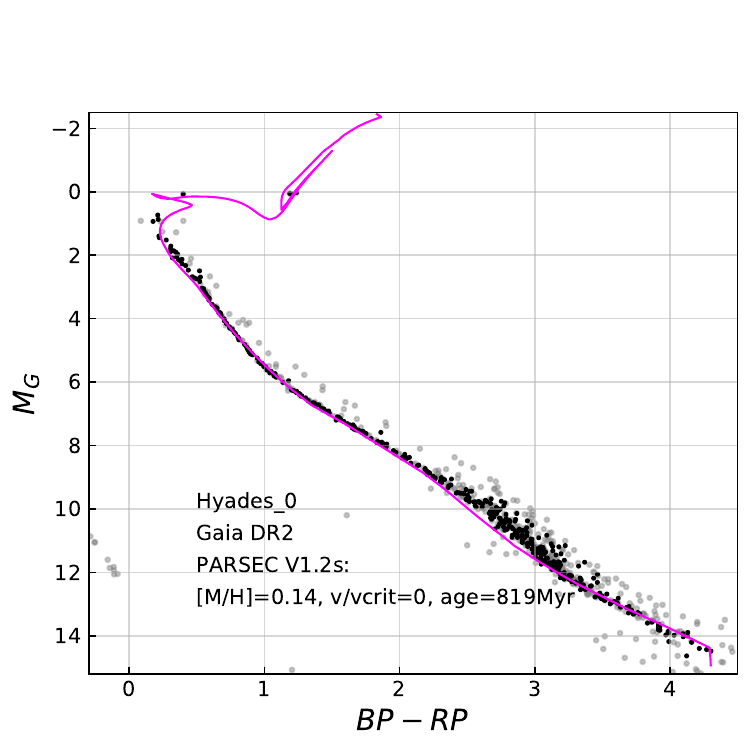}\quad
    \includegraphics[width=0.3\textwidth, trim=1.5cm 1.22cm 0.2cm 1.7cm, clip]{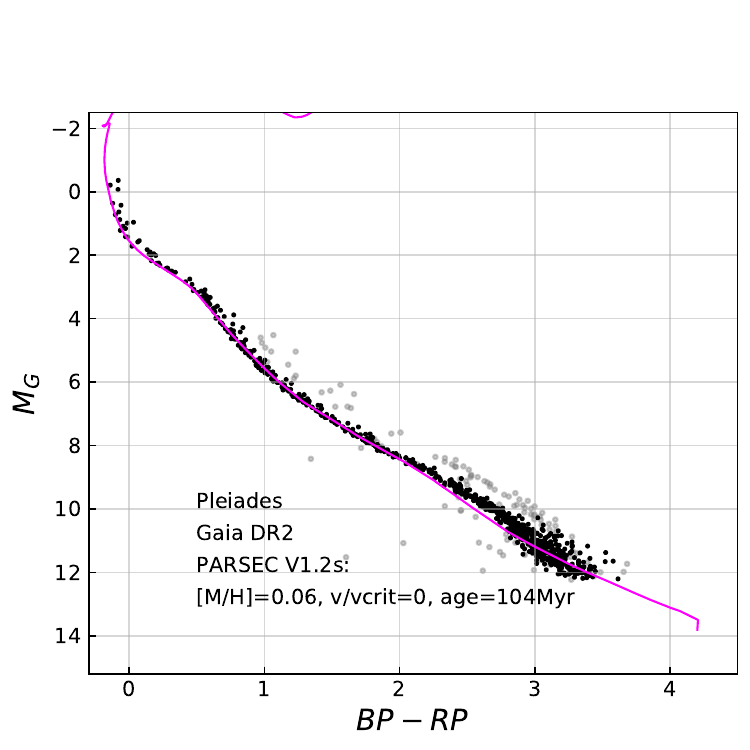}\quad
    \includegraphics[width=0.3\textwidth, trim=1.5cm 1.22cm 0.2cm 1.7cm, clip]{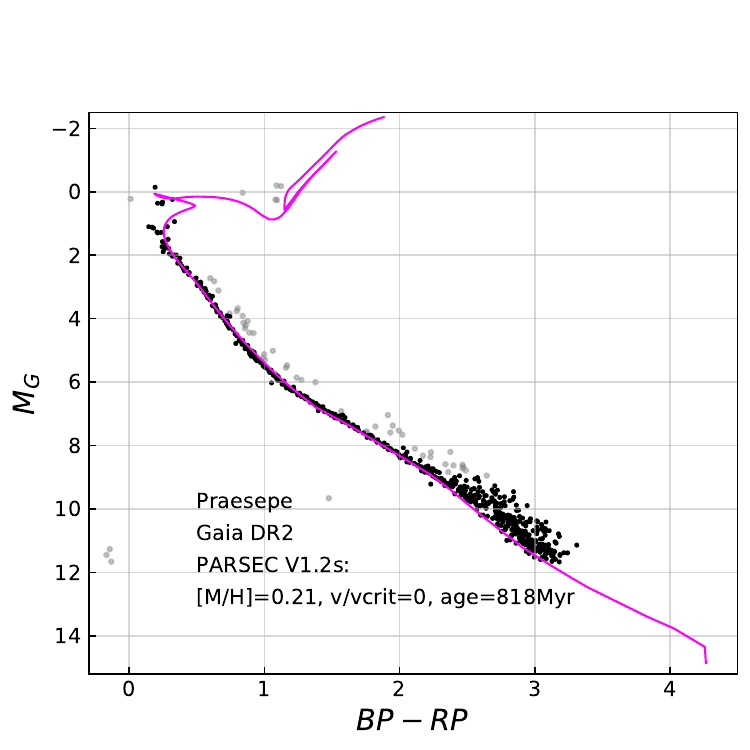}\quad}
    {%
    \includegraphics[width=0.343\textwidth, trim=0.00cm 0cm 0.2cm 1.92cm, clip]{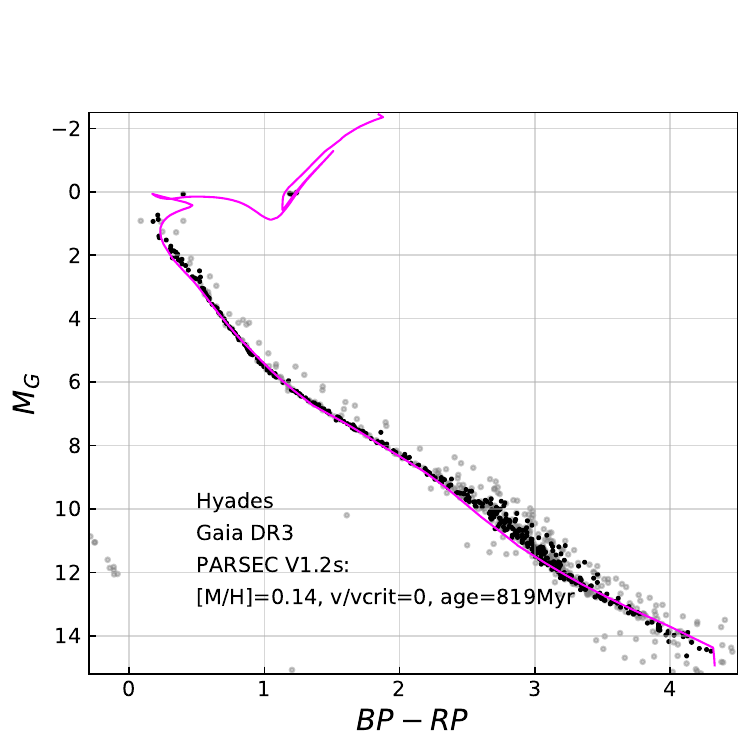}\quad
    \includegraphics[width=0.301\textwidth, trim=1.5cm 0cm 0.2cm 1.92cm, clip]{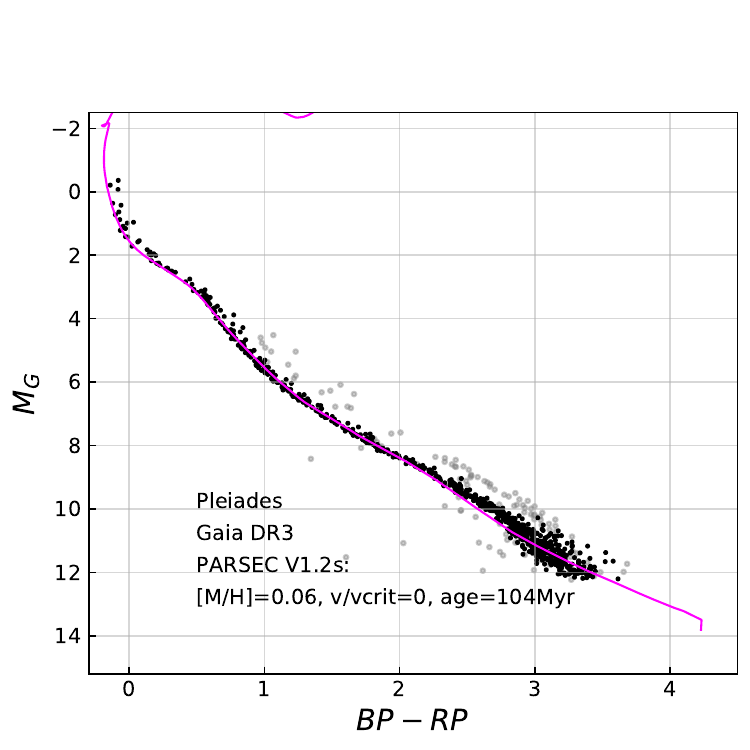}\quad
    \includegraphics[width=0.301\textwidth, trim=1.5cm 0cm 0.22cm 1.92cm, clip]{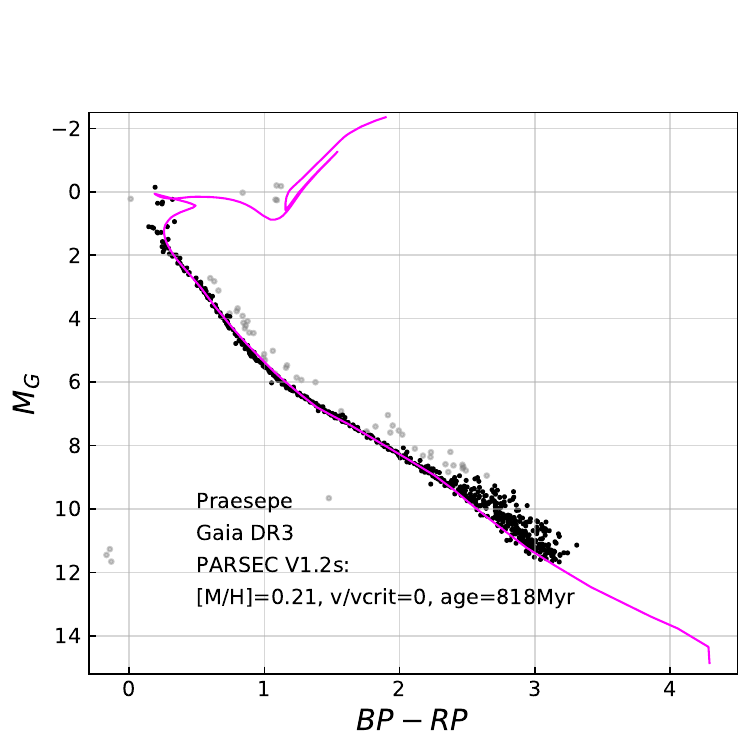}\quad}

    \caption{The CMDs of the clusters Hyades, Pleiades, and Praesepe (Upper panels: Gaia DR2; Lower panels: Gaia DR3). The magenta solid lines represent the isochrones from the PARSEC 1.2S model. Black dots represent cluster member stars obtained according to CG20 after removing potential field star contamination and binaries (described in section~\ref{sec:benchmark_isochrones}).
    }
    \label{fig:3OCs_literas}
\end{figure*}

\begin{figure}
    \centering
    {%
    \includegraphics[width=0.45\textwidth]{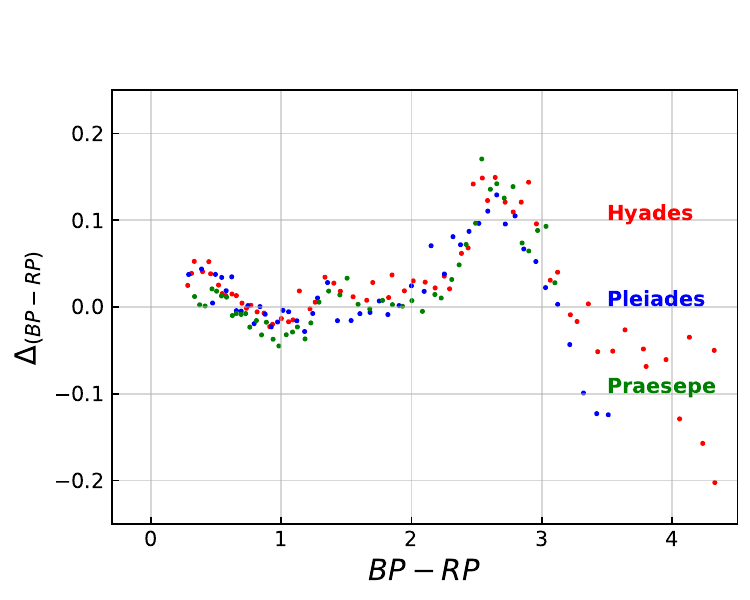}}
    
    \caption{The distribution of the color deviation between the model isochrones and the Gaia~DR3 data for the three clusters (red for Haydes, blue for Pleiades, and green for Praesepe).
    }
    \label{fig:delta_3OCs_literas}
\end{figure}

\section{Data and models}
\label{sec:Data}
In this section, we briefly introduce the data and stellar isochrones we use in this work, including the Gaia\,DR3 \citep{Gaia2023} data, the MIST \citep{Dotter2016ApJS, Choi2016ApJ} and PARSEC 1.2S \citep{Bressan2012MN, Tang2014MN, Chen2015MN} isochrone models.

\subsection{Gaia DR3}
\label{sec:gaiadr3}

The third release of the Gaia Data (DR3) \citep{Gaia2023} provides astrometric information for about 1.8 billion sources across the sky, along with near-mmag precision photometric data in G, $G_{BP}$ and $G_{RP}$ bands. The typical parallax uncertainty is 0.07\,mas at $G\approx17$\,mag and 0.5\,mas at $G=20$\,mag. The mean G-band photometry uncertainty is 1\,mmag at $G\approx17$\,mag and 6\,mmag at $G=20$\,mag. This very high precision in astrometry and photometry allows us to identify discrepancies between observations and isochrone model predictions, even when the differences are small.

The open cluster member stars we use in this work are adopted from \citet[][hereafter CG20]{CantatGaudin2020AA}. Since CG20 membership is based on Gaia\,DR2, we cross-matched all their member stars with Gaia\,DR3 using a $1^{\prime\prime}$ radius. For each member star, we use the code provided by the Gaia team\footnote{\url{https://gitlab.com/icc-ub/public/gaiadr3_zeropoint}} to correct the parallax zero point. We conducted 100,000 Gaussian distributed random samplings using the corrected-parallax and its error. Each sampled corrected-parallax was used to calculate distances through $\text{Distance}/\text{pc} = 1000/\text{parallax}$. We then computed the median and standard deviation of these 100,000 distance values, which were taken as the distance and its error of the star, respectively. 

\subsection{Isochrone models}
\label{sec:models_information}

Isochrones from two stellar evolutionary models are used in this work, they are both publicly available from their webpages. Here we introduce the settings we adopt to query the isochrones.

The MIST\footnote{\url{http://waps.cfa.harvard.edu/MIST/index.html}} model, developed by the Modules for Experiments in Stellar Astrophysics \citep[MESA,][]{Paxton2011ApJS, Paxton2013ApJS, Paxton2015ApJS} project, encompasses stellar evolutionary tracks and isochrones. In this study, we utilize MIST isochrones without rotation, adopting solar-scaled abundance ratios with a metallicity of $Z_{\sun}=0.0142$ \citep{Asplund2009ARAA}. These isochrones cover a wide range of ages ($5\,\leq\,\lg(\tau/\text{yr})\,\leq\,10.3$), masses ($0.1\,\leq\,M/M_{\sun}\,\leq\,300$), and metallicities ($-2.0\,\leq\,[Z/H]\,<\,0.5$). The bolometric corrections used by MIST models are those derived from the PHOENIX models \citep{Hauschildt1999a, Hauschildt1999b} (span the effective temperature $2000\,K \leq \text{$T_{eff}$} \leq 10000\,K$ and the gravity $-0.5 \leq \text{log($g$)} \leq 5.5$) and the COND model \citep{Allard2001} (span the effective temperature $\text{$T_{eff}$} < 3000\,K$).

The PARSEC (the PAdova \& TRieste Stellar Evolution Code) isochrones used in this work are retrieved from the CMD web interface v3.7\footnote{\url{http://stev.oapd.inaf.it/cgi-bin/cmd}}. These isochrones are produced from PARSEC version 1.2S stellar tracks without rotation, which covers the metallicities [M/H] range from $-2.2$ to 0.5\,dex and the mass range from 0.1 to 350\,\Msun. The corresponding bolometric corrections are produced with the default settings of YBC\footnote{\url{https://sec.center/YBC/}} \citep{Chen2019}. Concerning the intermediate and low mass stars focused by this work, ATLAS9 spectral libraries \citep{Castelli2003} are used for the temperature range of $\text{$T_{eff}$} > 6500\,K$, PHOENIX BT-Settl models \citep{Allard2012} are used for cooler stars with $\text{$T_{eff}$} < 5500\,K$, a smooth interpolation of these two sets of atmosphere models are applied for stars with $6500\,K >\text{$T_{eff}$} > 5500\,K$).

\begin{figure} 
    \centering
    {%
    \includegraphics[width=0.22\textwidth]{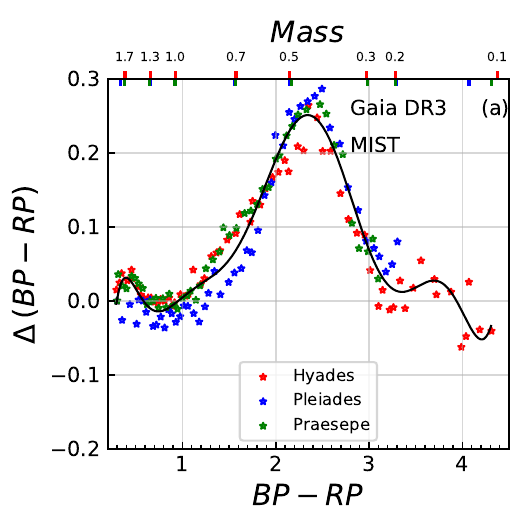}\quad
    \includegraphics[width=0.22\textwidth]{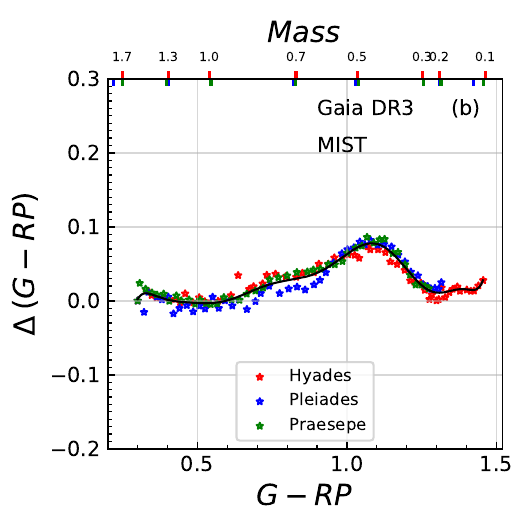}}

    {%
    \includegraphics[width=0.22\textwidth]{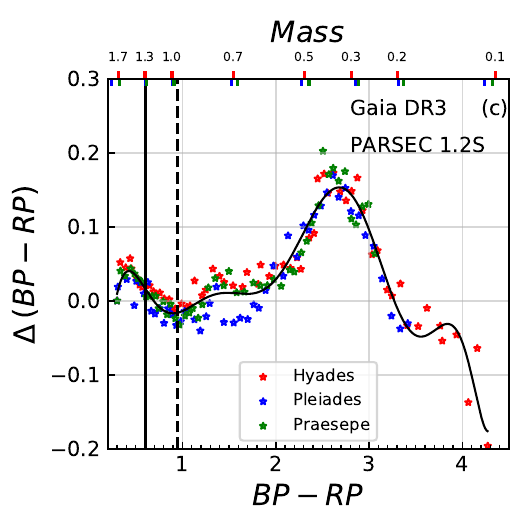}\quad
    \includegraphics[width=0.22\textwidth]{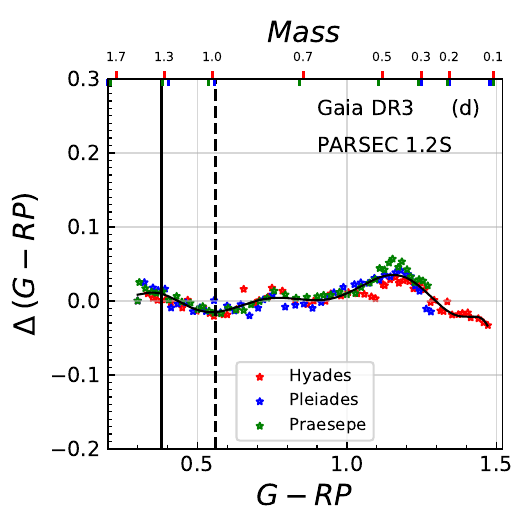}}
    
    \caption{The distribution of the color deviation between the median color value in each color bin and that of the best fitted isochrones obtained from models MIST and PARSEC 1.2S Red. The color points correspond to the three bench-marked clusters. Color coding is identical to Figure~\ref{fig:delta_3OCs_literas}. The solid black lines represent the trends of color deviations fitted using polynomial regression. Additionally, the colors corresponding to a temperature of approximately 5500 and 6500\,K are indicated by the dashed and solid line in (c) and (d) panels, respectively.
    }
    \label{fig:delta_3ocs_bp-g_rp_dr3_mist/parsec}
\end{figure}

In this work, the ranges of age and metallicity for the isochrones used are as follows: 
For both MIST and PARSEC models, we select isochrones with age ranging from $\lg(\tau/\text{yr})=6$ to $10$ and a step of $\lg(\Delta\tau/\text{yr})=0.07$, as well as metallicitiy ($[Fe/H]$ or $[M/H]$) ranging from $-0.05$ to $0.25$\,dex. and a step of $\Delta[M/H]$=0.05 dex.

\section{Benchmark clusters} 
\label{sec:benchmark_isochrones}
As a well-defined single stellar population that formed in the same molecular cloud at the same time, open clusters have  long been considered a good calibrator for stellar evolution and atmosphere models. The loci of an open cluster's member stars on the CMD provide a laboratory for testing the agreement between theoretical predictions and observations.

We select Hyades, Pleiades, and Praesepe as our benchmark clusters to quantify the color deviation on CMD between Gaia DR3 observations and isochrone model predictions.
These  clusters are located relatively close to the Sun, with distances of approximately 40$-$50\,pc, 130$-$140\,pc, and 180$-$190\,pc, respectively. They all show a clear single star sequence on CMD and are minimally affected by extinction \citep[e.g., ][]{Bell2014MN3496B, Kounkel2019AJ122K, CantatGaudin2020AA}. Previous literature provides age and metallicity estimates obtained through various calculation methods, as well as multi-band photometry and high-precision spectroscopic data. Table\,\ref{table:summa_litare} includes the ages and metallicities of the three clusters. According to the literature, the metallicity of the Hyades cluster ranges from 0 to 0.28\,dex, with isochronal ages spanning from 500\,Myr to 1\,Gyr, and a Lithium Depletion Boundary (LDB) age of around 650\,Myr. The Pleiades cluster exhibits a metallicity range of $-0.039$ to 0.1\,dex, isochronal ages ranging from 50 to 141\,Myr, and LDB ages ranging from 112 to 151\,Myr. As for the Praesepe cluster, its metallicity falls between 0.08 and 0.27, with isochronal ages ranging from 600 to 794\,Myr, and an LDB age of around 700\,Myr. 

Member stars of the three benchmark clusters are from the CG20 catalog as described in Sec\,\ref{sec:gaiadr3}. To obtain a single stellar sequence, we identify the main sequence ridge line of the cluster through number density and remove these stars with low photometric quality, where the color errors ($\sigma_{BP-RP}$ and $\sigma_{G-RP}$) exceed  0.05 mag, and manually remove potential binaries, whose absolute magnitudes ($M_G$) differ from the main sequence ridge line by more than 0.4\,mag, as well as field stars that are more than 0.4\,mag below the ridge line. We also remove the white dwarfs. 
Using this method, we removed 239(25\%), 130(14\%), and 62(9\%) potential binaries, field stars and white drawfs from the three clusters, respectively. We also performed de-reddening to  cluster member stars. We utilized the three-dimensional extinction map STructuring by Inversion the Local Interstellar Medium (STILISM\footnote{https://stilism.obspm.fr}; \citet{Lallement2014AA91L, Capitanio2017AA65C, Lallement2018AA132L}) to calculate the color excess, E(B-V), and its error, $\sigma_{E(B-V)}$, for each member in the clusters, and then determined the extinction \( A_{V} = 3.1 \times E(B-V) \) and its propagated error \( \sigma_{A_V} \). Next, by using the spectral data of main sequence stars from the Pickles library \citep{Pickles1998PASP} and the extinction law proposed by \citet{Cardelli1989ApJ} (with \( R_V = 3.1 \)), we obtain the extinction in bands ($A_{BP}$, $A_{RP}$, $A_G$) to calculate the de-reddened colors (BP$-$RP) and (G$-$RP), along with their errors $\sigma_{\text{(BP$-$RP)}}$ and $\sigma_{\text{(G$-$RP)}}$, and the de-reddened absolute magnitude $M_G$ and its error $\sigma_{\text{$M_G$}}$. The extinctions from the clusters are independent on the colors and absolute magnitudes.   

We calculate the median ages and metallicities provided in the references for the three clusters (see Table~\ref{table:summa_litare}), respectively.
We then compare the CMDs of the clusters with the isochrones from the PARSEC 1.2S model, as depicted in Figure\,\ref{fig:3OCs_literas}. From this figure, the discrepancy between the isochrones and the CMD of the clusters at the very low-mass part is evident. Such discrepancies have also been noted in \citet{Brandner2023AJ108B,Brandner2023b, Brandner2023MN}. The reason for the discrepancy is that the model either underestimates the observed stellar luminosity or predicts a bluer stellar color. For the former, \citet{Jeffries2017MN1456J} proposed that for low-mass stars, magnetic field suppression of convection and the presence of numerous cool starspots could lead to stellar radius inflation, resulting in a systematic underestimation of stellar luminosity by the models. They further suggested that incorporating stellar radius inflation in the models could potentially reconcile the isochronal ages with the LDB ages. To address this issue, it may be necessary to reevaluate the stellar evolution models, a task that falls beyond the scope of this work. As for the latter, using empirical color corrections on the model isochrones may be a better approach. This approach allows the model isochrones to match the observations without including complicated physics, i.e., magnetic field and cool starspots in stellar evolution models discussing in \citet{Jeffries2017MN1456J}. In the CMDs (see Figure\,\ref{fig:3OCs_literas}), we binned the $M_{G}$ in intervals of 0.2\,mag. For each bin, we calculated the median color and its difference from the best fitted isochrones. Figure\,\ref{fig:delta_3OCs_literas} shows that the color deviations for the three clusters exhibit a similar trend. Compared with the observations, the model PARSEC 1.2S colors are bluer for 2$\leq$(BP$-$RP)$\leq$3 and redder for (BP$-$RP)$>$3. Though the deviation generally small but much larger than the photometric uncertainties (0.002~mag, \citealt{Gaia2023}). Therefore, empirically correcting the colors can improve the fit between the model isochrones and the CMDs of the clusters.

\section{Empirical Color Corrections}
\label{sec:process_color_correction}
In this section, we demonstrate how to perform empirical color (BP$-$RP and G$-$RP) corrections on the MIST and PARSEC 1.2S models using the three benchmark clusters. 

In previous works, the $M_G$ versus (BP$-$RP) CMD is widely used to fit model isochrones \citep[see e.g.,][]{CantatGaudin2020AA, Kounkel2019AJ122K, Dias2021MN356D, Hunt2024}. However, the Gaia BP band is slightly less reliable compared to the other two bands. As discussed in the Gaia DR2 photometry validation paper \citep{Evans2018}, the Gaia BP magnitude band show a larger calibration uncertainty with respect to RP and G magnitude bands. From the Early DR3 of Gaia, the overall photometry uncertainty for the three passbands is improved  to an even better level compared to the previous data releases, but the BP band is still more affected by calibration issues and is the less good one among the three bands. In this study, we work on CMD based on the widely used colors (BP$-$RP) and (G$-$RP).

We use the model isochrones to fit CMDs of the benchmark clusters. To avoid possible fitting uncertainties introduced by the color deviations on the low-mass part as described before, we select the blue and bright members for the CMD fittings.
As shown in Figures~\ref{fig:delta_3OCs_literas} or \ref{fig:delta_3ocs_bp-g_rp_dr3_mist/parsec}(c), the color deviations between the PARSEC 1.2S isochrones and the cluster are relatively small within the range of (BP$-$RP$)\leq$2\,mag. Similarly, as shown in Figure~\ref{fig:delta_3ocs_bp-g_rp_dr3_mist/parsec}(d), the color deviations are small within the range of (G$-$RP$)\leq$1\,mag. To maintain consistency in the color correction range, we selected cluster members within the color ranges of (BP$-$RP$)\leq$2\,mag and (G$-$RP$)\leq$1\,mag for the first iteration and fitted them with the PARSEC 1.2S and MIST models to obtain the best-fitted isochrones. Although, as shown in Figures~\ref{fig:delta_3ocs_bp-g_rp_dr3_mist/parsec}(a) and (b), the MIST model exhibits noticeable deviations in these color ranges, the effects from these deviations are effectively mitigated after multiple iterations.
Using the method as described in \cite{Liu(2019)}, for each cluster member, we calculate the smallest departure to a model isochrone, and compute the mean departure of all members to that isochrone. Note that we don't apply any weighting during the fitting. The isochrone with the shortest mean departure is taken as the best-fit one, and its age is considered as the age of the cluster. On the CMD, we bin the $M_{G}$ in intervals of 0.2\,mag. For each bin, we calculate the median color and its difference between the cluster members and the best-fitting isochrones. The overall trend of these color differences is consistent for the three clusters. We use the polynomial fitting method to fit the color differences of the clusters and obtain an initial color correction function. 

Next, we apply the initial color correction function obtained above to  the model isochrones and use isochrones with corrected color to refit the CMDs of the clusters. The initial color correction function reduces the color deviations between the model isochrones and the benchmark clusters. Therefore, the range of colors (BP$-$RP) and (G$-$RP) used in this fitting process includes the entire color range of the clusters. We recalculate the mean distances between the color-corrected model isochrones and the cluster members and determine the best-fitting isochrone by the shortest mean distance. We then bin the $M_{G}$ with a 0.2\,mag interval in the CMD. For each bin, we calculate the median color and its difference between the clusters and the best-fitting isochrones. In Figure\,\ref{fig:delta_3ocs_bp-g_rp_dr3_mist/parsec}, we show the color differences between the three clusters and the best-fitting isochrones from the MIST and PARSEC 1.2S models, respectively. In these four subplots, we observe that the trends of the color differences are generally consistent. We then use the polynomial fitting method to refit the differences and obtain an empirical color correction function for the differences. We represent the color-corrected function with black solid lines in Figure\,\ref{fig:delta_3ocs_bp-g_rp_dr3_mist/parsec}. The color-correction function obtained the color deviation ($f$ ) between the model isochrone and the observation is given as follows:
\begin{equation}\label{eq:colors_dr3_models}
f(x) = \sum_{i=0}^{N} k_i x^{N-i},
\end{equation}
where $x$ represents the color (BP$-$RP) or (G$-$RP). The values of the coefficients $k_i$ and the polynomial degree $N$ for MIST and PARSEC 1.2S models are all listed in Table~\ref{table:coefficients_color_corrections}. When fitting the overall trend of color deviation, we determine the best-fitted value of $N$ by comparing the goodness of fit between the original data and the fitted curve data, choosing the smallest integer where the goodness of fit exceeds 98\%.

\setcounter{table}{0}
\setcounter{table}{0}
\begin{table*}[htbp!]
\begin{center}
\caption{The coefficients for the color correction functions in Section\,\ref{sec:process_color_correction}. \label{table:coefficients_color_corrections} }

     \begin{tabular}{ccccc}
        \hline
        \hline 
         
        \multicolumn{1}{c}{\multirow{2}{*}{}} &  \multicolumn{2}{c}{\multirow{1}{*}{MIST}} & \multicolumn{2}{c}{\multirow{1}{*}{PARSEC 1.2S}} \\
        \cline{2-3}
        \cline{4-5}
        
        \multicolumn{1}{c}{} & \multicolumn{1}{c}{\textbf{BP$-$RP}} & \multicolumn{1}{c}{\textbf{G$-$RP}} & \multicolumn{1}{c}{\textbf{BP$-$RP}} & \multicolumn{1}{c}{\textbf{G$-$RP}} \\        
        \hline  $k_0$ & $-$4.84109156662e-03 & 1.66041686674e+03 & 4.43946646259e-03 & $-$1.55281212469e+02   \\
        \cline{1-1}
        \cline{2-5}
                $k_1$ & 1.18790335297e-01 & $-$1.56566075272e+04 & $-$8.99126898611e-02 & 1.19242974533e+03   \\
        \cline{1-1}
        \cline{2-5}
                $k_2$ & $-$1.24788718579e+00 & 6.55134171293e+04 & 7.57462148449e-01 & $-$3.93129271756e+03   \\
        \cline{1-1}
        \cline{2-5}
                $k_3$ & 7.34202485899e+00 & $-$1.60375747677e+05  & $-$3.41507655298e+00 & 7.27822027924e+03   \\
        \cline{1-1}
        \cline{2-5}
                $k_4$ & $-$2.65953606912e+01 & 2.54904982070e+05  & 8.75609327100e+00 & $-$8.30782701688e+03   \\
        \cline{1-1}
        \cline{2-5}
                $k_5$ & 6.14897637911e+01 & $-$2.75935718076e+05 & $-$1.21266013380e+01 & 6.03969724150e+03   \\
        \cline{1-1}
        \cline{2-5}
                $k_6$ & $-$9.11778295307e+01 & 2.07427644403e+05 & 6.13317544359e+00 & $-$2.78554182219e+03   \\
        \cline{1-1}
        \cline{2-5}
                $k_7$ & 8.49214572624e+01 & $-$1.08227497419e+05 & 5.20844674997e+00 & 7.83342648627e+02   \\
        \cline{1-1}
        \cline{2-5}
                $k_8$ & $-$4.69874927204e+01 & 3.84024814235e+04 & $-$8.76070696284e+00 & $-$1.21713434987e+02   \\
        \cline{1-1}
        \cline{2-5}
                $k_9$ & 1.37158168882e+01 & $-$8.82591877952e+03 & 4.12464395718e+00 & 7.97636109336e+00   \\
        \cline{1-1}
        \cline{2-5}
                $k_{10}$ & $-$1.56992702563e+00 & 1.18259672423e+03 & $-$6.05928788398e-01 & 0   \\
        \cline{1-1}
        \cline{2-5}
                $k_{11}$ & 0              & $-$6.99859722054e+01 &   0                  & 0   \\
        \hline
        \hline
    \end{tabular}
\end{center}       
\footnotesize{ Column\,1: the coefficients of the color correction functions. Column\,2$-$3: the coefficients of the color correction functions for the (BP$-$RP) and (G$-$RP) colors, respectively, based on the MIST model. Column\,4$-$5 the coefficients of the color correction functions for the (BP$-$RP) and (G$-$RP) colors, respectively, based on the PARSEC 1.2S model.
} 
\end{table*}

\begin{figure*} 
    \centering
    {%
    \includegraphics[width=0.45\textwidth, trim=0.1cm 0.2cm 0.0cm 1.7cm, clip]{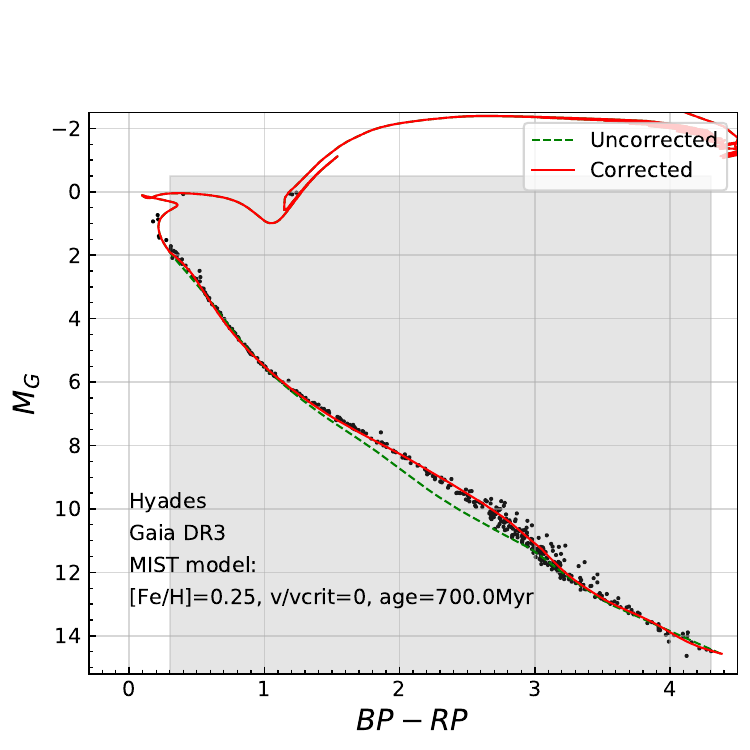}\quad
    \includegraphics[width=0.45\textwidth, trim=0.1cm 0.2cm 0.0cm 1.7cm, clip]{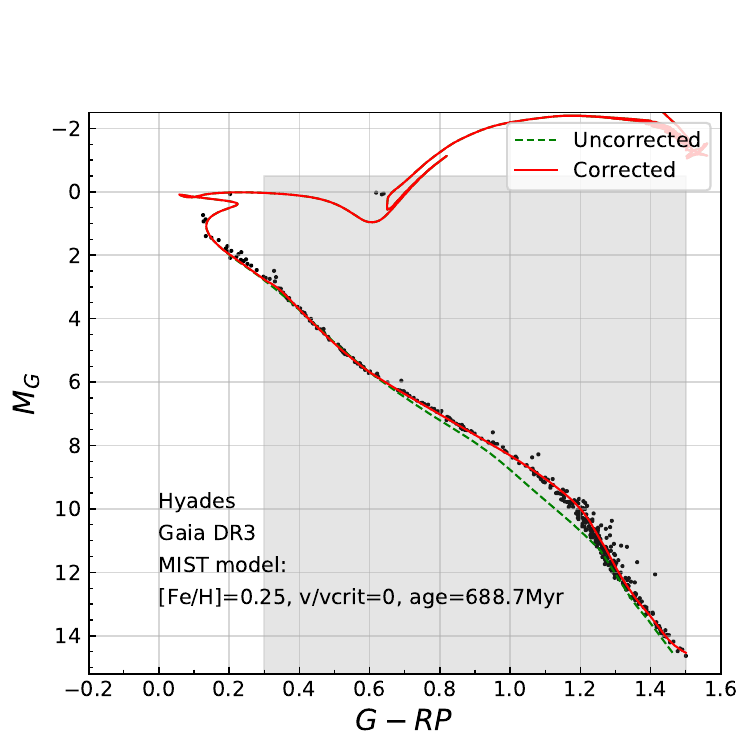}}
    \caption{CMDs of Hyades open cluster used by MIST model. The data are from Gaia\,DR3\citep{Gaia2023}. The member candidates are manually selected because of the existence of likely binary or multiple stars and white dwarf stars, in short, most of them are bonafide single stars (according to the work provided by \citet{Brandner2023MN}).}
    \label{fig:Hyades_bp-g_rp_mist}
\end{figure*}

\begin{figure*} 
    \centering
    {%
    \includegraphics[width=0.45\textwidth, trim=0.1cm 0.2cm 0.0cm 1.7cm, clip]{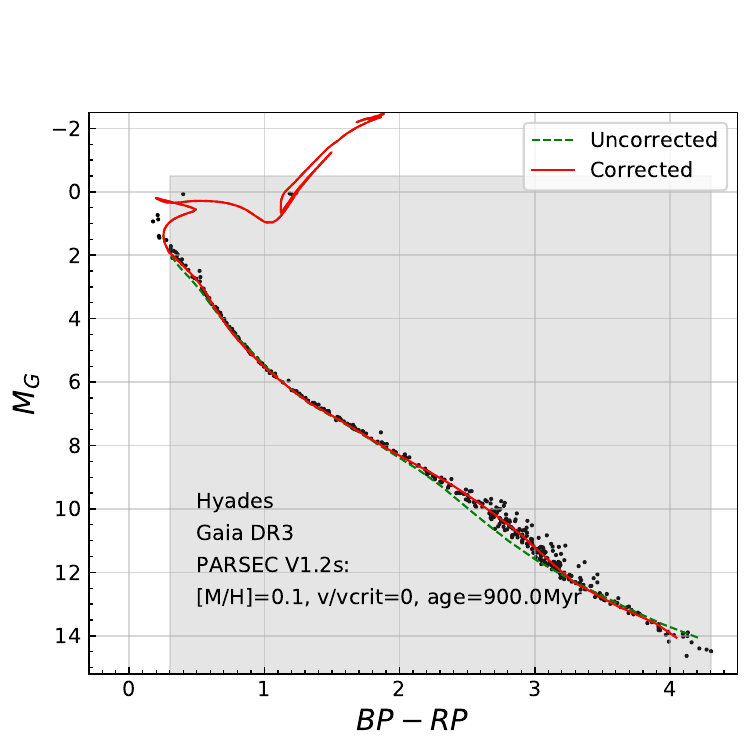}\quad
    \includegraphics[width=0.45\textwidth, trim=0.1cm 0.2cm 0.0cm 1.7cm, clip]{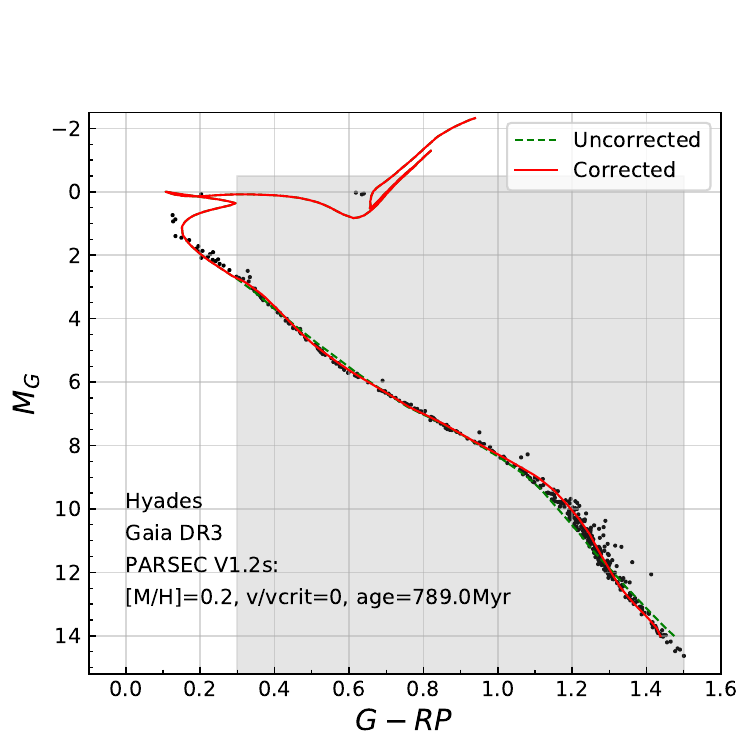}}
    \caption{CMDs of Hyades open cluster used by PARSEC model. Similar to Figure \ref{fig:Hyades_bp-g_rp_mist}.
    }
    \label{fig:Hyades_bp-g_rp_parsec}
\end{figure*}

To correct the model isochrones, we calculate the color correction functions and add them to the model isochrone colors. In Figures\,\ref{fig:Hyades_bp-g_rp_mist} and \ref{fig:Hyades_bp-g_rp_parsec}, as well as Figures\,\ref{fig:Pleiades_bp-g_rp_mist/parsec} to \ref{fig:Praesepe_bp-g_rp_mist/parsec}, we display the comparison of the best-fitting isochrones for the three clusters in CMDs before and after color correction. The uncertainty in age is obtained by performing bootstrap resampling on the colors (BP$-$RP and G$-$RP) and the absolute magnitude ($M_G$), selecting only 50\% of the data in each sample. This process is repeated 1000 times, and the 68\% quantile of the results is calculated. We summarize these results in Table\,\ref{table:bp-g_rp_mist/parsec}.

\begin{table*}[htbp!]
\begin{center}
\caption{Ages and metallicities of the Hyades, Pleiades and Praesepe open clusters calculated from the perspective of $M_{G}$ versus (BP$-$RP) and (G$-$RP) based on the MIST and PARSEC 1.2S isochrone models.} 
\label{table:bp-g_rp_mist/parsec} 

    \begin{tabular}{ccccccccc}
        \hline
        \hline 
         
        \multicolumn{1}{c}{\multirow{3}{*}{Cluster}} &  \multicolumn{4}{c}{\multirow{1}{*}{MIST}} & \multicolumn{4}{c}{\multirow{1}{*}{PARSEC 1.2S}} \\
        \cline{2-5}
        \cline{6-9}
        
        \multicolumn{1}{c}{} & \multicolumn{2}{c}{\textbf{BP$-$RP}} & \multicolumn{2}{c}{\textbf{G$-$RP}} & \multicolumn{2}{c}{\textbf{BP$-$RP}} & \multicolumn{2}{c}{\textbf{G$-$RP}} \\
        \cline{2-9}
        \multicolumn{1}{c}{} & $[Fe/H]$ & Age/Myr & $[Fe/H]$ & Age/Myr & $[M/H]$ & Age/Myr & $[M/H]$ & Age/Myr  \\
        
        \hline  Hyades  & $0.25\pm0.0$ & $700^{+75}_{-142}$ & $0.25\pm0.0$ & $688.7\pm100$ & $0.10^{+0.05}_{-0.0}$ & $900^{+15}_{-162}$ & $0.20\pm0.05$ & $789^{+98}_{-62}$  \\
        \cline{1-1}
        \cline{2-9}
                Pleiades & $0.25\pm0.0$ & $206^{+16}_{-66}$ & $0.25\pm0.0$ & $202.7^{+13}_{-65}$ & $0.10^{+0.0}_{-0.05}$ & $152\pm27$ & $0.10^{+0.0}_{+0.05}$ & $133\pm13$  \\
        \cline{1-1}
        \cline{2-9}
                Praesepe & $0.25\pm0.05$ & $656^{+116}_{-90}$ & $0.25\pm0.0 $& 666.7 & $0.15\pm0.0$ & $738^{+47}_{-91}$ & $0.15\pm0.05$ & $776^{+73}_{-107}$  \\
        \hline
        \hline
    \end{tabular} 
\end{center}       
\footnotesize{Column\,1:the names of the open clusters. Columns 2-3: Best-fit age and metallicity obtained by fitting isochrones from the MIST model with (BP$-$RP) isochrone colors.
Columns 4-5: Best-fit age and metallicity obtained by fitting isochrones from the MIST model with (G$-$RP) isochrone colors.
Columns 6-7: Best-fit age and metallicity obtained by fitting isochrones from the PARSEC 1.2S model with (BP$-$RP) isochrone colors.
Columns 7-8: Best-fit age and metallicity obtained by fitting isochrones from the PARSEC 1.2S model with (G$-$RP) isochrone colors. 
} 
\end{table*}

\section{Application to other stellar systems}
\label{sec:result-disc}
To check whether the color deviation of the original isochrones affects age determinations, in this section we apply the empirical color correction functions obtained in Section \ref{sec:process_color_correction} to fit CMDs of other star clusters and moving groups.
We then compare and discuss these results with isochronal ages and LDB ages provided in the literature.

\subsection{Application for color correction}
\label{sec:application_color_correction}
We selected 31 open clusters from CG20 with low extinction $A_V$, located at relatively short distances (approximately less than 1 kpc), and with a substantial number of member stars (more than 150). Additionally, we chose three young moving groups, namely $\beta$ Pictoris(BPMG), 32 Ori(THOR) \citep{Gagne2018ApJ}, and Tucana-Horologium moving group(THMG) \citep{Galli2023MNRAS}, which are also nearby (within 200 pc). The ages and metallicities of these 31 clusters and 3 moving groups were obtained from these literature, where the ages were derived either from isochrone fitting across multiple photometric bands or from LDB, and the metallicities were either derived from isochrone fitting or from spectroscopic data. According to the literature, the ages of these clusters and movig groups range from a few million years to over 1 Gyr, with metallicities spanning from $-0.1$ to 0.15.

We cross-matched the CG20 catalog members of these 31 clusters and 3 moving groups with Gaia\,DR3 within a radius of 1$^{\prime\prime}$ and applied reddening correction. We obtained the dereddened colors (BP$-$RP) and (G$-$RP), along with their errors $\sigma_{\text{(BP$-$RP)}}$ and $\sigma_{\text{(G$-$RP)}}$, as well as the dereddened absolute magnitude $M_G$ and its error $\sigma_{\text{$M_G$}}$ for each member. Following the method outlined in Section\,\ref{sec:benchmark_isochrones}, we used the empirically color-corrected isochrones obtained from Section\,\ref{sec:process_color_correction} to fit the CMDs of these clusters and moving groups and calculated the mean distances between the model isochrones and the members, and then determine the best-fitting isochrone by the shortest mean distance. 
For each star cluster and moving group, we perform same method described in Section~\ref{sec:process_color_correction} to calculate age and its uncertainty.

In Figures\,\ref{fig:IC2391_bp/g-rp_dr3_parsec/mist} and others in Appendix\,\ref{sec:addtional CMDs}, we show the isochrones before and after color correction with the distributions of stars in the clusters and associations in CMDs. From these figures, we can see that the isochrones after the empirical color correction align significantly  better with the observational CMDs of the clusters and moving groups. Only a few clusters exhibit less noticeable contrasts due to the lack of member stars in the low-mass regime. The ages and metallicities recomputed in this study for the clusters and associations are summarized in Table\,\ref{table:OCs_ThisWork}.

\setcounter{table}{0}
\setcounter{table}{2}
\begin{table*}[htbp!]
\begin{center}
\caption{Ages and metallicities of the 31 clusters and 3 moving groups calculated from the perspective of $M_{G}$ versus (BP$-$RP) and (G$-$RP) based on the MIST and PARSEC 1.2S isochrone models.}
\label{table:OCs_ThisWork} 

    \begin{tabular}{ccccccccc}
        \hline
        \hline 
         
        \multicolumn{1}{c}{\multirow{3}{*}{Cluster}} &  \multicolumn{4}{c}{\multirow{1}{*}{MIST}} & \multicolumn{4}{c}{\multirow{1}{*}{PARSEC 1.2S}} \\
        \cline{2-5}
        \cline{6-9}
        
        \multicolumn{1}{c}{} & \multicolumn{2}{c}{\textbf{BP$-$RP}} & \multicolumn{2}{c}{\textbf{G$-$RP}} & \multicolumn{2}{c}{\textbf{BP$-$RP}} & \multicolumn{2}{c}{\textbf{G$-$RP}} \\
        \cline{2-9}
        \multicolumn{1}{c}{} & $[Fe/H]$ & Age/Myr & $[Fe/H]$ & Age/Myr & $[M/H]$ & Age/Myr & $[M/H]$ & Age/Myr  \\
        
        \hline  
                NGC\,752  & $0.10^{+0.05}_{-0.0}$ & $1176^{+163}_{-19}$ & $0.10\pm0.10$ & $1235^{+529}_{-83}$ & $-0.05^{+0.1}_{-0.05}$ & $1382^{+110}_{-151}$ & $-0.05^{+0.1}_{-0.05}$ & $1405^{+359}_{-378}$  \\
        \cline{1-1}
        \cline{2-9}
                NGC 7092  & $0.10\pm0.0$ & $332^{+120}_{-11}$ & $0.10\pm0.10$ & $297^{+141}_{-94}$ & $0.00^{+0.05}_{-0.05}$ & $335^{+46}_{-69}$ & $0.00^{+0.0}_{-0.05}$ & $340^{+118}_{-6}$  \\
        \cline{1-1}
        \cline{2-9}
                NGC 2516  & $0.25\pm0.05$ & $656^{+58}_{-86}$ & $0.25\pm0.05$ & $667\pm28$ & $0.15\pm{0.05}$ & $738^{+24}_{-52}$ & $0.15^{+0.05}_{-0.0}$ & $776^{+105}_{-46}$  \\
        \cline{1-1}
        \cline{2-9}  
                Blanco 1  & $0.15\pm0.05$ & $101^{+17}_{-21}$  & $0.15\pm0.10$  & $99\pm32$  & $0.00^{+0.05}_{-0.00}$  & $135^{+14}_{-15}$  & $0.00^{+0.05}_{-0.0}$  & $117^{+19}_{-7}$   \\
        \cline{1-1}
        \cline{2-9}
                NGC 2451B  & $0.25^{+0.0}_{-0.05}$  & $38^{+4}_{-1}$  & $0.20\pm0.05$  & $36\pm12$  & $0.00^{+0.1}_{-0.05}$  & $41^{+9}_{-7}$  & $0.00^{+0.1}_{-0.05}$  & $33^{+5}_{-3}$   \\
        \cline{1-1}
        \cline{2-9}
                IC 2602  & $0.25^{+0.0}_{-0.05}$  & $40^{+8}_{-1}$  & $0.25\pm0.05$  & $43\pm14$  & $0.10^{+0.1}_{-0.05}$  & $59^{+8}_{-1}$  & $0.20\pm0.05$  & $61^{+1}_{-6}$   \\
        \cline{1-1}
        \cline{2-9}
                IC 2391  & $0.20\pm0.05$  & $44^{+8}_{-1}$  & $0.20\pm0.05$  & $43\pm14$  & $0.00^{+0.15}_{-0.05}$  & $59^{+11}_{-3}$  & $0.10\pm0.05$  & $54^{+4}_{-9}$   \\
        \cline{1-1}
        \cline{2-9}
                NGC 2232  & $0.15^{+0.10}_{-0.0}$  & $27^{+4}_{-1}$  & $0.15\pm0.10$  & $26\pm9$  & $-0.05\pm{0.05}$  & $32^{+1}_{-2}$  & $-0.05\pm0.05$  & $28^{+1}_{-3}$   \\
        
        \cline{1-1}
        \cline{2-9}
                Pozzo 1  & $0.25^{+0.0}_{-0.05}$  & $16\pm1$  & $0.25\pm0.05$  & $15\pm5$  & $-0.05^{+0.0}_{-0.1}$  & $14\pm1$  & $-0.05\pm0.05$  & $12\pm1$   \\
        \cline{1-1}
        \cline{2-9}
                Alessi 3  & $0.10^{+0.05}_{-0.0}$  & $760^{+12}_{-228}$  & $0.1\pm0.10$  & $615^{+249}_{-222}$  & $0.00^{+0.0}_{-0.05}$  & $738^{+77}_{-70}$  & $0.00^{+0.1}_{-0.0}$  & $703^{+73}_{-189}$   \\
        \cline{1-1}
        \cline{2-9}
                ASCC 101  & $0.15\pm0.05$  & $397^{+20}_{-70}$  & $0.20\pm0.05$  & $366^{+165}_{-72}$  & $0.00^{+0.05}_{-0.0}$  & $450^{+61}_{-127}$  & $0.05\pm0.05$  & $450^{+72}_{-68}$   \\
        \cline{1-1}
        \cline{2-9}
                Alessi 9  & $0.15\pm0.05$  & $283^{+5}_{-67}$  & $0.20\pm0.05$  & $372^{+177}_{-66}$  & $0.00\pm0.05$  & $340^{+132}_{-17}$  & $0.10^{+0.0}_{-0.05}$  & $382^{+141}_{-48}$   \\
        \cline{1-1}
        \cline{2-9}
                ASCC 41  & $0.15\pm0.05$  & $227^{+11}_{-24}$  & $0.15\pm0.10$  & $227\pm85$  & $0.00^{+0.0}_{-0.05}$  & $218^{+48}_{-71}$  & $0.00^{+0.05}_{-0.0}$  & $357^{+57}_{-100}$   \\
        \cline{1-1}
        \cline{2-9}
                NGC 2422  & $0.25^{+0.0}_{-0.05}$  & $104^{+1}_{-14}$  & $0.25\pm0.05$  & $104\pm34$  & $0.10\pm0.05$  & $125^{+04}_{-10}$  & $0.15^{+0.0}_{-0.05}$  & $165^{+5}_{-50}$   \\
        \cline{1-1}
        \cline{2-9}
                Teutsch 35  & $0.25^{+0.0}_{-0.05}$  & $113^{+36}_{-30}$  & $0.25\pm0.05$  & $200^{+2}_{-3}$  & $0.10^{+0.05}_{-0.0}$  & $127^{+58}_{-10}$  & $0.15^{+0.05}_{-0.0}$  & $222^{+108}_{-15}$   \\
        \cline{1-1}
        \cline{2-9}
                UPK 612  & $0.20^{+0.05}_{-0.0}$  & $113^{+14}_{-22}$  & $0.25\pm0.05$  & $142^{+54}_{-46}$  & $0.00^{+0.05}_{-0.05}$  & $123^{+28}_{-11}$  & $0.10^{+0.05}_{-0.0}$ &  $176^{+73}_{-32}$  \\
        \cline{1-1}
        \cline{2-9}
                Roslund 6  & $0.15\pm0.05$  & $99^{+2}_{-20}$  & $0.20\pm0.05$  & $187^{+13}_{-9}$  & $-0.05^{+0.05}_{-0.05}$  & $106^{+36}_{-17}$  & $0.05^{+0.05}_{-0.0}$  & $157^{+72}_{-19}$   \\
        \cline{1-1}
        \cline{2-9}
                Platais 9  & $0.20^{+0.05}_{-0.0}$  & $59^{+2}_{-6}$  & $0.25\pm0.05$  & $65\pm21$  & $0.05^{+0.05}_{-0.0}$  & $75^{+5}_{-7}$  & $0.10^{+0.05}_{-0.0}$  & $71^{+5}_{-6}$   \\
        \cline{1-1}
        \cline{2-9}
                NGC 2451A  & $0.20^{+0.05}_{-0.0}$  & $52^{+2}_{-30.1}$  & $0.20\pm0.05$  & $49\pm16$  & $0.00^{+0.05}_{-0.0}$  & $62^{+4}_{-3}$  & $0.05^{+0.10}_{-0.0}$  & $59\pm5$   \\
        \cline{1-1}
        \cline{2-9}
                RSG 5   & $0.25\pm0.0$  & $49^{+1}_{-2}$  & $0.25\pm0.05$  & $49\pm16$  & $0.00^{+0.1}_{-0.0}$  & $49^{+9}_{-2}$  & $0.10\pm0.05$  &  $53^{+4}_{-7}$  \\
        \cline{1-1}
        \cline{2-9}
                Trumpler 10  & $0.25\pm0.0$  & $52\pm1$  & $0.25\pm0.05$  & $51\pm16$  & $0.10^{+0.05}_{-0.0}$  & $62^{+2}_{-3}$  & $0.15^{+0.0}_{-0.05}$  & $62^{+2}_{-7}$   \\
        \cline{1-1}
        \cline{2-9}
                NGC 2547  & $0.25^{+0.0}_{-0.05}$  & $42^{+1}_{-5}$  & $0.20\pm0.05$  & $36\pm12$  & $-0.05^{+0.05}_{-0.0}$  & $37\pm3$  & $0.00^{+0.10}_{-0.0}$  &  $35^{+5}_{-2}$  \\
        \cline{1-1}
        \cline{2-9}
                BH 164  & $0.25^{+0.05}_{-0.0}$  & $42^{+1}_{-5}$  & $0.25\pm0.05$  & $37\pm12$  & $0.00^{+0.1}_{-0.0}$  & $40^{+7}_{-2}$  & $0.10^{+0.1}_{-0.0}$  & $40^{+8}_{-3}$   \\
        \cline{1-1}
        \cline{2-9}
                NGC 3228  & $0.20\pm0.05$  & $34^{+3}_{-2}$  & $0.25\pm0.05$  & $36\pm12$  & $0.05^{+0.1}_{-0.05}$  & $36\pm6$  & $0.15^{+0.0}_{-0.05}$  & $33\pm3$   \\
        \cline{1-1}
        \cline{2-9}
                Col 140  & $0.25^{+0.0}_{-0.05}$  & $42^{+1}_{-4}$  & $0.20\pm0.05$  & $38\pm12$  & $-0.05^{+0.15}_{-0.0}$  & $37^{+13}_{-3}$  & $-0.05^{+0.15}_{-0.0}$  &  $35^{+14}_{-3}$  \\
        \cline{1-1}
        \cline{2-9}
                Col 135  & $0.25^{+0.0}_{-0.05}$  & $44^{+1}_{-5}$  & $0.20\pm0.05$  & $38\pm12$  & $0.00^{+0.1}_{-0.0}$  & $44^{+6}_{-4}$  & $0.10^{+0.05}_{-0.0}$  & $47^{+6}_{-5}$   \\
        \cline{1-1}
        \cline{2-9}
                UBC 17b  & $0.15\pm0.05$  & $10\pm3$  & $0.25\pm0.05$  & $12\pm4$  & $-0.05^{+0.05}_{-0.0}$  & $10^{+1}_{-2}$  & $0.10^{+0.10}_{-0.15}$  & $9^{+3}_{-2}$   \\
        \cline{1-1}
        \cline{2-9}
                ASCC 21  & $0.25^{+0.0}_{-0.1}$  & $14^{+1}_{-3}$  & $0.25\pm0.05$  & $13\pm4$  & $-0.05\pm0.0$  & $12\pm1$  & $0.00\pm0.05$  & $13^{+1}_{-4}$   \\
        \cline{1-1}
        \cline{2-9}
                NGC 2281  & $0.20^{+0.05}_{-0.0}$  & $490^{+87}_{-158}$  & $0.20\pm0.05$  & $423^{+191}_{-108}$  & $0.00^{+0.05}_{-0.0}$  & $637^{+43}_{-155}$  & $0.05^{+0.0}_{-0.05}$  & $616^{+67}_{-135}$   \\
        \cline{1-1}
        \cline{2-9}
                NGC 1960  & $0.20^{+0.05}_{-0.1}$  & $21^{+1}_{-2}$  & $0.20\pm0.05$  & $21\pm7$  & $0.15^{+0.0}_{-0.05}$  & $21^{+1}_{-2}$  & $0.15\pm0.10$  &  $21^{+1}_{-4}$  \\
        \cline{1-1}
        \cline{2-9}
                IC 4665  & $0.25\pm0.0$  & $41^{+2}_{-5}$  & $0.25\pm0.05$  & $38\pm13$  & $0.15^{+0.1}_{-0.05}$  & $47^{+11}_{-7}$  & $0.25^{+0.0}_{-0.05}$  &  $48^{+2}_{-9}$  \\
        \cline{1-1}
        \cline{2-9}
                BPMG  & $0.10\pm0.05$  & $21\pm3$  & $0.25^{+0.0}_{-0.05}$  & $25\pm5$  & $-0.05^{+0.10}_{-0.0}$  & $23^{+2}_{-3}$  & $-0.05^{+0.05}_{-0.0}$  &  $19\pm3$  \\
        \cline{1-1}
        \cline{2-9}
                THOR  & $0.25^{+0.0}_{-0.05}$  & $22\pm4$  & $0.25^{+0.0}_{-0.05}$  & $19^{+1}_{-3}$  & $-0.05^{+0.05}_{-0.0}$  & $20\pm3$  & $-0.05^{+0.05}_{-0.0}$  & $15^{+4}_{-3}$   \\
        \cline{1-1}
        \cline{2-9}
                THMG  & $0.25^{+0.0}_{-0.05}$  & $48^{+1}_{-5}$  & $0.25^{+0.0}_{-0.05}$  & $50\pm5$  & $0.15\pm0.05$  & $55^{+2}_{-6}$  & $0.15^{+0.05}_{-0.0}$  &  $55\pm5$  \\
        \hline
        \hline
    \end{tabular} 
\end{center}       
\footnotesize{Column\,1:the names of the open clusters. Columns 2-3: Best-fit age and metallicity obtained by fitting isochrones from the MIST model with (BP$-$RP) isochrone colors.
Columns 4-5: Best-fit age and metallicity obtained by fitting isochrones from the MIST model with (G$-$RP) isochrone colors.
Columns 6-7: Best-fit age and metallicity obtained by fitting isochrones from the PARSEC 1.2S model with (BP$-$RP) isochrone colors.
Columns 7-8: Best-fit age and metallicity obtained by fitting isochrones from the PARSEC 1.2S model with (G$-$RP) isochrone colors. 
} 
\end{table*}

\begin{figure} 
    \centering
    {%
    \includegraphics[width=0.45\textwidth,trim=0.1cm 0.2cm 0.0cm 1.0cm, clip]{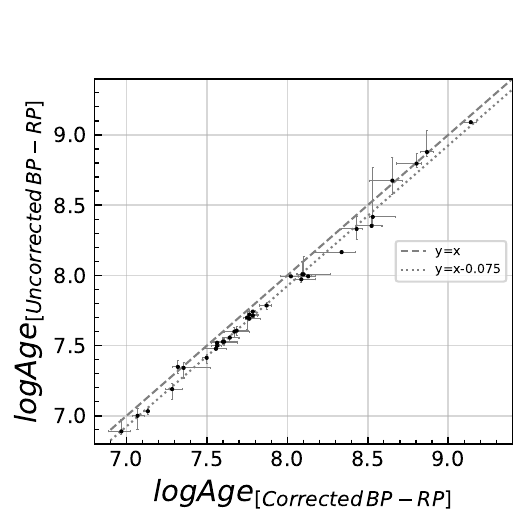}
    }
    \caption{Comparison between the ages derived from the corrected color (BP$-$RP) model PARSEC 1.2S isochrones and the original color model isochrones for 31 star clusters and 3 moving groups. The dashed line represents a 1:1 ratio, while the dot-dashed lines are offset by 0.075~dex from the dashed line.
    }
    \label{fig:34OCs_uncorrect-correct}
\end{figure}

We derive the cluster ages from isochrones without corrections using the method described in Section\,\ref{sec:process_color_correction}, and compare them with the ones obtained with the corrections in Figure~\ref{fig:34OCs_uncorrect-correct}. The ages obtained without correction are generally younger. This is especially evident for the clusters with the low-mass stars ($\leq$0.6\,\Msun). As shown in Figure~\ref{fig:delta_3OCs_literas}, When (BP$-$RP)$\geq$2mag, the isochrones tend to be bluer than the observations. Due to this effect, the isochrone fits including the low-mass stars result in the younger ages by 0.075~dex in log(age) than the ones from the corrected isochrone. For the clusters without the low-mass stars, the two sets of ages argee with each other within 0.01~dex.

\begin{figure} 
    \centering

    {%
    \includegraphics[width=0.22\textwidth, trim=0.1cm 0.9cm 0.0cm 1.6cm, clip]{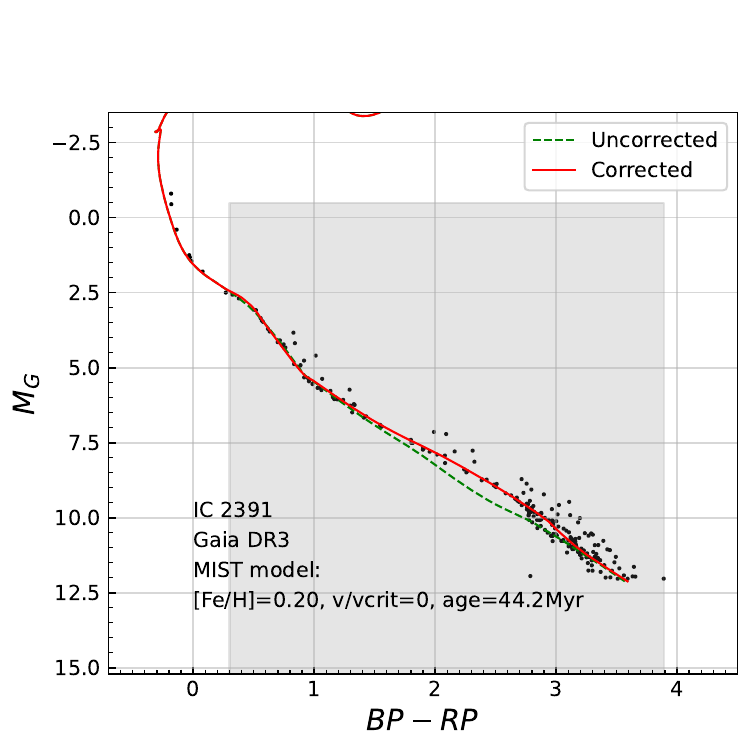}\quad
    \includegraphics[width=0.22\textwidth, trim=0.1cm 0.9cm 0.0cm 1.6cm, clip]{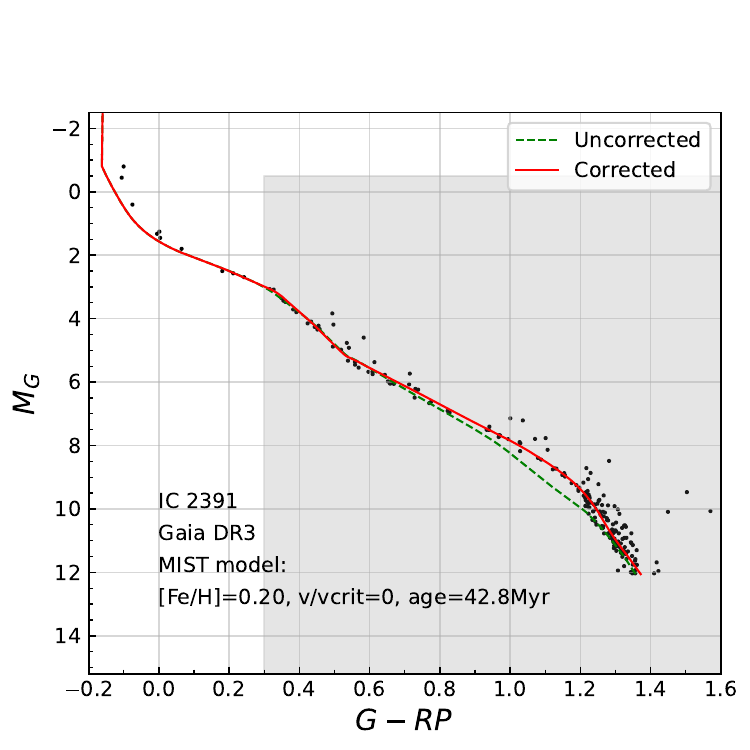}
    }
    {%
    \includegraphics[width=0.22\textwidth, trim=0.1cm 0.2cm 0.0cm 1.7cm, clip]{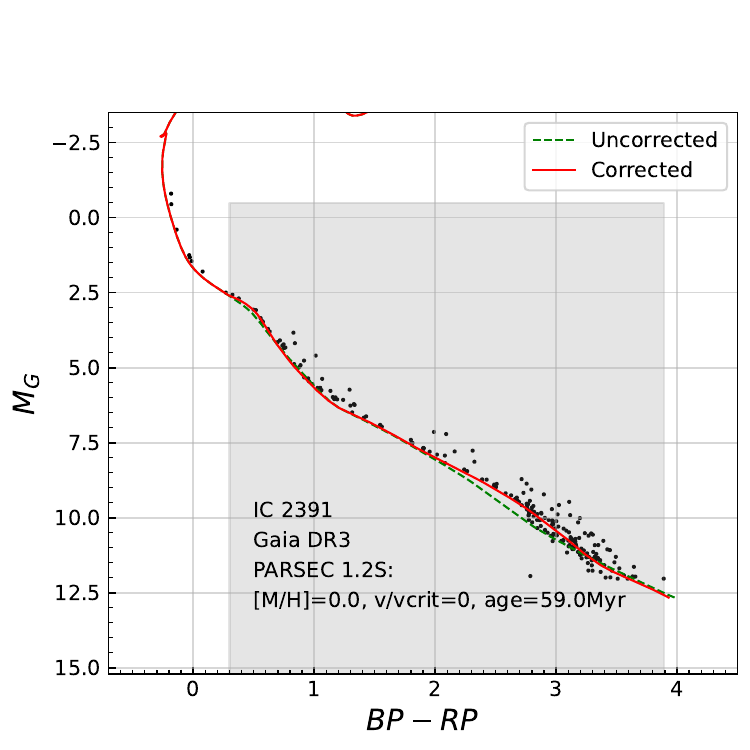}\quad
    \includegraphics[width=0.22\textwidth, trim=0.1cm 0.2cm 0.0cm 1.7cm, clip]{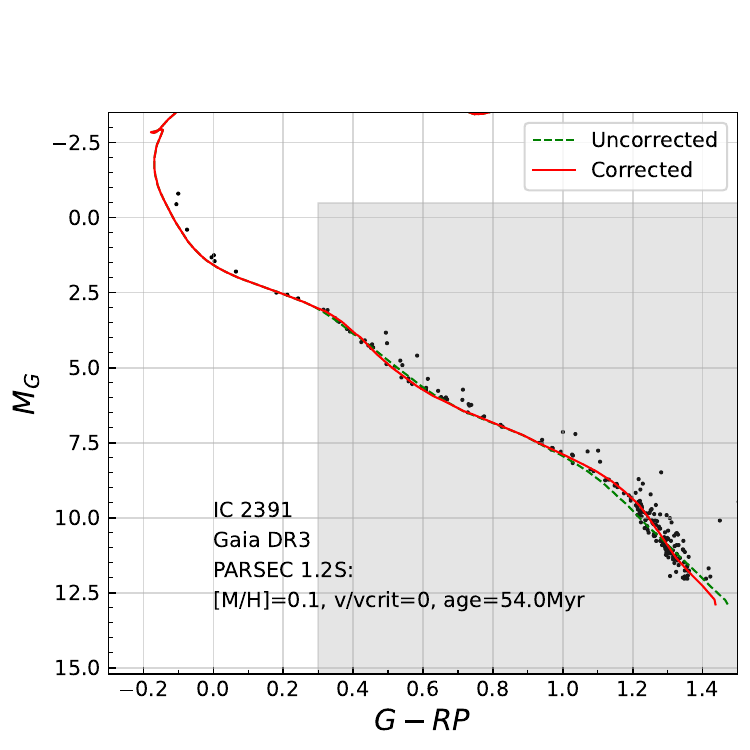}
    }
    \caption{CMDs of IC\,2391.
    }
    \label{fig:IC2391_bp/g-rp_dr3_parsec/mist}
\end{figure}

It worth noting that the age determination, no matter which method is applied, is highly sensitive to the quality of data and the physics detail of models. The isochrone fitting method on CMDs is particularly good with non-negligible uncertainties, especially for young stellar populations with $\lesssim$ 50\,Myr. For stars with this young age, it is very challenging to model their magnetic field, as well as the bolometric corrections with starspots and accretion disc, which affect the stellar color and brightness. Therefore, in this work we can only provide a rough estimate of the age parameter and use it to assess the formation and evolutionary status of star clusters or moving groups with very young age.

\subsection{Comparison with Ages from Other Literature}
\label{sec:Compa_with_others}

In this subsection, we compare the ages and metallicities of the clusters and associations obtained from this work with those provided in the literature (see Table \ref{table:summa_litera_10OCs}).

\begin{figure} 
    \centering
    {%
    \includegraphics[width=0.45\textwidth, trim=0.1cm 0.2cm 0.0cm 1.0cm, clip]{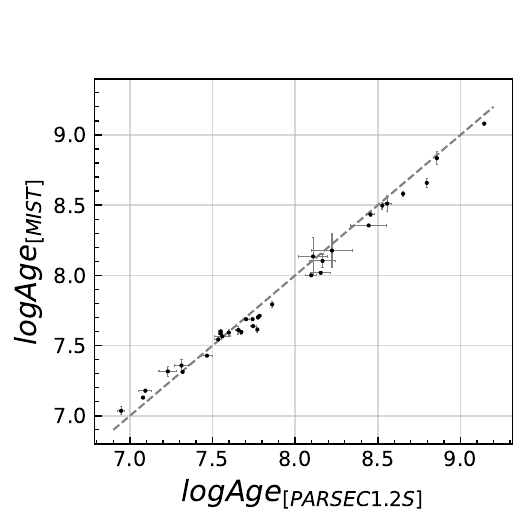}
    }
    \caption{Comparison between the ages obtained from the PARSEC 1.2S model and those obtained from the MIST model for the 31 clusters and 3 moving groups in this work. The grey dashed line represents the ratio 1:1, as in Figure~\ref{fig:34OCs_uncorrect-correct}.
    }
    \label{fig:thiswork_parsec-vs-mist}
\end{figure}

In Figure \ref{fig:thiswork_parsec-vs-mist}, we compare the ages from the MIST and PARSEC 1.2S isochrones. The ages and their uncertainties for each model are  the mean and the dispersion of the isochronal ages from the two CMDs (see Section \ref{sec:application_color_correction}), G vs. (G$-$RP) and G vs. (BP$-$RP). From this figure, we observe that for clusters and associations with $\log\text{Age} \geq 7.5$, the PARSEC isochronal ages tend to be slightly older, while for those with $\log\text{Age} \leq 7.5$, they appear slightly younger. However, almost all clusters and associations exhibit isochronal age distributions that fall within their respective 1$\sigma$ error ranges.

\begin{figure}
    \centering
    {%
    \includegraphics[width=0.4\textwidth, trim=0.1cm 0.3cm 0.0cm 0.8cm, clip]{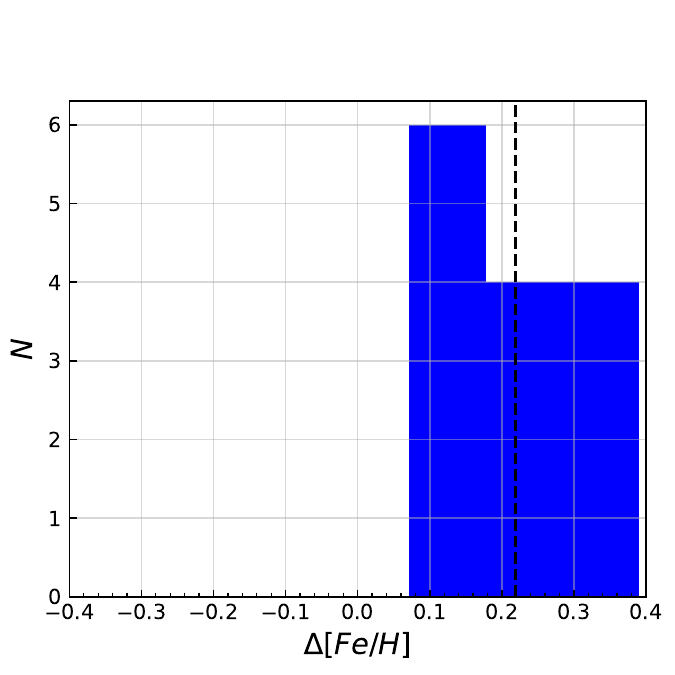}\quad 
    \includegraphics[width=0.4\textwidth, trim=0.1cm 0.3cm 0.0cm 0.8cm, clip]{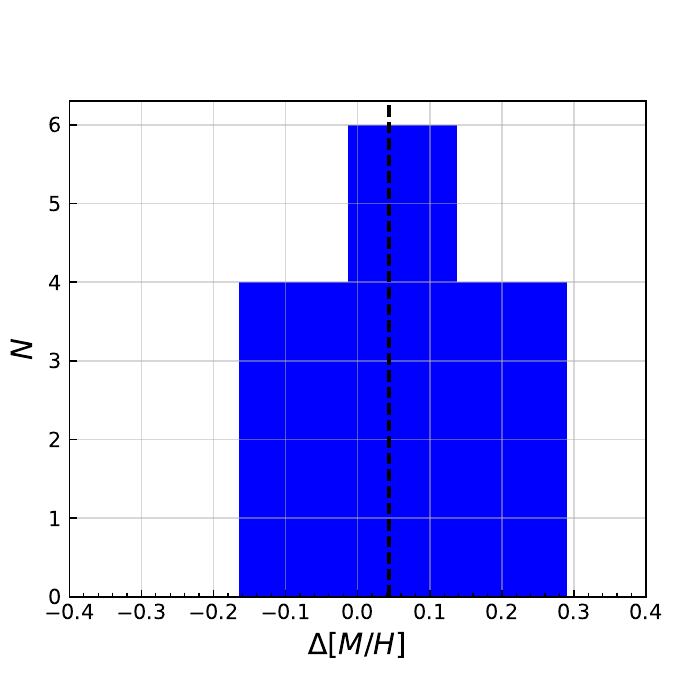}  
    }
    \caption{The distribution of differences between the metallicities obtained for the 31 clusters and 3 moving groups in this work and those obtained from spectroscopic data in the literature. The upper panel illustrates the difference distribution between the metallicities obtained from the MIST model in this work and those derived from spectroscopic data in the literature. The lower panel shows the difference distribution between the metallicities obtained from the PARSEC 1.2S model in this work and those derived from spectroscopic data in the literature. The black solid line represents the median difference.}
    \label{fig:feh_parsec-vs-mist}
\end{figure}

In addition to comparing ages, we also assessed the differences in metallicity obtained from the empirical color corrections applied to the two models in Figure\,\ref{fig:feh_parsec-vs-mist}. Based on the metallicities provided by spectroscopic data for each cluster and moving group (see Table \ref{table:summa_litera_10OCs}) and those obtained from fitting with the empirically color-corrected MIST and PARSEC 1.2S models, we calculated the median values and their differences, yielding values of 0.21$\pm$0.10 dex and 0.03$\pm$0.13 dex, respectively. From the distribution in this figure, it can be observed that when fitting for the best ages and metallicities using the empirically color-corrected MIST and PARSEC 1.2S models separately, the metallicities derived from the MIST model tend to be slightly more enriched, while those from the PARSEC 1.2S model are closer to the spectroscopic metallicities. We also obtained the ages and metallicities of the clusters from fitting their CMDs with \(M_G < 8\) (i.e., excluding low-mass stars),
and compared these metallicities with those from spectroscopic data. The median differences and standard deviations are \(0.21 \pm 0.10\) dex and \(0.04 \pm 0.14\) dex, respectively, which are consistent with the results in the above (including low-mass stars).

\begin{figure*} 
    \centering
    {%
    \includegraphics[width=0.45\textwidth,trim=0.1cm 0.2cm 0.0cm 1.0cm, clip]{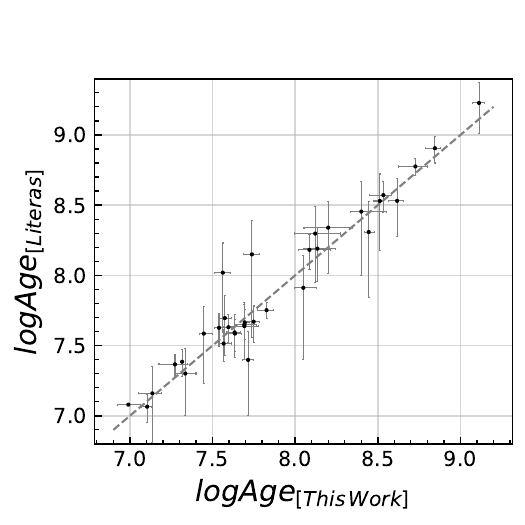}\quad
    \includegraphics[width=0.45\textwidth, trim=0.1cm 0.2cm 0.0cm 1.0cm, clip]{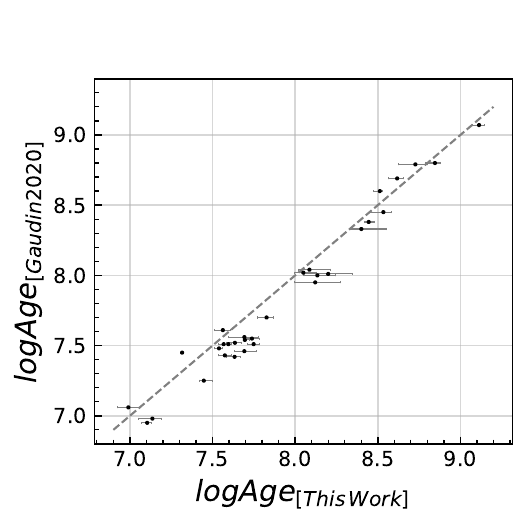}
    }
    \caption{Comparison of the ages obtained from the PARSEC model for the 31 clusters and 3 moving groups in this work with those in the literature. Left panel: Comparison with ages from the literature. Right panel: Comparison with ages provided by CG20. The grey dashed line represents the ratio 1:1, the same as in the preceding Figure~\ref{fig:34OCs_uncorrect-correct} and \ref{fig:thiswork_parsec-vs-mist}. }
    \label{fig:thiswork-vs-literas}
\end{figure*}

As shown in the left panel of Figure \ref{fig:thiswork-vs-literas}, we compare the ages of clusters and moving groups calculated using the PARSEC 1.2 model in this work (see Section \ref{sec:application_color_correction}) with those provided in the literature by taking their median and error values. In this subplot, we observe that the age distributions of almost all clusters and moving groups are consistent within the 1$\sigma$ error range. In the right panel, we further compare the ages of clusters calculated using the PARSEC 1.2 model in this work, considering their median and error values, with those provided by CG20. From the plot, it is evident that for clusters with $\log \text{Age} \geq 8$, which are relatively older, the age distributions are nearly identical within the 1$\sigma$ error range. However, for clusters with $\log\text{Age} \leq 8$, which are relatively younger, CG20's isochronal ages tend to be younger. This suggests that the ages recalculated after empirical color corrections tend to be slightly older. This difference may arise because CG20's age calculations involve fitting the entire CMD of the cluster, including the very low-mass part with color deviations. Whilst  our empirical color corrections to the very low-mass part of the isochrones have 
a better agreement to the observations with a slightly redder color.

\begin{figure} 
    \centering
    {%
    \includegraphics[width=0.45\textwidth,trim=0.1cm 0.2cm 0.0cm 1.0cm, clip]{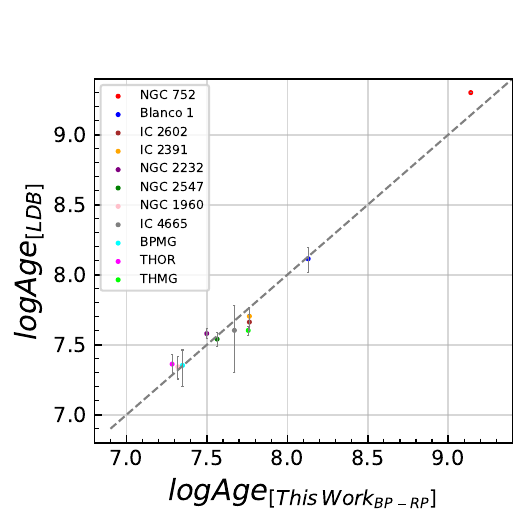}
    }
    \caption{Comparison of ages for 8 star clusters and 3 moving groups. Ages derived from the best-fitting isochrones using color (BP$-$RP) and its color-corrected version are compared with the LDB ages. The grey dashed line represents the ratio 1:1, the same as in the preceding Figures~\ref{fig:34OCs_uncorrect-correct}, \ref{fig:thiswork_parsec-vs-mist}, and \ref{fig:thiswork-vs-literas}.
    }
    \label{fig:8OCs_thiswork-vs-literas/LDB}
\end{figure}

The Lithium Depletion Boundary (LDB) method is an independent method comparing with other model-dependant techniques, such as isochrone fitting, for determinating the ages of stellar groups and is based on measuring the strength of the lithium line in low-mass stars and brown dwarfs \citep{Basri1996ApJ600B, Rebolo1996ApJ, Burke2004ApJ272B}. In the low mass range, the lithium burning process is very rapid and highly dependent on mass, leading to a distinct boundary between lithium-rich and lithium-poor cluster members \citep[e.g.,][]{DAntona1994ApJS, Bildsten1997ApJ}. By identifying the brightest lithium-rich star as the upper boundary and the faintest lithium-poor star as the lower boundary, and comparing these with theoretical models, the cluster's age can be determined. As shown in Figure \ref{fig:8OCs_thiswork-vs-literas/LDB}, to compare the LDB ages provided in the literature with the ages obtained in this work after empirical color correction using the PARSEC 1.2S model, we calculated the median and standard deviation of the LDB ages for the 8 open clusters and 3 moving groups listed in Table \ref{table:summa_litera_10OCs}, respectively.
From the distribution in this plot, it can be seen that the ages of the vast majority of clusters and moving groups are nearly identical. Although there are a few clusters and groups whose ages obtained in this study slightly differ from the LDB ages, such as NGC\,2232 and THMG, they still fall within the 3$\sigma$ range of LDB age errors.

\section{Summary and Discussions} 
\label{sec:summary}

In this work, we quantify the color deviations between  Gaia DR3 CMD of open clusters and model isochrones, focusing on the low-mass part (approximately M $\approx$ 0.3 $\sim$ 0.6 M$_\odot$).  Using the empirical color corrections on (BP$-$RP) and (G$-$RP) to the currently available MIST and PARSEC 1.2S isochrones, we re-determine the age of 31 open clusters and 3 moving groups.

Three nearby open clusters with low extinction and well-constrained age  serve as the benchmark clusters in this study: Hyades, Pleiades, and Praesepe. We  quantify the color discrepancies between these benchmark clusters and their best-fitted isochrones. Subsequently, we derive empirical color correction functions via polynomial fitting, particularly at low masses.  The (BP$-$RP) color correction has a maximum value of $\sim$0.25 mag for MIST and $\sim$0.15 mag for PARSEC 1.2S. The (G$-$RP) color correction is slightly smaller, with a maximum value of $\sim$0.08 mag for MIST and $\sim$0.04 mag for PARSEC 1.2S.

Based on the isochrones with empirical color correction functions, we recalculate the age of these open clusters and moving groups. The recalculated ages are consistent with the literature values based on the spectral LDB method within the 3$\sigma$ error range. Additionally, the ages obtained from original isochrones are younger by $\sim$0.075 dex in $\log\text{Age}$ compared to those from empirically color-corrected isochrones. However, it's important to acknowledge that isochrone fitting for very young stellar populations carries larger uncertainties compared to older stars.  For the very young clusters and moving groups we can only offer a rough age estimate using the homogeneous age determination method.

The empirical color corrections we present in this work, can serve the community as a tool for reliable age determination using the current available isochrone models. However, the underlying cause of the model-observation color discrepancies in this mass/color/temperature regime remains elusive. Here we list several possible factors that may contribute to the observed color discrepancy.

\begin{itemize}
    \item Limited understanding of the Gaia photometry:
    With Gaia being the sole source for such high-precision photometry data for a large number of low-mass stars in open clusters, the color discrepancy may have gone unnoticed previously. Additionally, the possibility of unknown issues with Gaia passband bolometric corrections or photometry itself cannot be entirely ruled out.
    \item Stellar rotation:
    Rotational speed and inclination angle can influence stellar color \citep[see e.g.][]{Costa2019, Nguyen2022, Girardi2019}. While rotation has minimal impact on the structure of the low-mass stars we focus on (being fully convective, see discussions in \citet{Bressan2012MN, fu2015}, and geometrical distortion discussed in \citet{Girardi2019}), 
    the inclination angle can affect their bolometric correction. As shown by \citet{Girardi2019}, fast rotation stars appear redder when viewed from the equator compared to the pole. 
    \item Magnetic fields: 
    As mentioned earlier for young stellar populations, magnetic fields can also affect stellar color. However, constraining and modeling magnetic fields remains a significant challenge observationally and theoretically.
\end{itemize}
Future research efforts  focus on a deeper understanding of these potential contributors may help to unfold the hidden nature of the color discrepancy we discussed in this work.

It is worth noting that in this work we choose to use the trigonometric parallax method to calculate the distances of star cluster and moving group members. This method has a distance error of approximately 1 pc within a range of 500 pc, which is relatively small and can be neglected. However, the method is not suitable to the distant cluster members because of the larger distance errors.

\acknowledgments
This work is supported by the National Key Research and Development Program of China (2023YFA1608100).
X.F. thanks the support of the National Natural Science Foundation of China (NSFC) No. 12203100 and the China Manned Space Project with No. CMS-CSST-2021-A08. Y.C. acknowledges National Natural Science Foundation of China (NSFC) No. 12003001. Xiaoying Pang acknowledges the financial support of the National Natural Science Foundation of China through grants 12173029 and 12233013.
L.Li thanks the support of NSFC No. 12303026 and the Young Data Scientist Project of the National Astronomical Data Center. W.Cui thanks the support of NSFC No. 12173013. Z. Li thanks the support of NSFC No. 12473029. This work has made use of data from the European Space Agency (ESA) mission Gaia \url{(http://www.cosmos.esa.int/gaia)}, processed by the Gaia Data Processing and Analysis Consortium \url{(DPAC, http://www.cosmos.esa.int/web/gaia/ dpac/consortium)}. 
This research has made use of the TOPCAT catalogue handling and plotting tool\citep{topcat}; of the Simbad database and the VizieR catalogue access tool, CDS, Strasbourg, France; and of NASA’s Astrophysics Data System.

%

\vspace{5mm}
\facilities{Topcat}


\newpage

\bibliographystyle{aasjournal}
\bibliography{references}

\appendix

\section{The metallicities and ages from literature for Hyades, Pleiades and Praesepe.}
\label{sec:coeff_color_corrections}
As shown in the table, we show the metallicities $[Fe/H]$ and Ages of Hyades, Pleiades and Praesepe from the literature.

\counterwithin{table}{section} 

\setcounter{table}{0}
\setcounter{table}{0}
\startlongtable
\begin{deluxetable*}{ccccccc}
\tablecaption{Data from the literature for the Hyades, Pleiades and Praesepe open clusters with metallicies and ages. \label{table:summa_litare}}
\tabletypesize{\scriptsize}
\tablehead{
\colhead{Clusters} & \colhead{$[Fe/H]$} & \colhead{R}  & \colhead{ref} & \colhead{Age/Myr} & \colhead{method} & \colhead{ref} }
\startdata
\hline
\hline
Hyades  & 0.28$\pm$0.03   &1800       &1 &650$\pm$70         &LDB        &11\\
        & 0.05$\pm$0.08   &1800       &2 &750$\pm$100        &LDB        &12\\
        & 0.135$\pm$0.005 &           &3 &625$\pm$25         &iso        &13 \\
        &0.13$\pm$0.02    &           &4 &690$\pm$160        &iso        &4\\
        &0.10             &           &5 &650$\pm$70         &LDB        &14\\
        &0.11$\pm$0.01    &30000      &6 &580$-$950          &iso        &15\\
        &0.0$\pm$0.02     &60000      &7 &$640^{+67}_{-49}$  &iso        &16\\
        &0.13$\pm$0.05    &$\geq$25000&8 &720$\pm$130        &iso        &8\\
        &0.127$\pm$0.022  &           &9 &500$-$1000         &iso        &17\\
        &0.05$\pm$0.05    &           &10&648$\pm$45         &iso        &18\\
        &0.12$\pm$0.04    &25000      &56&$794^{+161}_{-102}$&iso        &19\\ 
        &0.145$\pm$0.063  &28000      &57&650$\pm$70         &LDB        &11\\ 
        &0.131$\pm$0.015  &           &58&$794^{+161}_{-102}$&iso        &19\\
\hline
Pleiades & 0.10$\pm$0.002  &1800       &1 & 130$\pm$20        &LDB       &27\\
         & 0.03$\pm$0.04   &           &20&112$\pm$5          &LDB       &28\\
         & $-$0.034$\pm$0.024&           &9 &120$-$130          &LDB       &29\\
         &0.0$\pm$0.01     &           &21&135                &iso       &30\footnote{http://www.univie.ac.at/webda}\\
         &0.06$\pm$0.05    &           &22&120                &iso       &31\\
         & 0.06$\pm$0.23   &           &23&$\geq$120         &iso       &32\\
         & 0.03$\pm$0.02   &           &24&150                &iso       &33\\
         &$-$0.039$\pm$0.014 &           &25&78                 &iso       &34\\
         &0.06$\pm$0.06    &           &26&50                 &iso       &35\\
         &$-$0.037$\pm$0.026&           &4 &135$\pm$25         &iso       &4\\
         &$-$0.01$\pm$0.05  &$\geq$25000&8 &100$\pm$40         &iso       &8\\
         &0.0$\pm$0.05     &25000     &56&$134^{+9}_{-10}$   &LDB       &36\\
         &$-$0.010$\pm$0.049&28000      &57&$115^{+3}_{-11}$   &iso       &37\\
         &                 &           &  &132$\pm$2          &iso       &38\\
         &                 &           &  &$86^{+6}_{-3}$     &iso       &39\\
         &                 &           &  &$132^{+26}_{-27}$  &iso       &16\\
         &                 &           &  &126$\pm$11         &LDB       &40\\
         &                 &           &  &$\geq$1151        &LDB       &41\\
         &                 &           &  &1241               &iso       &42\\
         &                 &           &  &$141^{+29}_{-41}$  &iso       &43\\
         &                 &           &  &135                &iso       &44\\
\hline
Praesepe & 0.16$\pm$0.06   &           &45&750$\pm$7          &iso        &4\\     
         & 0.08$\pm$0.03   &           &46&730$\pm$190        &iso        &8\\     
         & 0.25$\pm$0.003  & 1800      &1 &700                &iso        &42\\   
         &  0.221$\pm$0.084& 1800      &53&$794^{+253}_{-270}$&iso        &43\\
         &0.16             &           &4 &700                &iso        &50\\
         &0.16$\pm$0.05    &30000      &47&760                &iso        &51\\
         &0.27$\pm$0.10    &100000     &48&$762^{+64}_{-59}$  &iso        &52\\
         &0.11$\pm$0.03    &55000      &49&750                &iso        &53\\
         &0.16$\pm$0.08    &$\geq$25000&8 &729                &iso        &54\\
         & 0.16$\pm$0.07   &25000      &56&600                &iso        &55\\ 
         &0.16             &           &50&                   &           &\\ 
         & 0.10$\pm$0.002  &           &27&                   &           &\\ 
         &0.133$\pm$0.041  &28000      &57&                   &           &\\ 
         & 0.196$\pm$0.039 &           &52&                   &           &\\ 
         & 0.118$\pm$0.014 &           &58&                   &           &\\ 
         &                 &           &  &                   &           &\\                  
\hline
\hline
\enddata
\tablerefs{1.\citet{Fu2022AA668A}, 2.\citet{Zhong2020AA127Z}, 3.\citet{Cummings2012AJ137C}, 4.\citet{Ilin2021AA42I}, 5.\citet{Taylor2005ApJS100T}, 6.\citet{Carrera2011AA30C}, 7.\citet{Liu2016MNRAS3934L}, 8.\citet{Netopil2016AA150N}, 9.\citet{Boesgaard1990ApJ467B}, 10.\citet{Gebran2010AA71G}, 11.\citet{Lodieu2020MmSAI84L}, 12.\citet{Brandt2015ApJ58B}, 13.\citet{Perryman1998AA81P}, 14.\citet{Martin2018ApJ40M}, 15.\citet{Lodieu2018AA12L}, 16.\citet{Lodieu2019AA66L}, 17.\citet{Eggen1998AJ284E}, 18.\citet{DeGennaro2009ApJ12D}, 19.\citet{GaiaCollaboration2018AA10G}, 20.\citet{Soderblom2009AJ1292S}, 21.\citet{Barrado2001ApJ452B}, 22.\citet{King2000ApJ944K}, 23.\citet{Groenewegen2007AA579G}, 24.\citet{An2007ApJ233A}, 25.\citet{Taylor2008AJ1388T}, 26. \citet{Gebran2008AA567G}, 27.\citet{Barrado2004ApJ386B}, 28.\citet{Dahm2015ApJ108D}, 29.\citet{Stauffer1998ApJ199S}, 30.Webda database, 31.\citet{Kharchenko2005AA1163K}, 32.\citet{Ventura1998AA1011V}, 33.\citet{Mazzei1989AA1M}, 34.\citet{Mermilliod1981AA235M}, 35.\citet{Patenaude1978AA225P}, 36.\citet{Cargile2014ApJ29C}, 37.\citet{Naylor2009MN432N}, 38.\citet{Bell2014MN3496B}, 39.\citet{Bossin2019AA108B}, 40.\citet{Burke2004ApJ272B}, 41.\citet{Basri1996ApJ600B}, 42.\citet{Pang2023AJ110P}, 43.\citet{Yen2018yCat12Y}, 44.\citet{Kounkel2019AJ122K}, 45.\citet{Yang2015AJ}, 46.\citet{Reddy2016MNRAS}, 47.\citet{Carrera2011AA30C}, 48.\citet{Pace2008AA403P}, 49.\citet{An2007ApJ233A}, 50.\citet{Godoy-Rivera2021Ap46G}, 51.\citet{Cordoni2023AA29C}, 52.\citet{Dias2021MN356D}, 53. \citet{Zhong2020AA127Z}, 54.\citet{Dias2002AA871D}, 55.\citet{Boudreaul2010AA27B}, 56.\citet{Netopil2022MN}, 57. \citet{Spina2021MN}, 58. \citet{Casamiquela2021AA}}
\tablecomments{spectral resolution(R), Lithum depletion boundary(LDB); isochrone fitting(iso)}
\end{deluxetable*}

\section{Addtional CMDs for the color correction of Pleiades and Praesepe}
\label{sec: addtional CMDs HPPs}
Similar to Figure\,\ref{fig:Hyades_bp-g_rp_mist} and \ref{fig:Hyades_bp-g_rp_parsec} for Hyades in Section\,\ref{sec:process_color_correction}, there are CMDs for the color correction of Pleiades and Praesepe.

\renewcommand{\thefigure}{\Alph{section}.\arabic{figure}}
\setcounter{figure}{0} 

\begin{figure}
    \centering
    {%
    \includegraphics[width=0.22\textwidth, trim=0.1cm 0.2cm 0.0cm 1.7cm, clip]{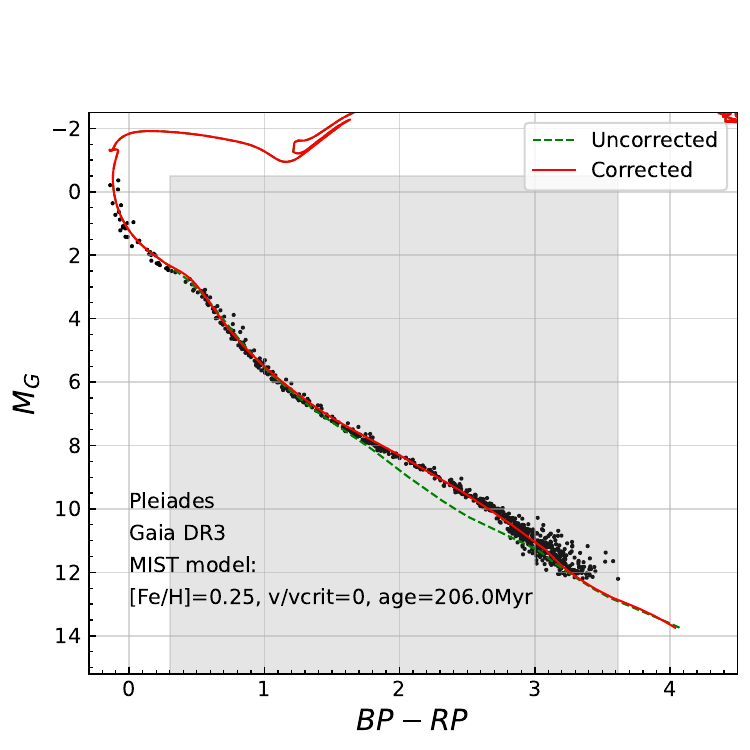}\quad
    \includegraphics[width=0.22\textwidth, trim=0.1cm 0.2cm 0.0cm 1.7cm, clip]{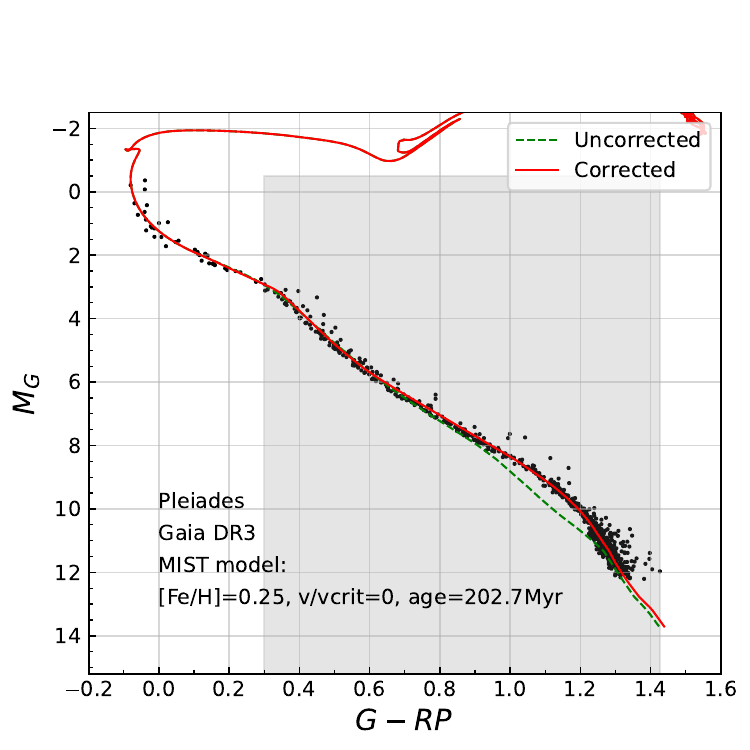}}
    {%
    \includegraphics[width=0.22\textwidth, trim=0.1cm 0.2cm 0.0cm 1.7cm, clip]{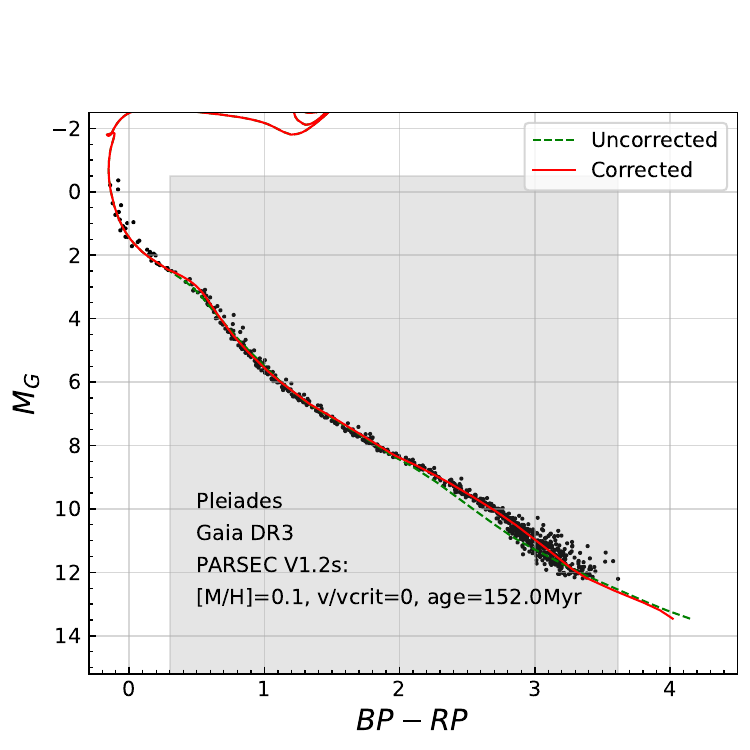}\quad
    \includegraphics[width=0.22\textwidth, trim=0.1cm 0.2cm 0.0cm 1.7cm, clip]{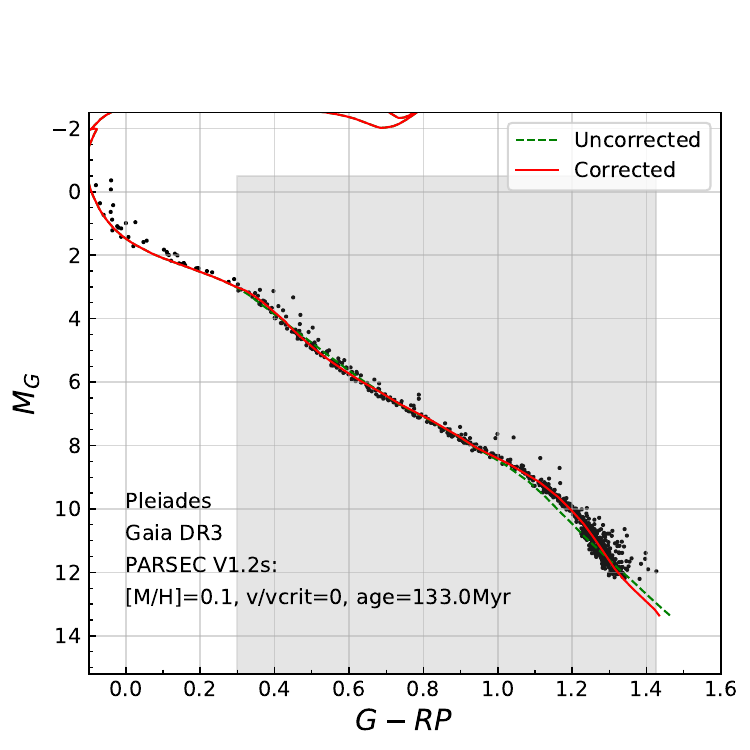}}
    \caption{CMDs of Pleiades open cluster used by MIST model and PARSEC 1.2S model, respectively, similiar to Figure\,\ref{fig:Hyades_bp-g_rp_mist} and \ref{fig:Hyades_bp-g_rp_parsec}.
    }
    \label{fig:Pleiades_bp-g_rp_mist/parsec}
\end{figure}
\begin{figure}
    \centering
    {%
    \includegraphics[width=0.22\textwidth, trim=0.1cm 0.2cm 0.0cm 1.7cm, clip]{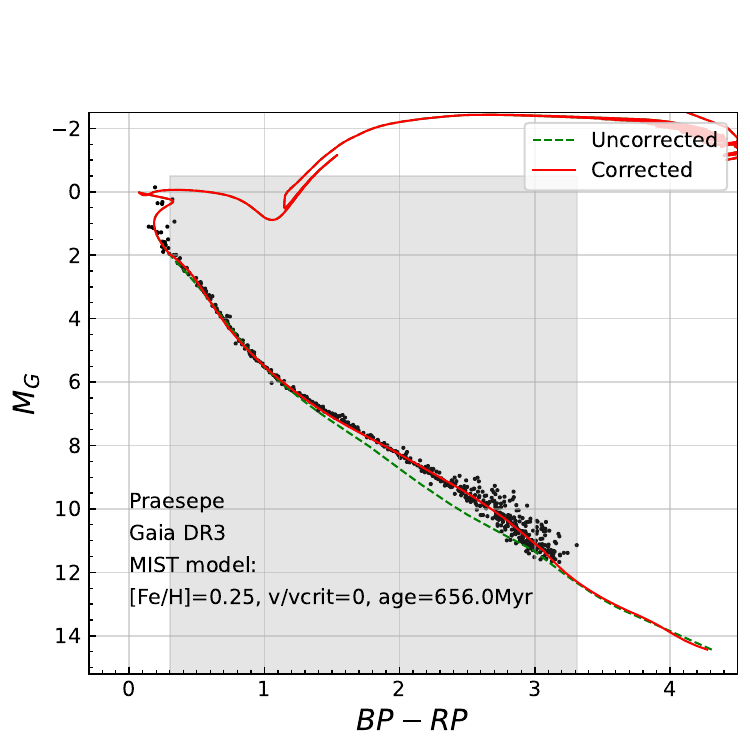}\quad
    \includegraphics[width=0.22\textwidth, trim=0.1cm 0.2cm 0.0cm 1.7cm, clip]{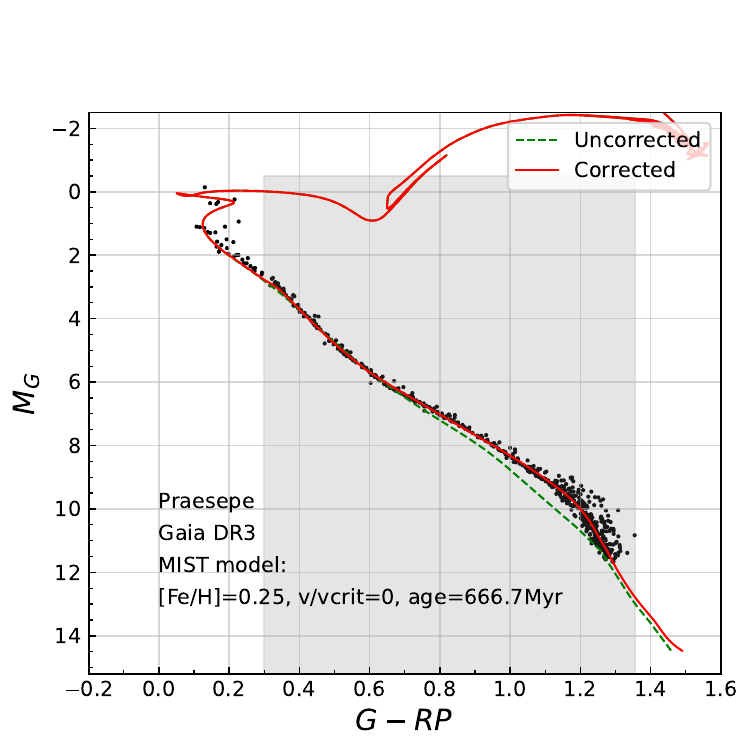}}
    {%
    \includegraphics[width=0.22\textwidth, trim=0.1cm 0.2cm 0.0cm 1.7cm, clip]{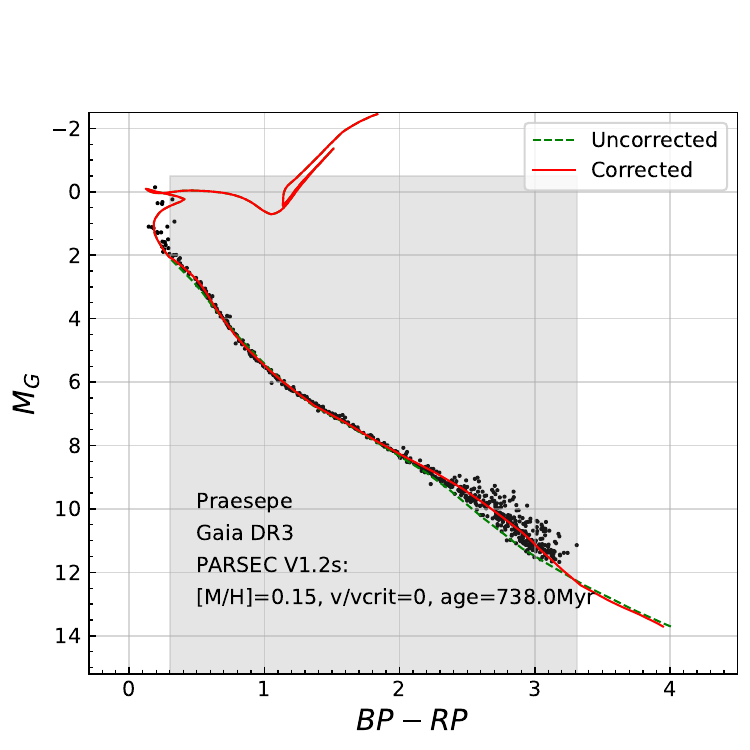}\quad
    \includegraphics[width=0.22\textwidth, trim=0.1cm 0.2cm 0.0cm 1.7cm, clip]{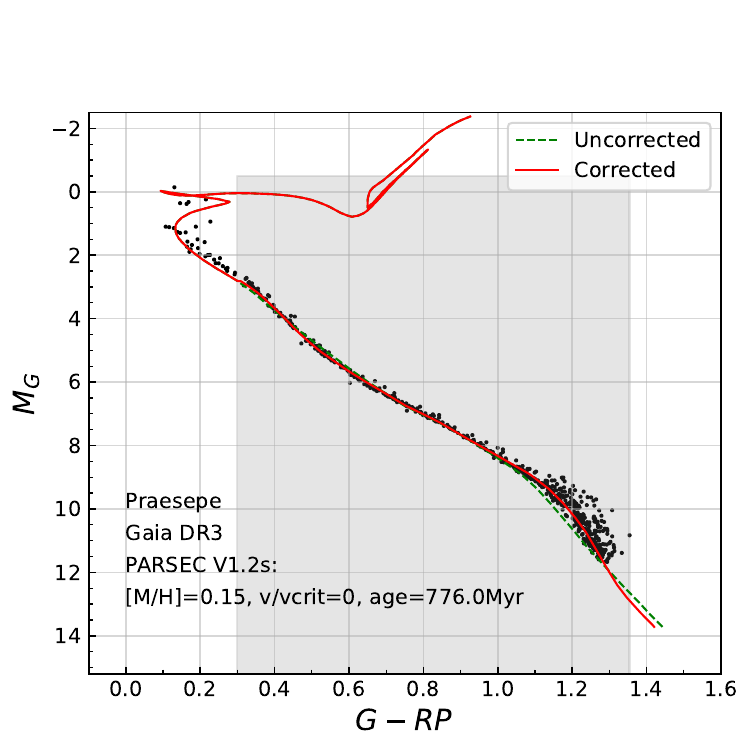}}
    \caption{CMDs of Praesepes open cluster, similiar to Figure\,\ref{fig:Hyades_bp-g_rp_mist}, \ref{fig:Hyades_bp-g_rp_parsec} and \ref{fig:Pleiades_bp-g_rp_mist/parsec}. 
    }
    \label{fig:Praesepe_bp-g_rp_mist/parsec}
\end{figure}

\section{Addtional information of 31 clusters and 3 moving groups}
\label{sec:addtional CMDs}
In this appendix, we present the CMDs of 30 clusters and 3 moving groups, mentioned in Section~\ref{sec:application_color_correction}, using the MIST model in Figure~\ref{fig:33_bp/g-rp_dr3_mist} and the PARSEC 1.2S model in Figure~\ref{fig:33_bp/g-rp_dr3_parsec}.

\newpage
\setcounter{figure}{0}
\begin{figure*}
\centering
    {%
    \includegraphics[width=0.22\textwidth, trim=0.1cm 0.9cm 0.0cm 1.6cm, clip]{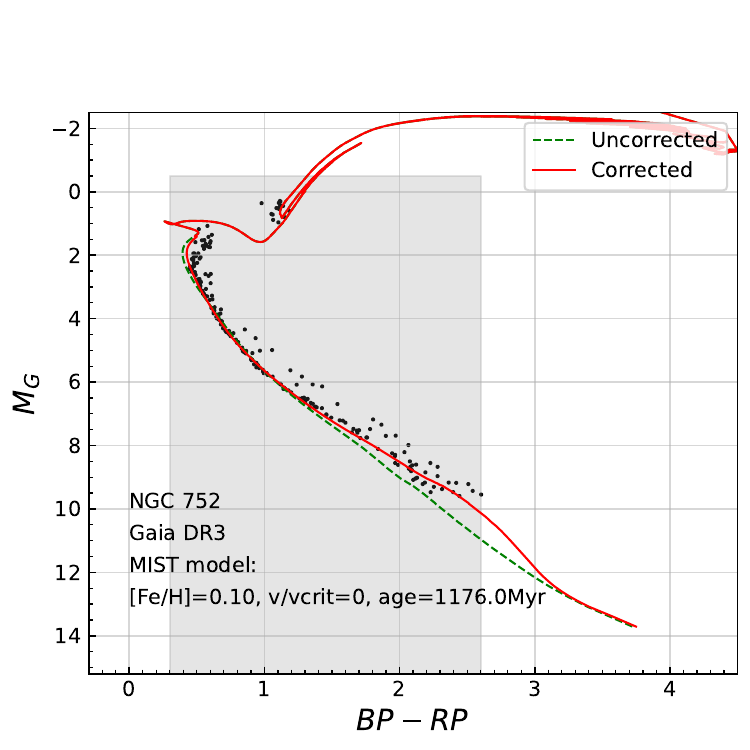}\quad
    \includegraphics[width=0.22\textwidth, trim=0.1cm 0.9cm 0.0cm 1.6cm, clip]{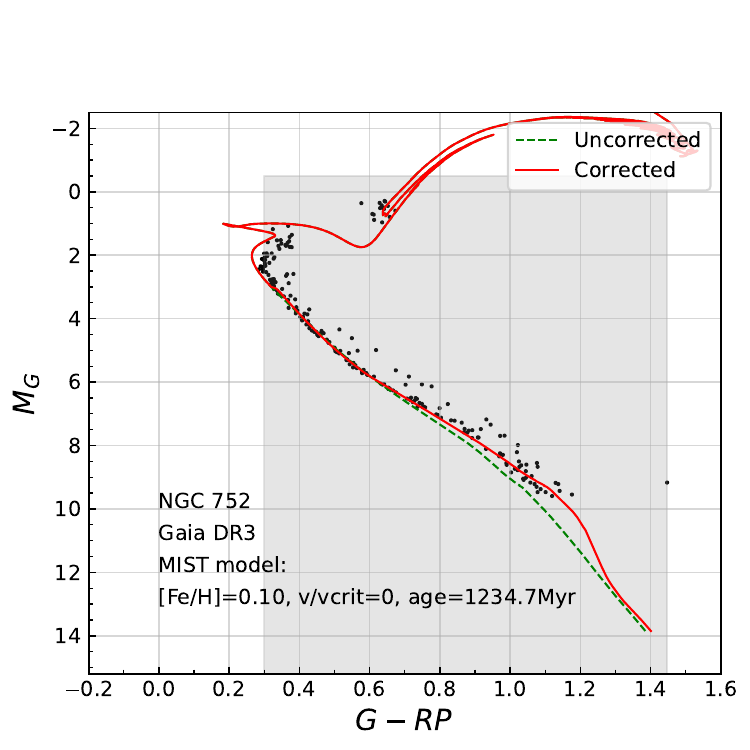}
    }
    {%
    \includegraphics[width=0.22\textwidth, trim=0.1cm 0.9cm 0.0cm 1.6cm, clip]{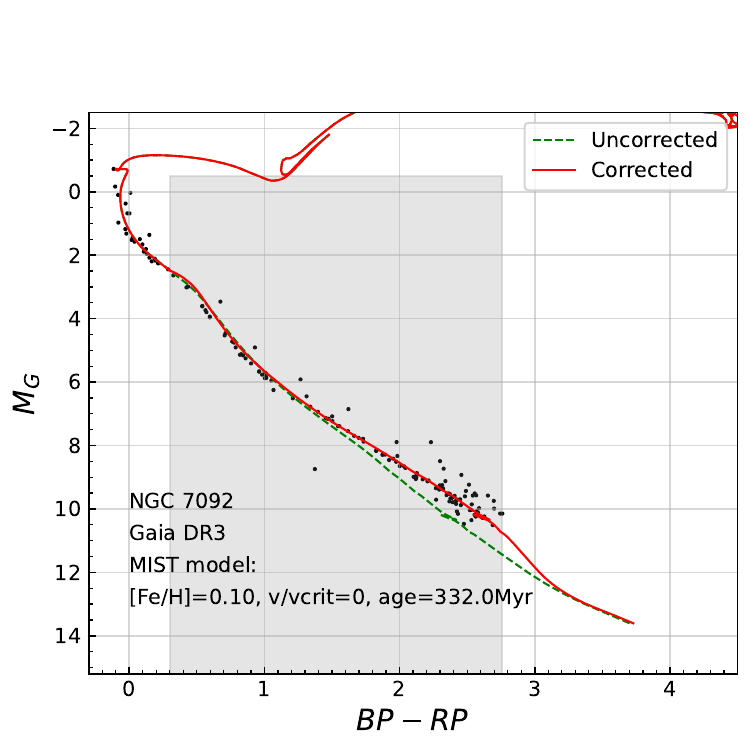}\quad
    \includegraphics[width=0.22\textwidth, trim=0.1cm 0.9cm 0.0cm 1.6cm, clip]{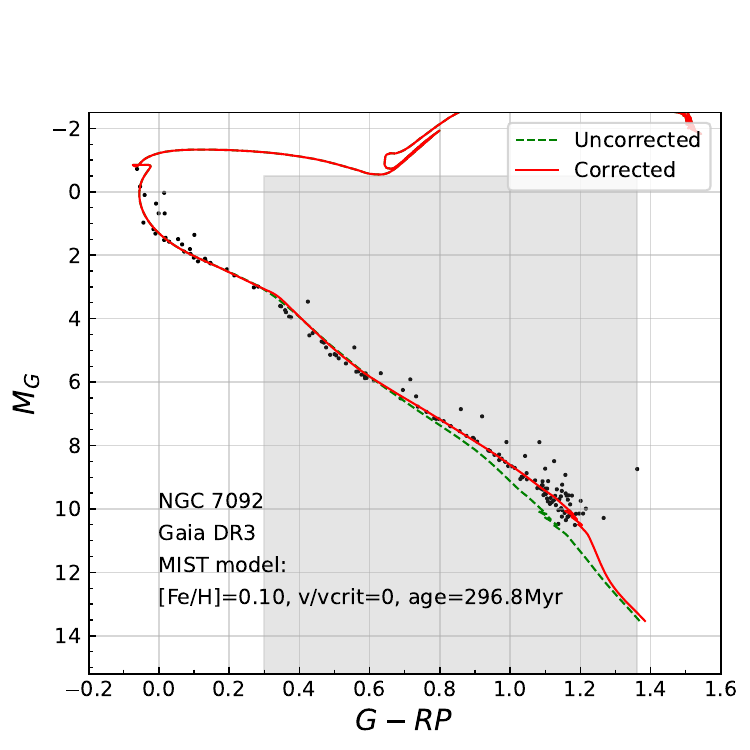}
    }
    {%
    \includegraphics[width=0.22\textwidth, trim=0.1cm 0.2cm 0.0cm 1.7cm, clip]{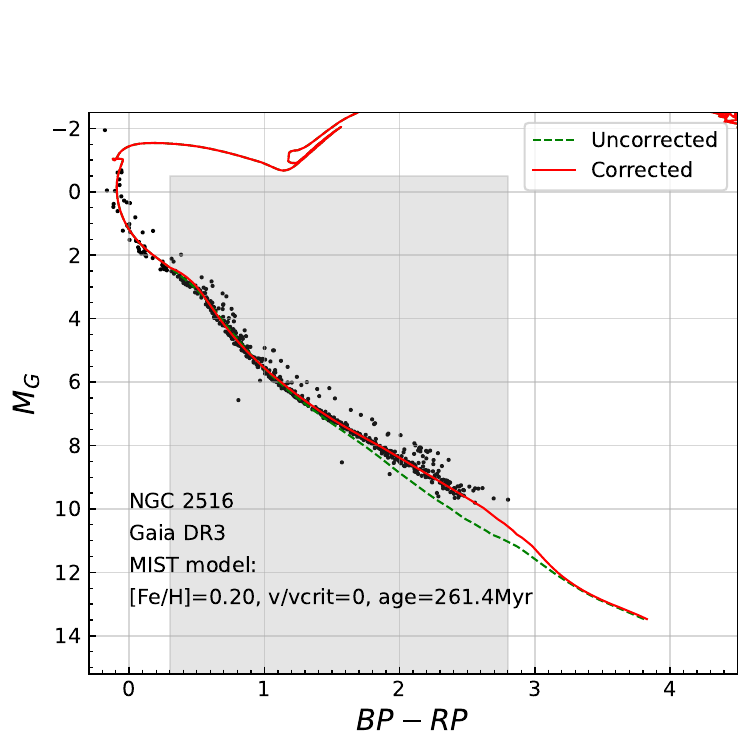}\quad
    \includegraphics[width=0.22\textwidth, trim=0.1cm 0.2cm 0.0cm 1.7cm, clip]{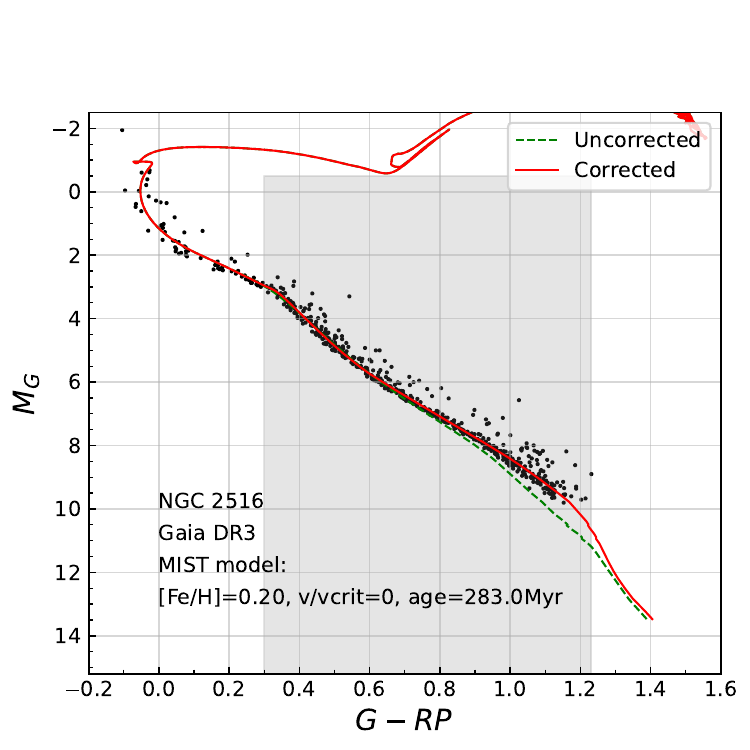}
    }
    {%
    \includegraphics[width=0.22\textwidth, trim=0.1cm 0.9cm 0.0cm 1.6cm, clip]{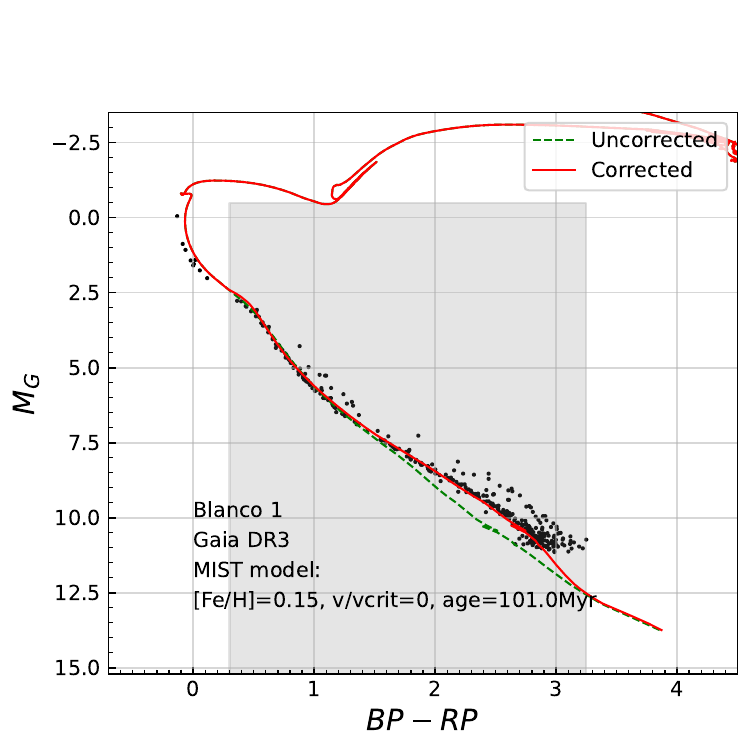}\quad
    \includegraphics[width=0.22\textwidth, trim=0.1cm 0.9cm 0.0cm 1.6cm, clip]{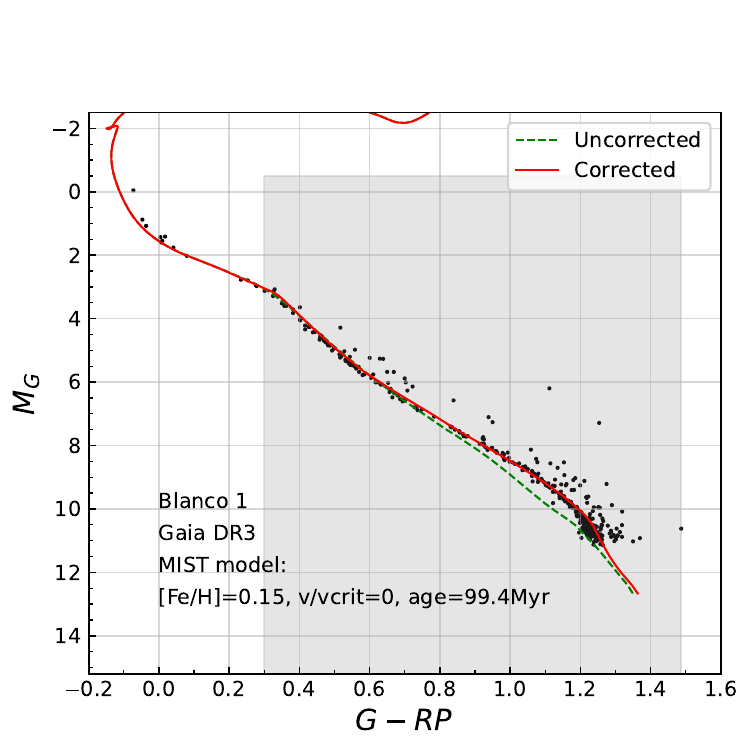}
    }
    {%
    \includegraphics[width=0.22\textwidth, trim=0.1cm 0.2cm 0.0cm 1.7cm, clip]{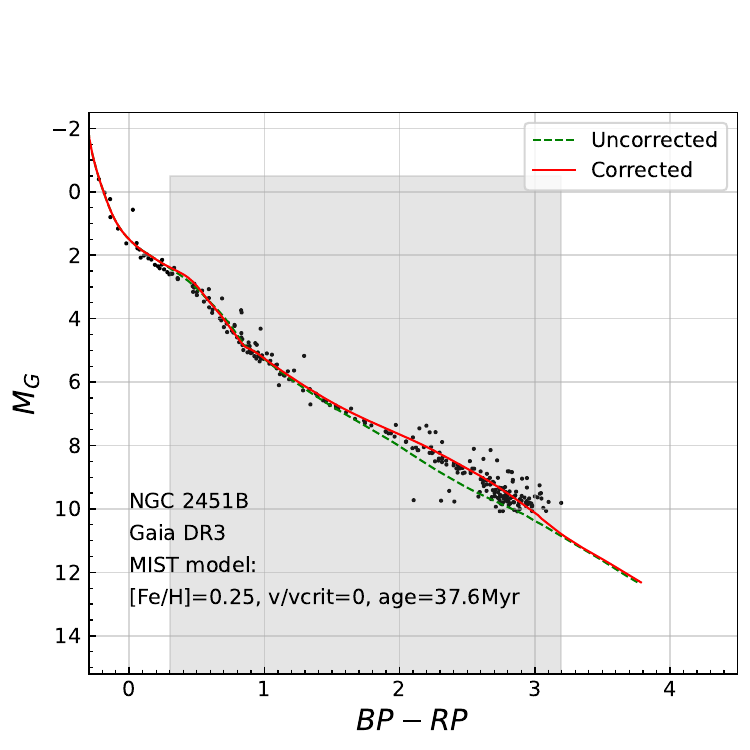}\quad
    \includegraphics[width=0.22\textwidth, trim=0.1cm 0.2cm 0.0cm 1.7cm, clip]{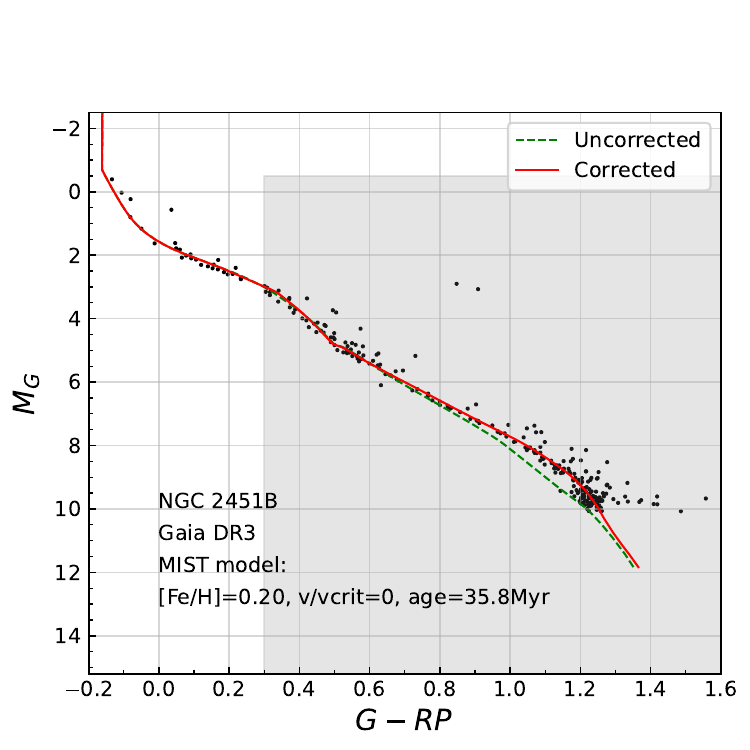}
    }
    {%
    \includegraphics[width=0.22\textwidth, trim=0.1cm 0.2cm 0.0cm 1.7cm, clip]{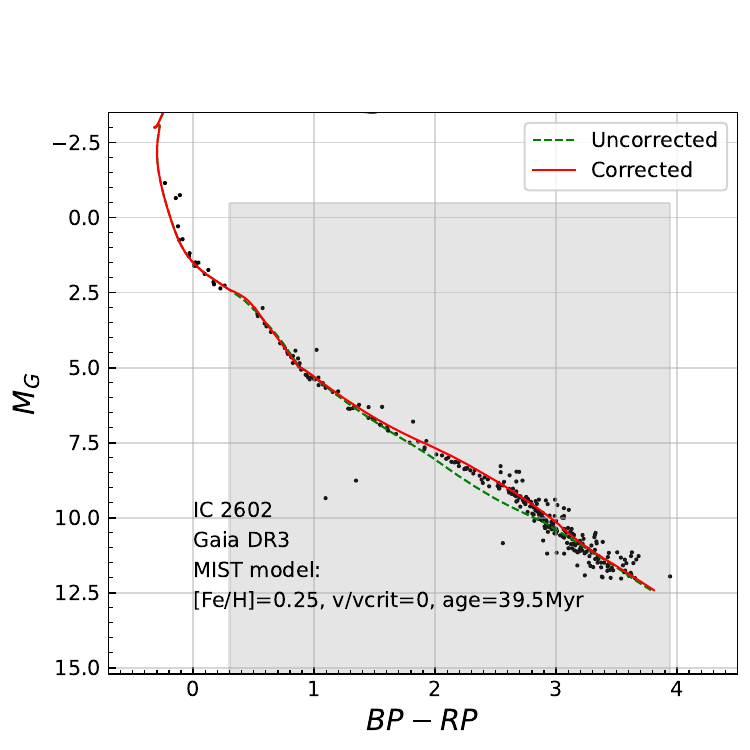}\quad
    \includegraphics[width=0.22\textwidth, trim=0.1cm 0.2cm 0.0cm 1.7cm, clip]{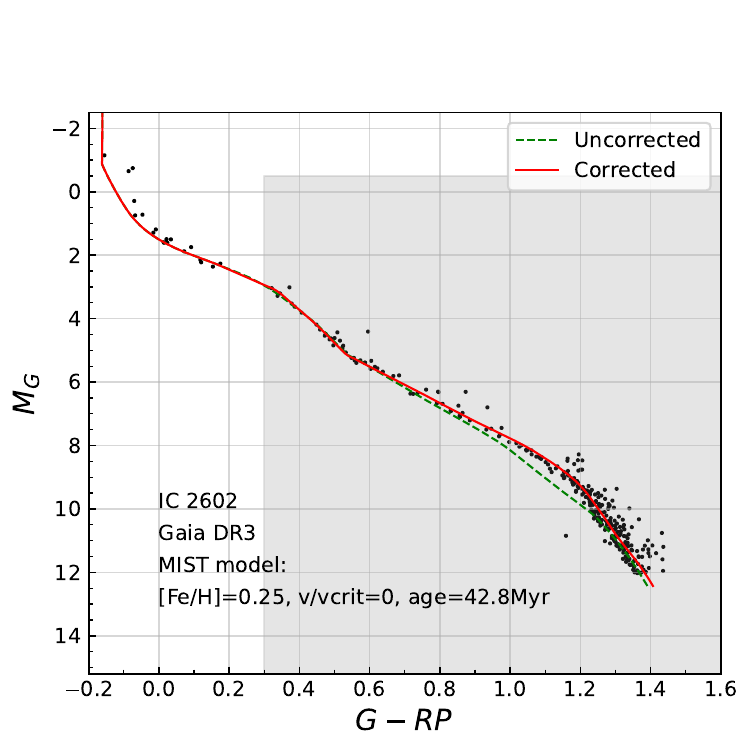}
    }
    {%
    \includegraphics[width=0.22\textwidth, trim=0.1cm 0.2cm 0.0cm 1.7cm, clip]{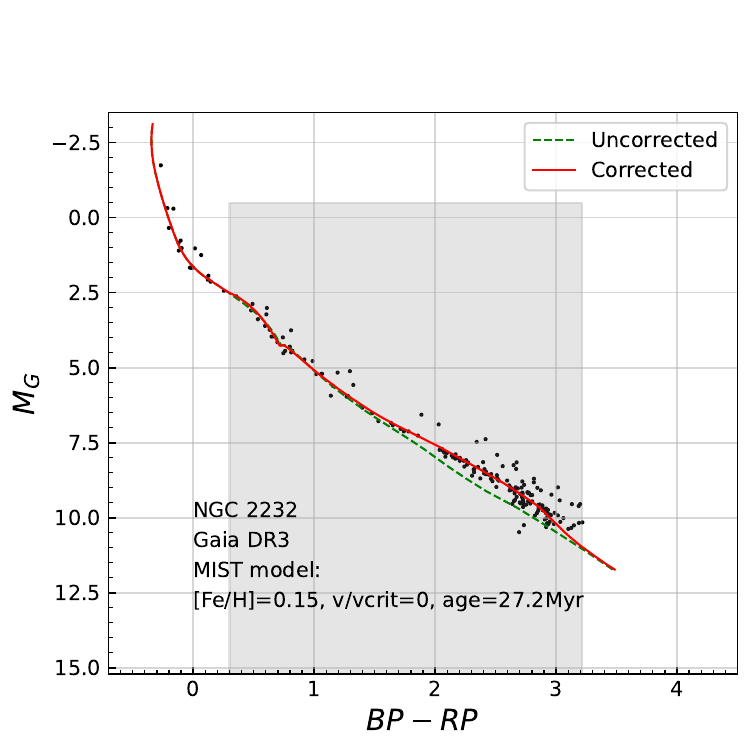}\quad
    \includegraphics[width=0.22\textwidth, trim=0.1cm 0.2cm 0.0cm 1.7cm, clip]{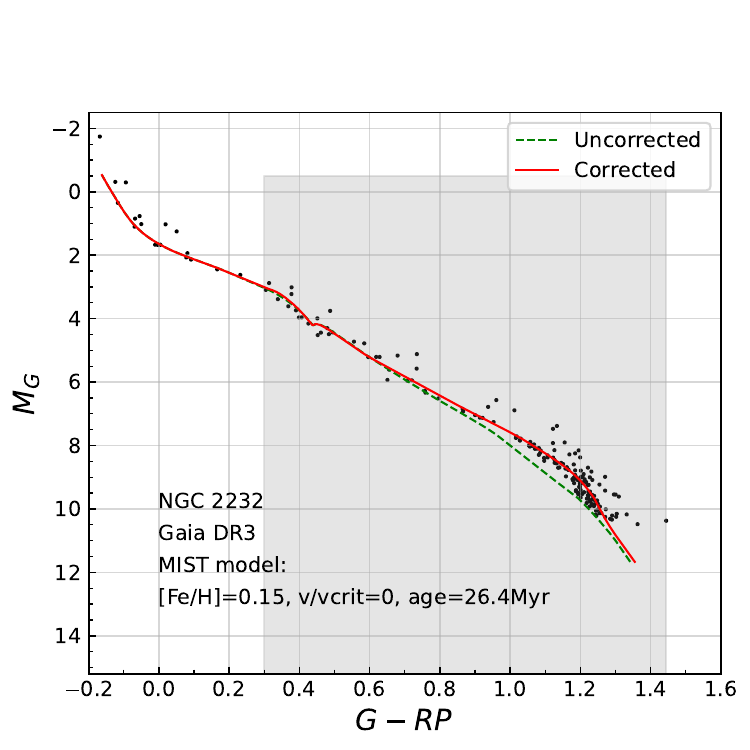}
    }
    {%
    \includegraphics[width=0.22\textwidth, trim=0.1cm 0.2cm 0.0cm 1.7cm, clip]{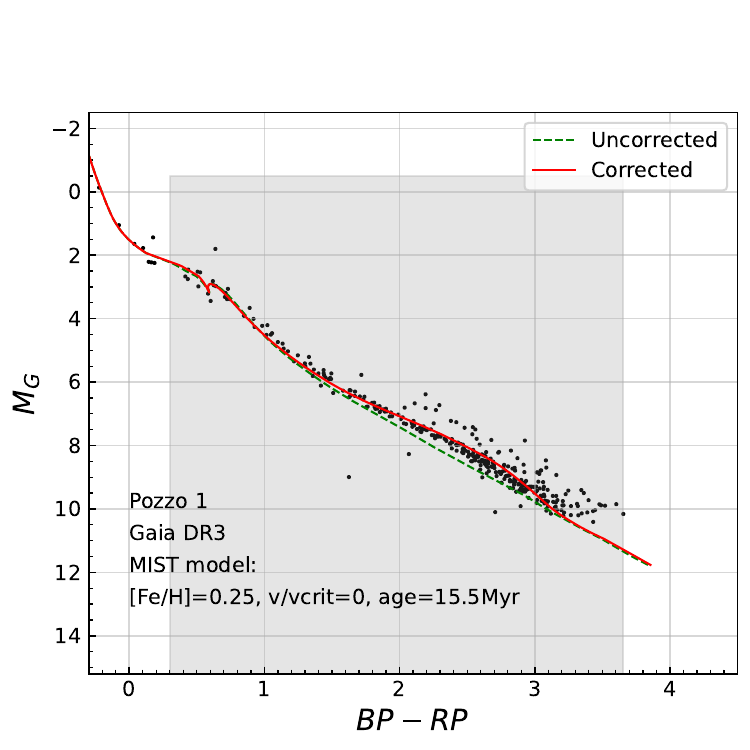}\quad
    \includegraphics[width=0.22\textwidth, trim=0.1cm 0.2cm 0.0cm 1.7cm, clip]{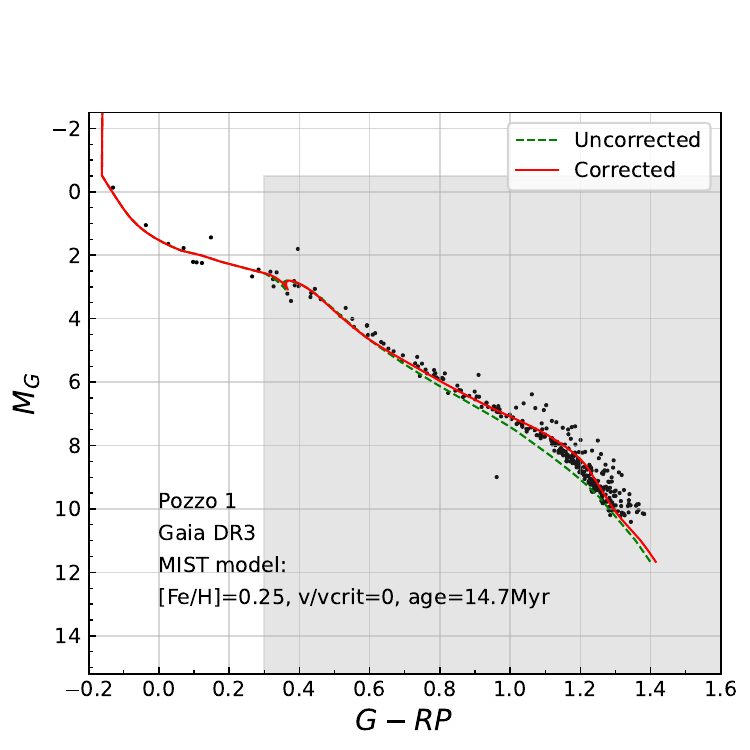}
    }
    {%
    \includegraphics[width=0.22\textwidth, trim=0.1cm 0.2cm 0.0cm 1.7cm, clip]{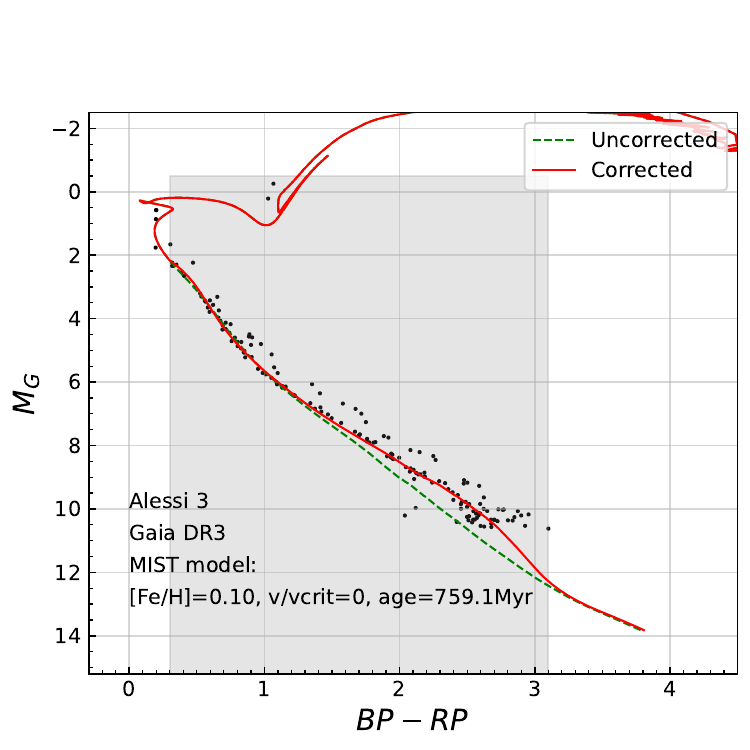}\quad
    \includegraphics[width=0.22\textwidth, trim=0.1cm 0.2cm 0.0cm 1.7cm, clip]{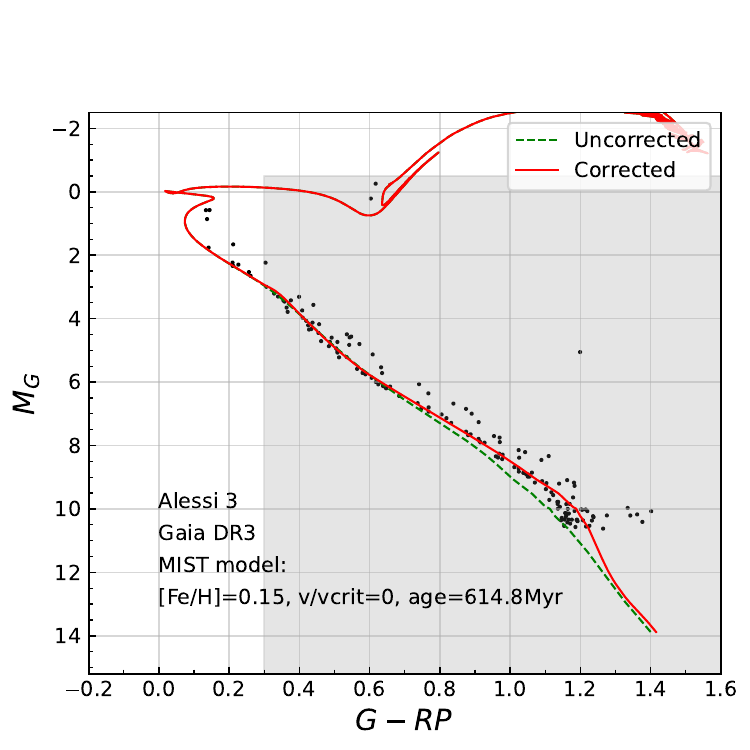}
    }
    {%
    \includegraphics[width=0.22\textwidth, trim=0.1cm 0.2cm 0.0cm 1.7cm, clip]{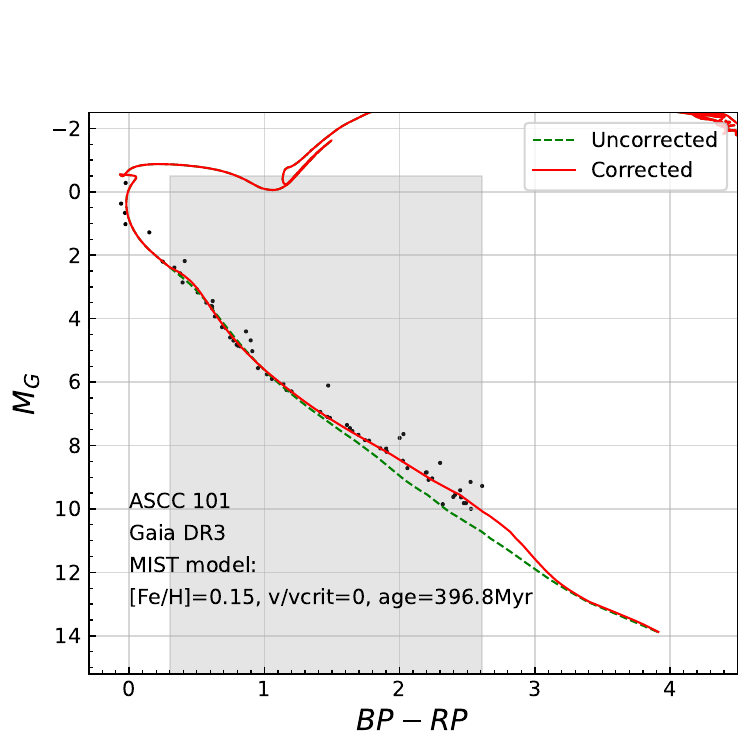}\quad
    \includegraphics[width=0.22\textwidth, trim=0.1cm 0.2cm 0.0cm 1.7cm, clip]{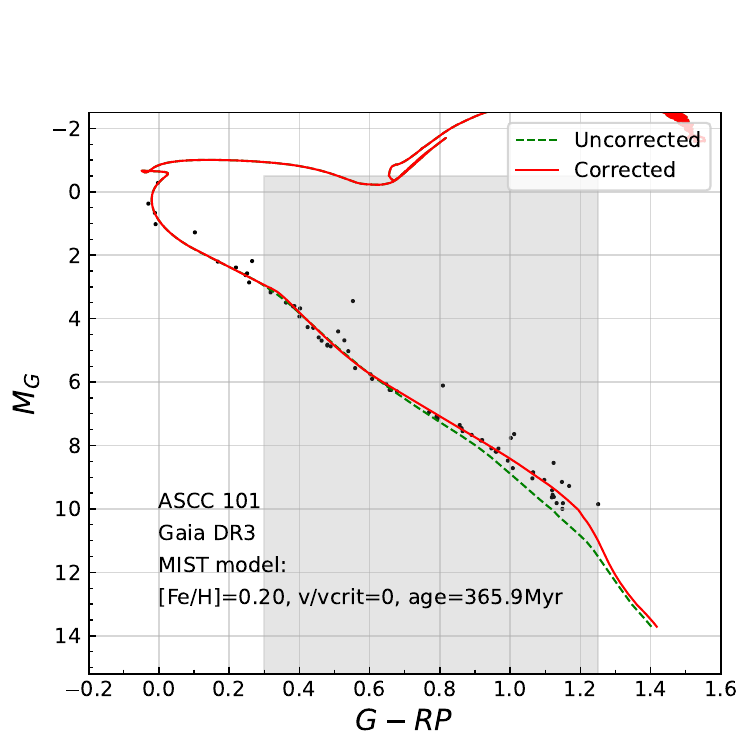}
    }
    {%
    \includegraphics[width=0.22\textwidth, trim=0.1cm 0.2cm 0.0cm 1.7cm, clip]{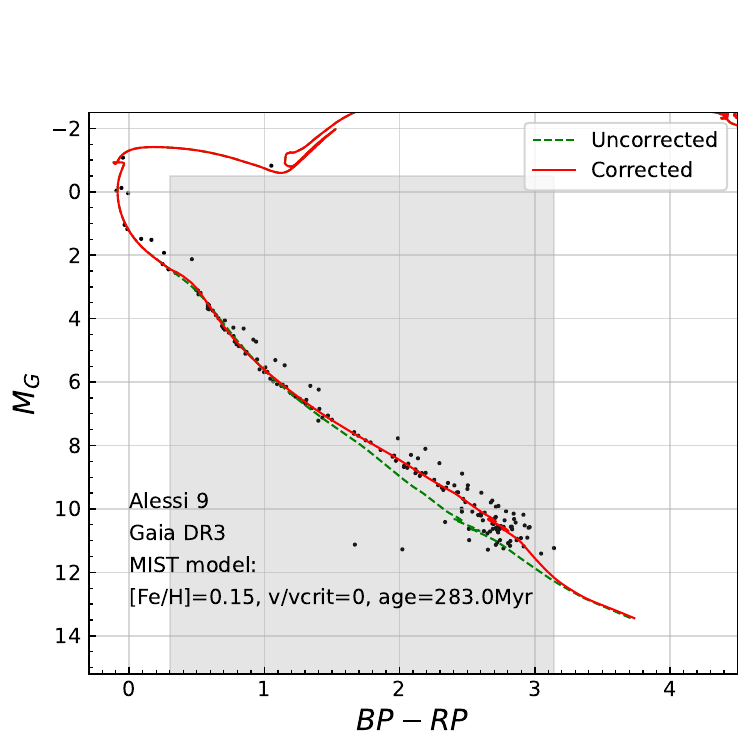}\quad
    \includegraphics[width=0.22\textwidth, trim=0.1cm 0.2cm 0.0cm 1.7cm, clip]{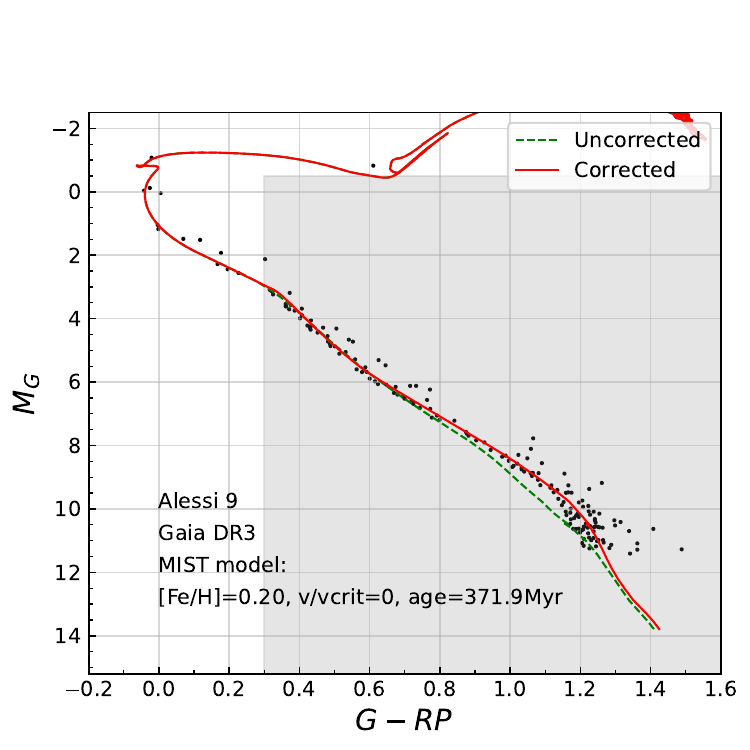}
    }
    {%
    \includegraphics[width=0.22\textwidth, trim=0.1cm 0.2cm 0.0cm 1.7cm, clip]{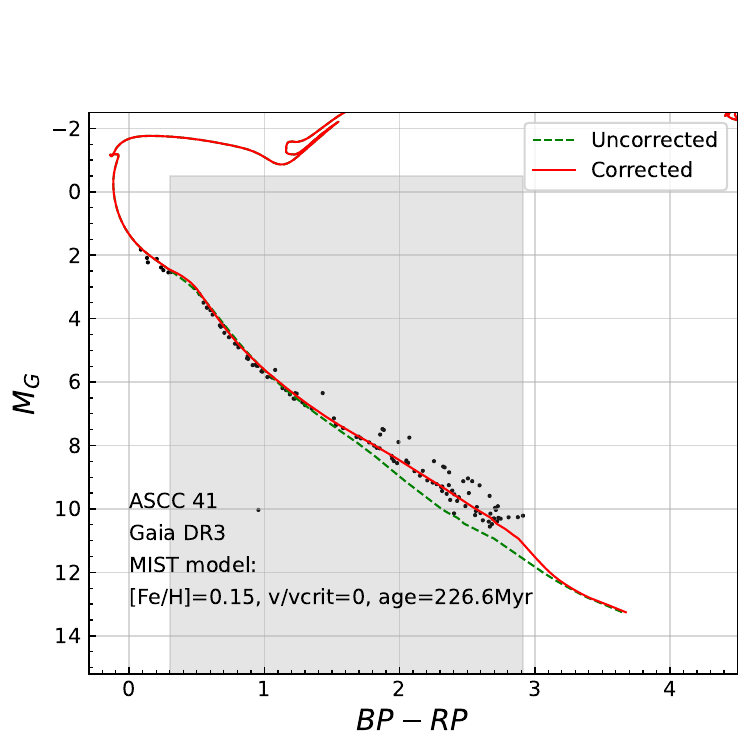}\quad
    \includegraphics[width=0.22\textwidth, trim=0.1cm 0.2cm 0.0cm 1.7cm, clip]{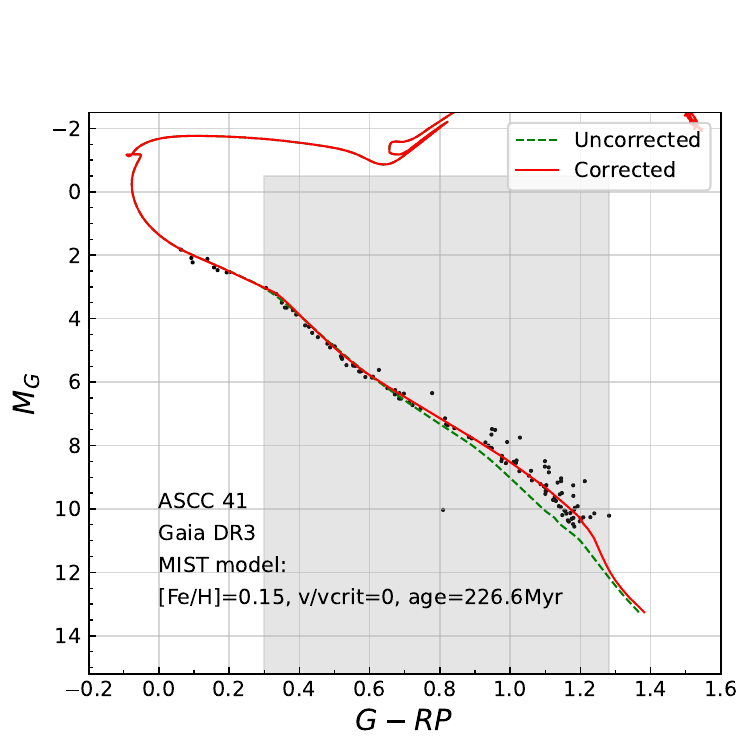}
    }
    {%
    \includegraphics[width=0.22\textwidth, trim=0.1cm 0.2cm 0.0cm 1.7cm, clip]{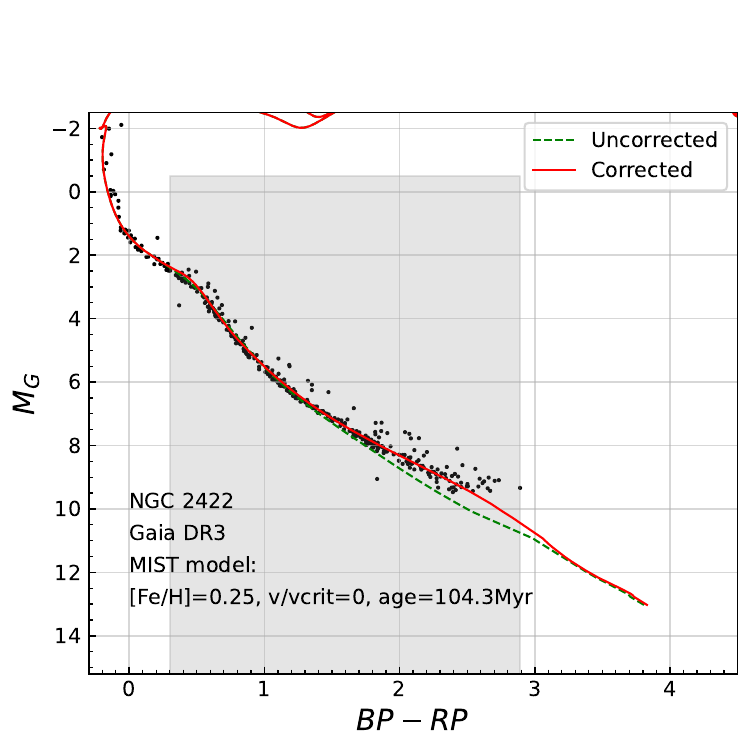}\quad
    \includegraphics[width=0.22\textwidth, trim=0.1cm 0.2cm 0.0cm 1.7cm, clip]{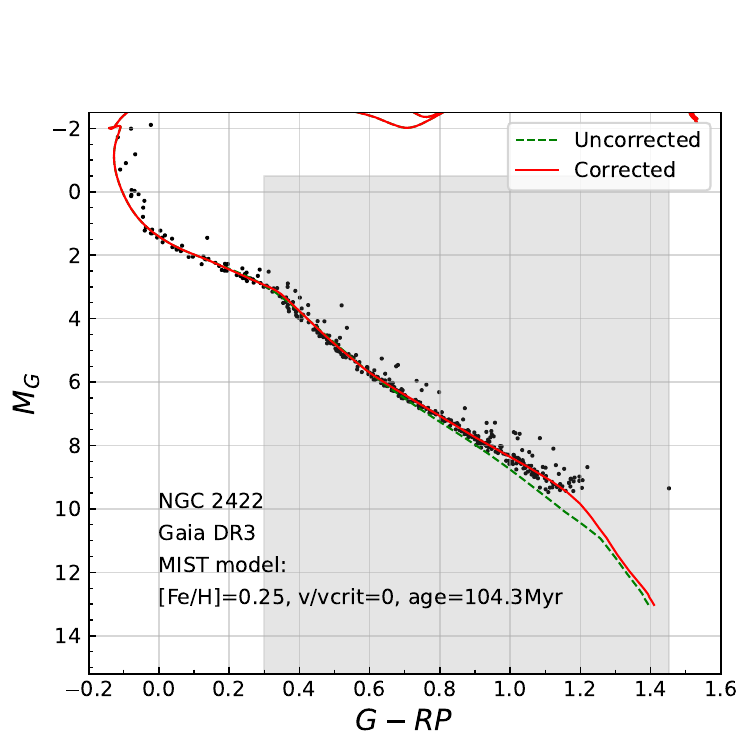}
    }
    {%
    \includegraphics[width=0.22\textwidth, trim=0.1cm 0.2cm 0.0cm 1.7cm, clip]{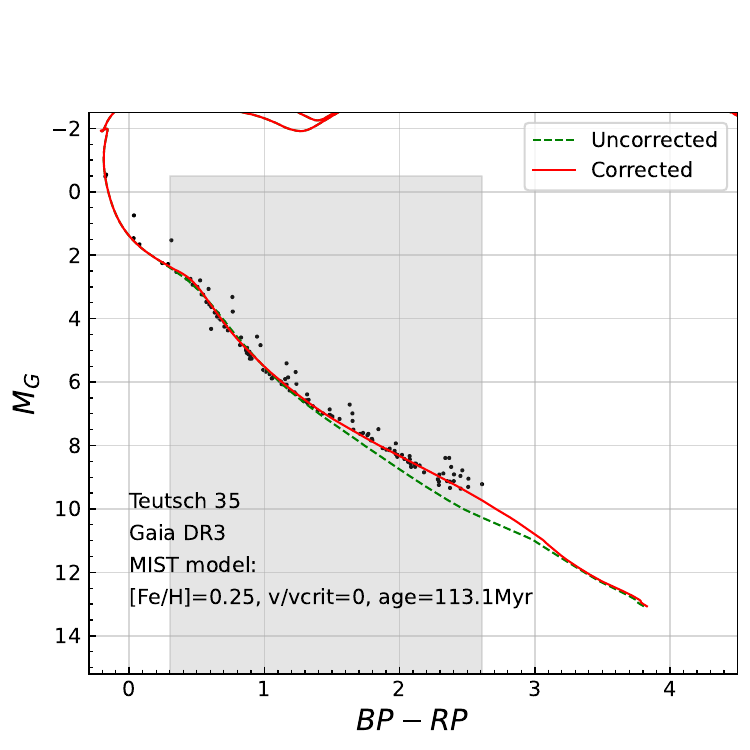}\quad
    \includegraphics[width=0.22\textwidth, trim=0.1cm 0.2cm 0.0cm 1.7cm, clip]{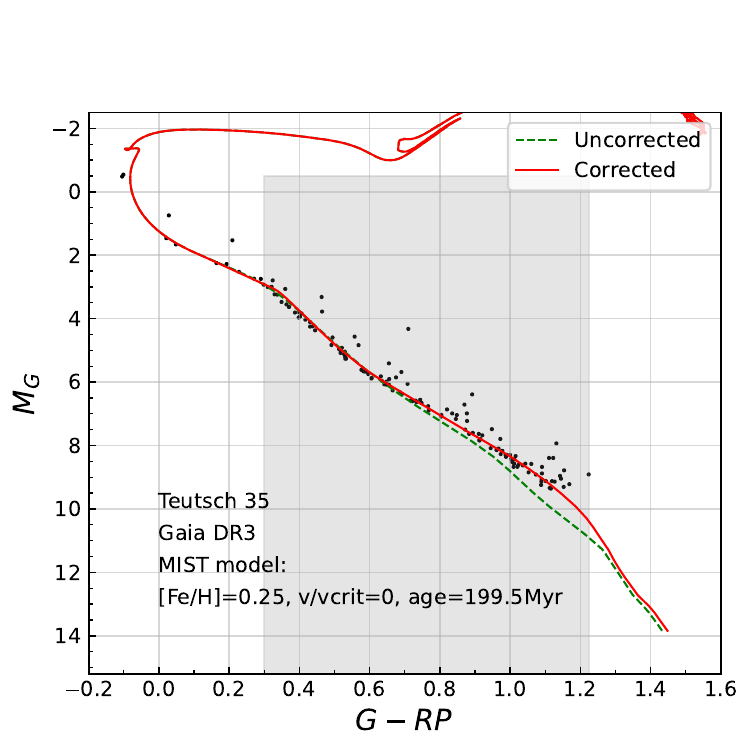}
    }
\caption{CMDs of open clusters used by MIST model.}
\label{fig:33_bp/g-rp_dr3_mist}
\end{figure*}

\setcounter{figure}{0}
\begin{figure*}
\centering
    {%
    \includegraphics[width=0.22\textwidth, trim=0.1cm 0.2cm 0.0cm 1.7cm, clip]{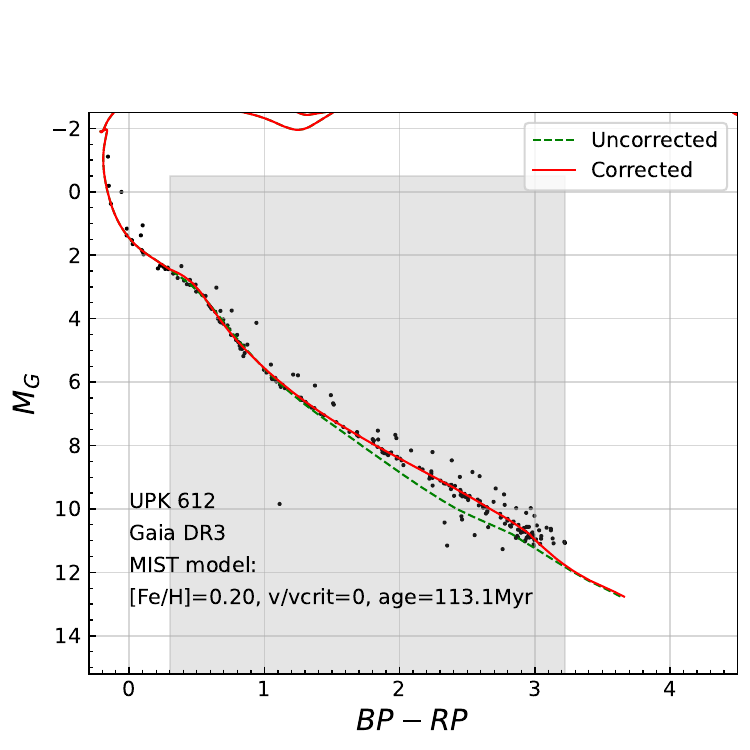}\quad
    \includegraphics[width=0.22\textwidth, trim=0.1cm 0.2cm 0.0cm 1.7cm, clip]{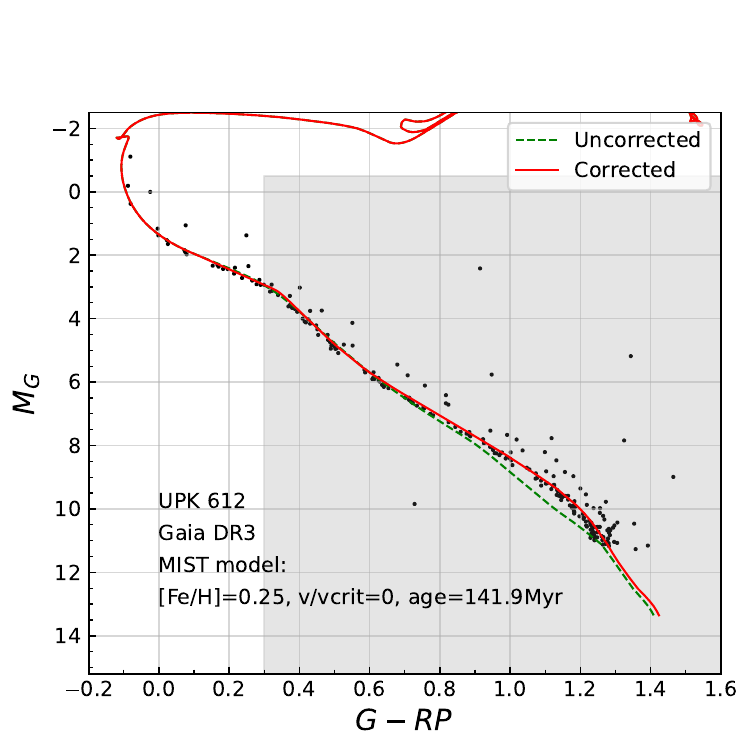}
    }
    {%
    \includegraphics[width=0.22\textwidth, trim=0.1cm 0.2cm 0.0cm 1.7cm, clip]{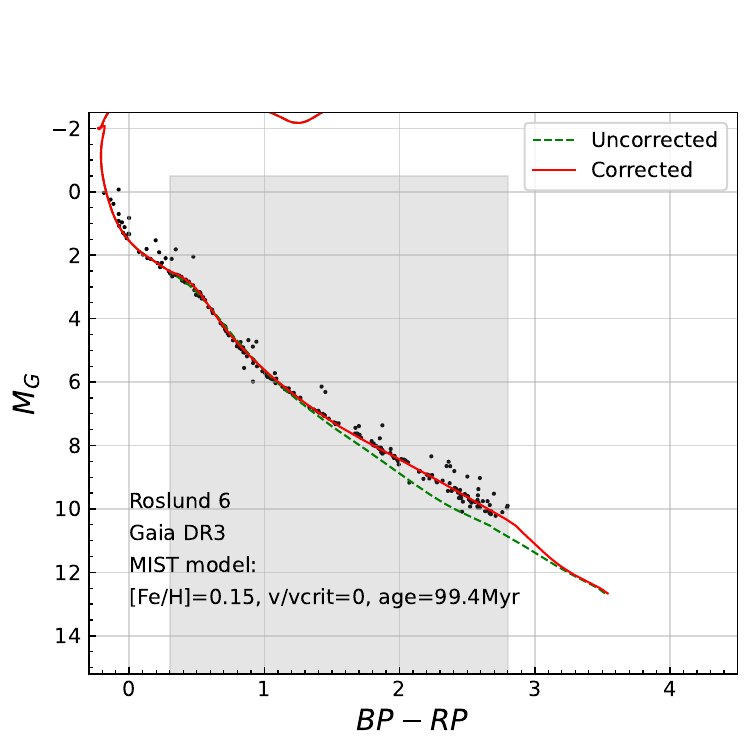}\quad
    \includegraphics[width=0.22\textwidth, trim=0.1cm 0.2cm 0.0cm 1.7cm, clip]{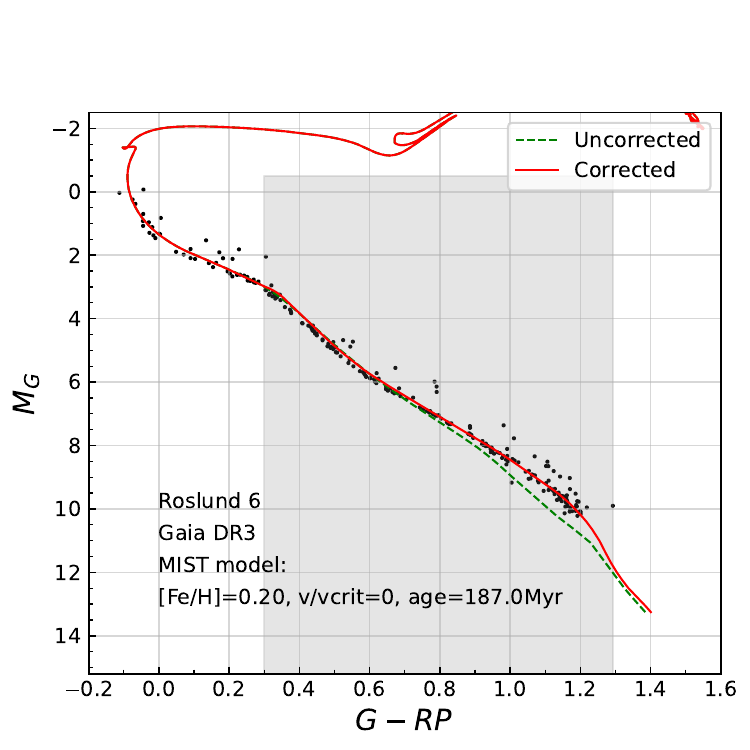}
    }
    {%
    \includegraphics[width=0.22\textwidth, trim=0.1cm 0.2cm 0.0cm 1.7cm, clip]{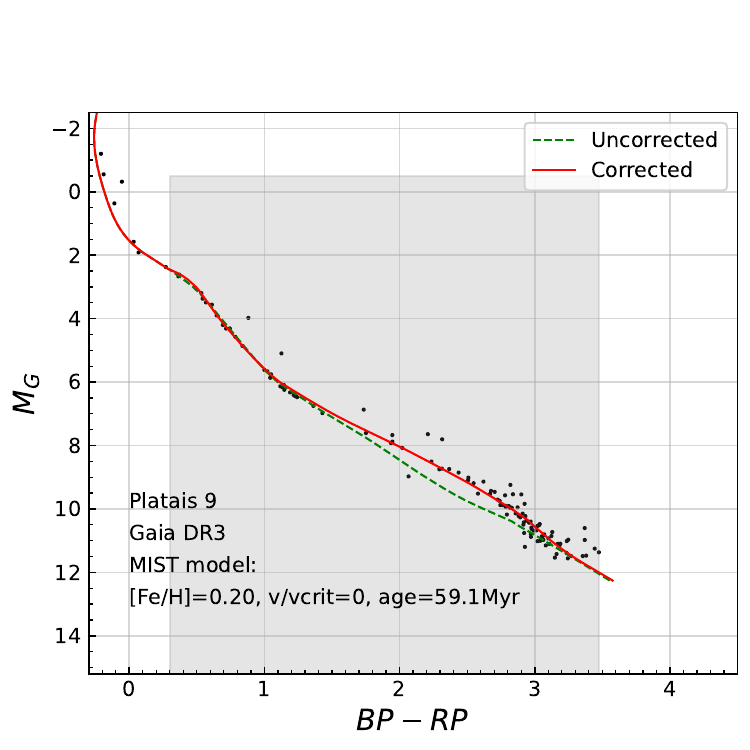}\quad
    \includegraphics[width=0.22\textwidth, trim=0.1cm 0.2cm 0.0cm 1.7cm, clip]{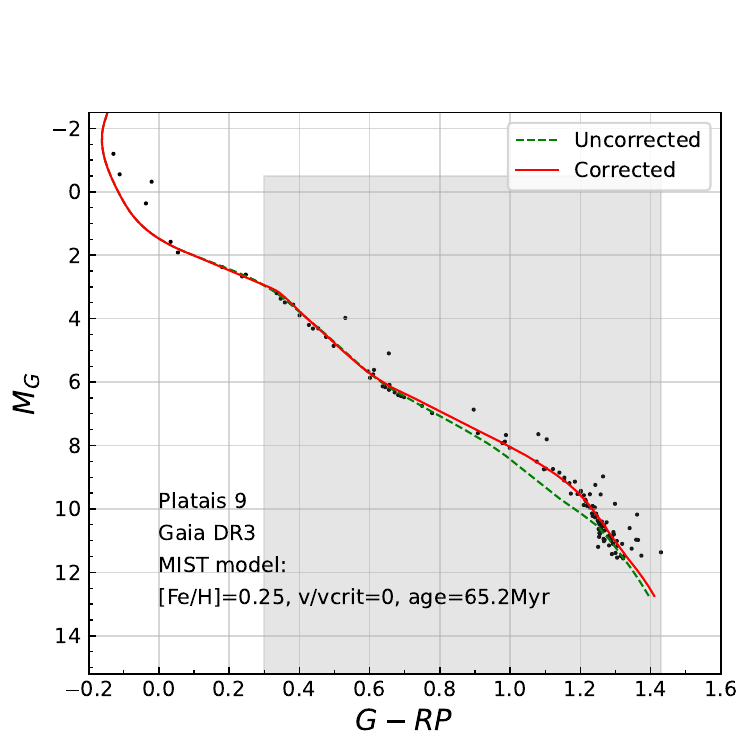}
    }
    {%
    \includegraphics[width=0.22\textwidth, trim=0.1cm 0.2cm 0.0cm 1.7cm, clip]{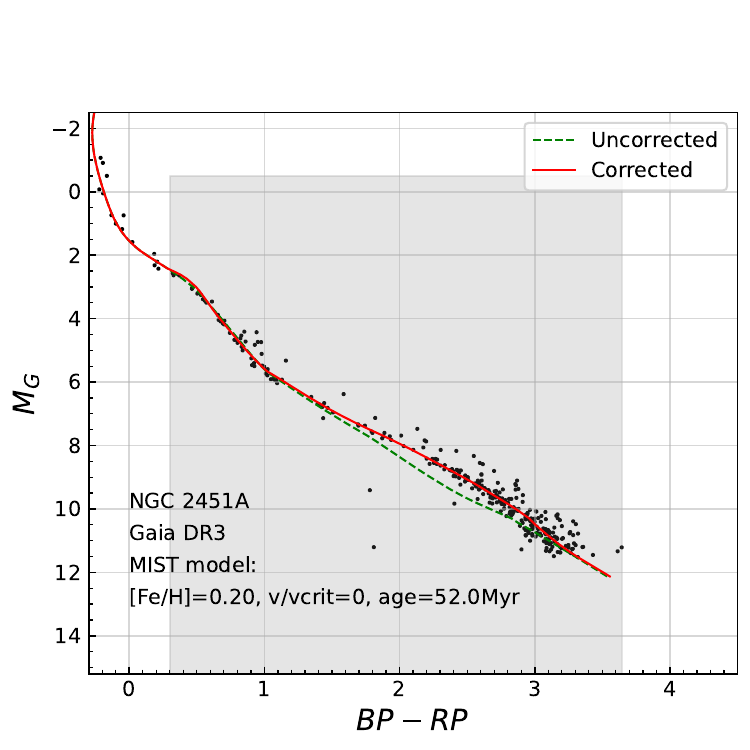}\quad
    \includegraphics[width=0.22\textwidth, trim=0.1cm 0.2cm 0.0cm 1.7cm, clip]{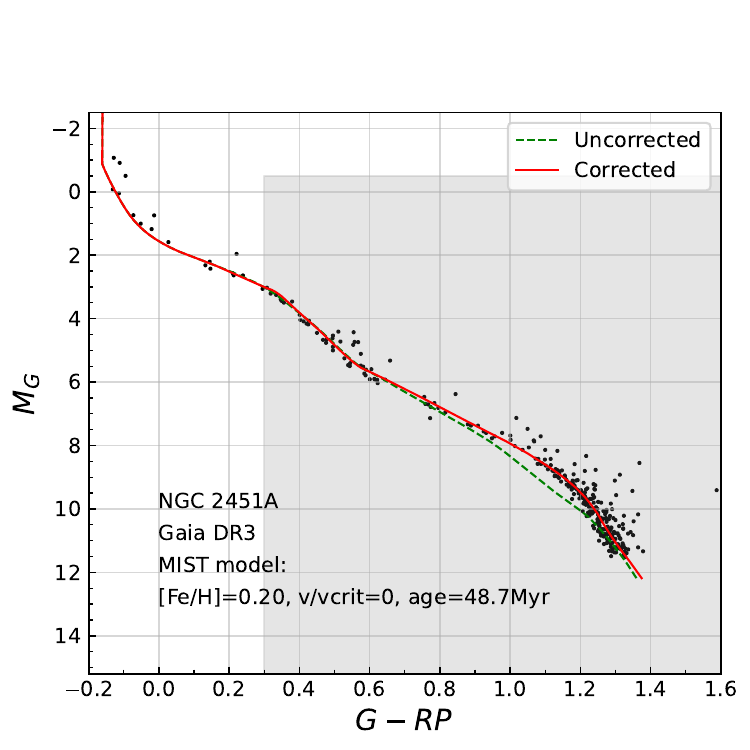}
    }
    {%
    \includegraphics[width=0.22\textwidth, trim=0.1cm 0.2cm 0.0cm 1.7cm, clip]{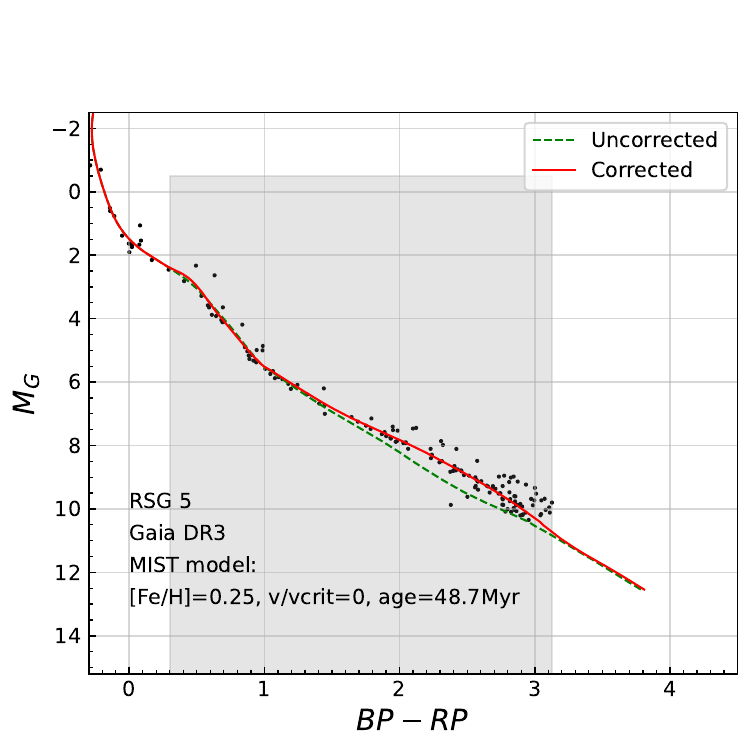}\quad
    \includegraphics[width=0.22\textwidth, trim=0.1cm 0.2cm 0.0cm 1.7cm, clip]{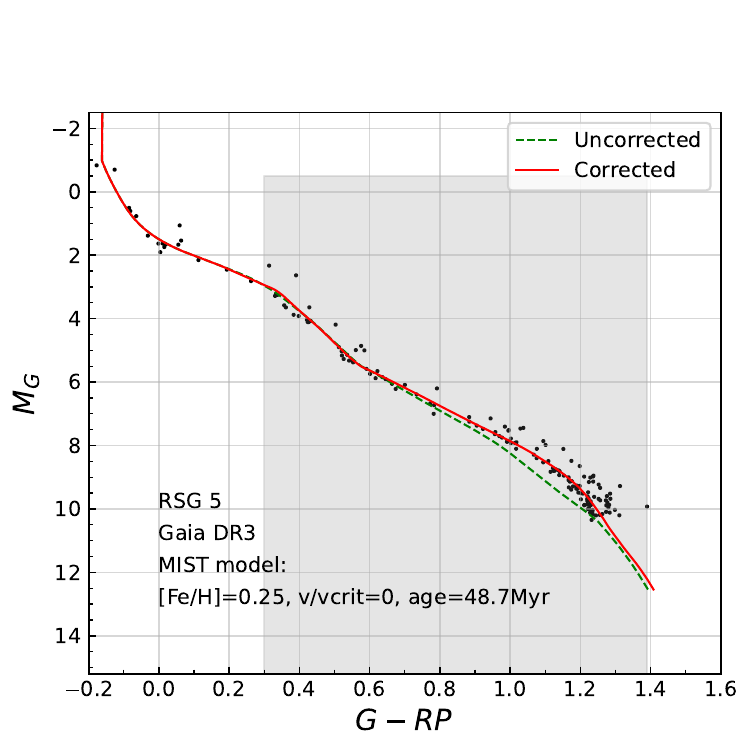}
    }
    {%
    \includegraphics[width=0.22\textwidth, trim=0.1cm 0.2cm 0.0cm 1.7cm, clip]{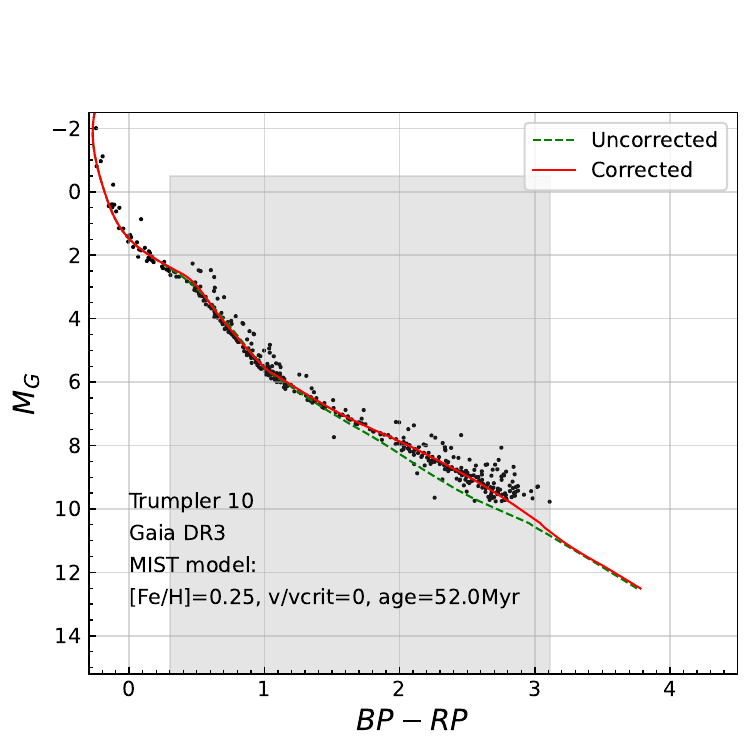}\quad
    \includegraphics[width=0.22\textwidth, trim=0.1cm 0.2cm 0.0cm 1.7cm, clip]{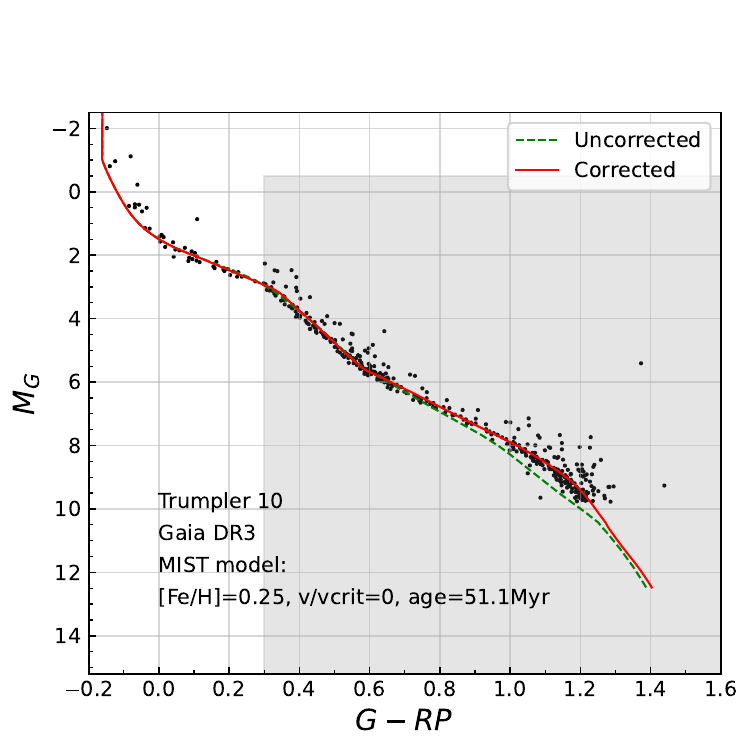}
    }
    {%
    \includegraphics[width=0.22\textwidth, trim=0.1cm 0.2cm 0.0cm 1.7cm, clip]{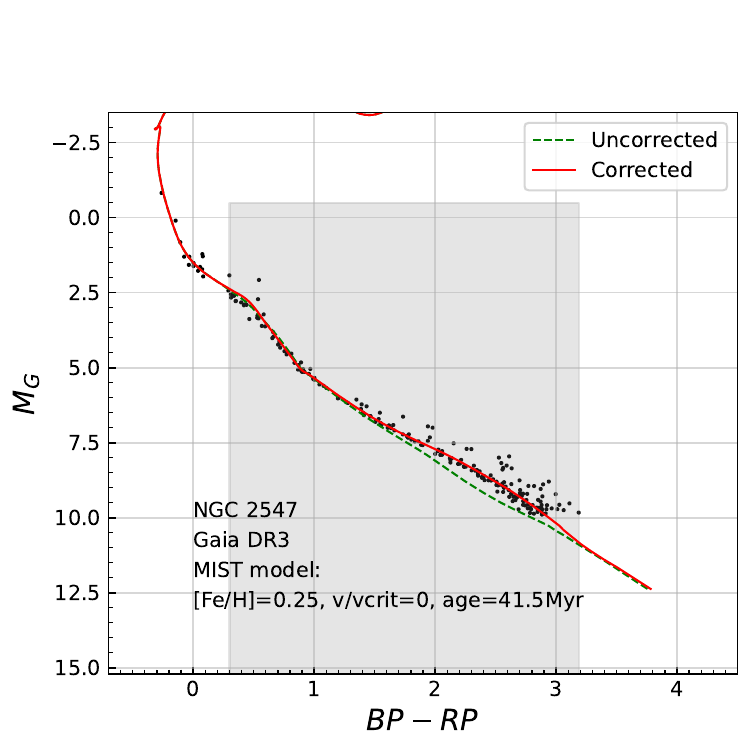}\quad
    \includegraphics[width=0.22\textwidth, trim=0.1cm 0.2cm 0.0cm 1.7cm, clip]{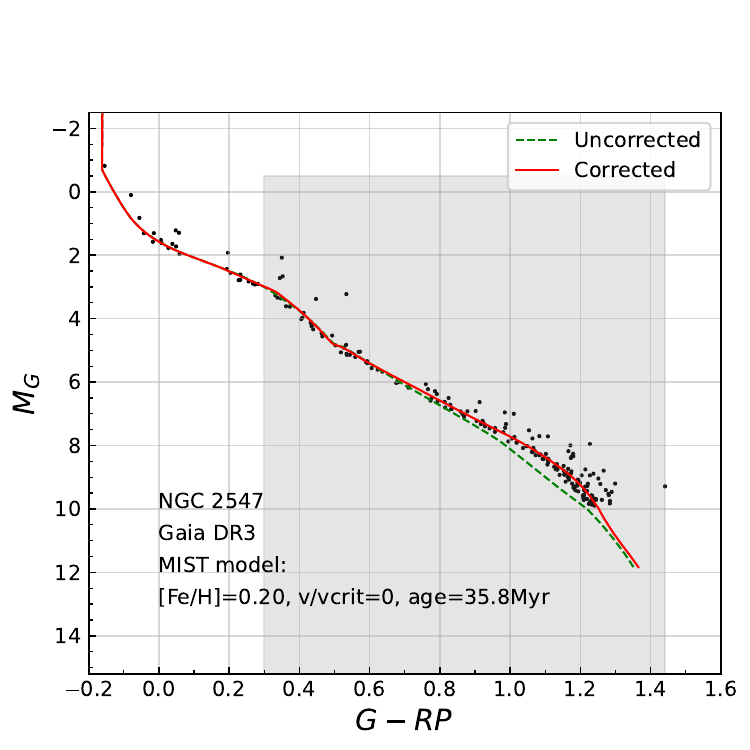}
    }
    {%
    \includegraphics[width=0.22\textwidth, trim=0.1cm 0.2cm 0.0cm 1.7cm, clip]{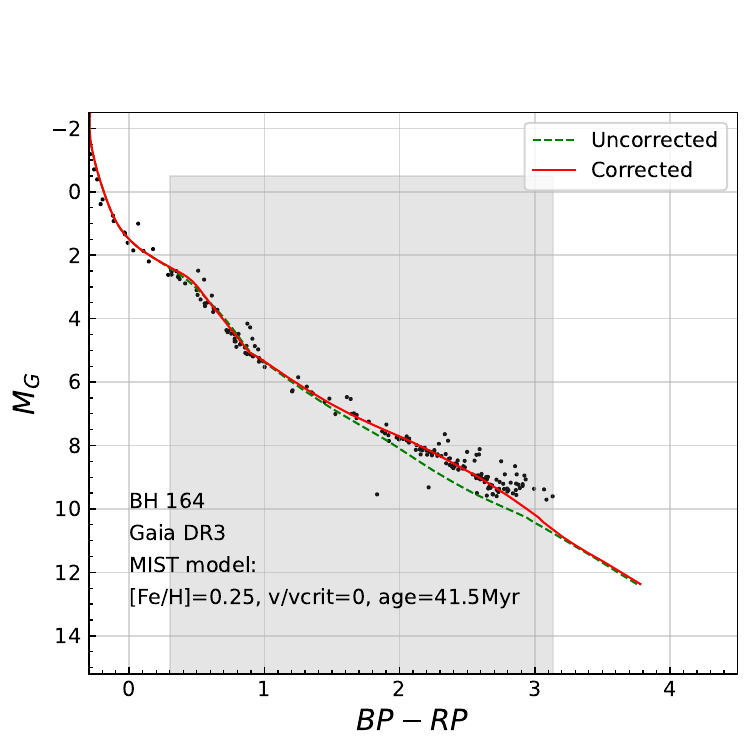}\quad
    \includegraphics[width=0.22\textwidth, trim=0.1cm 0.2cm 0.0cm 1.7cm, clip]{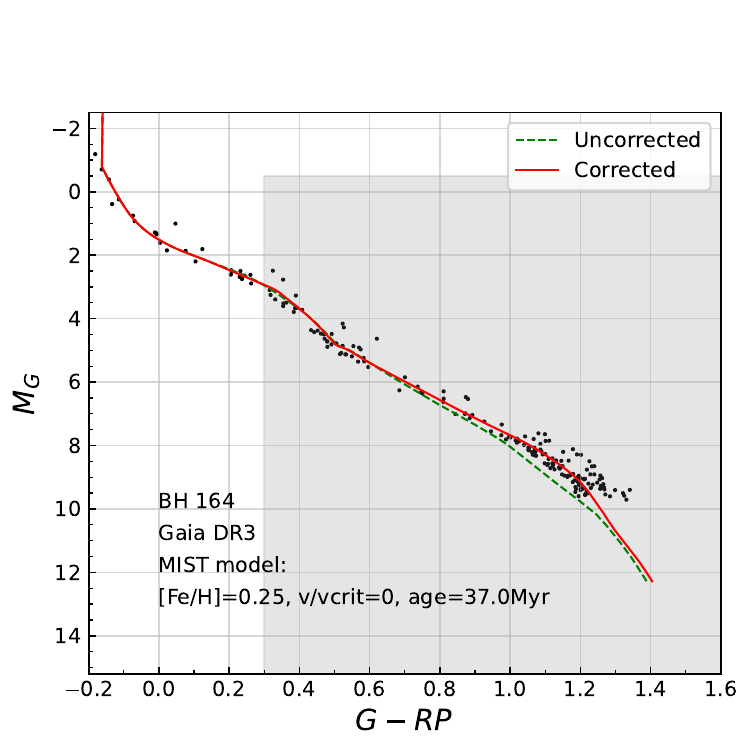}
    }
    {%
    \includegraphics[width=0.22\textwidth, trim=0.1cm 0.2cm 0.0cm 1.7cm, clip]{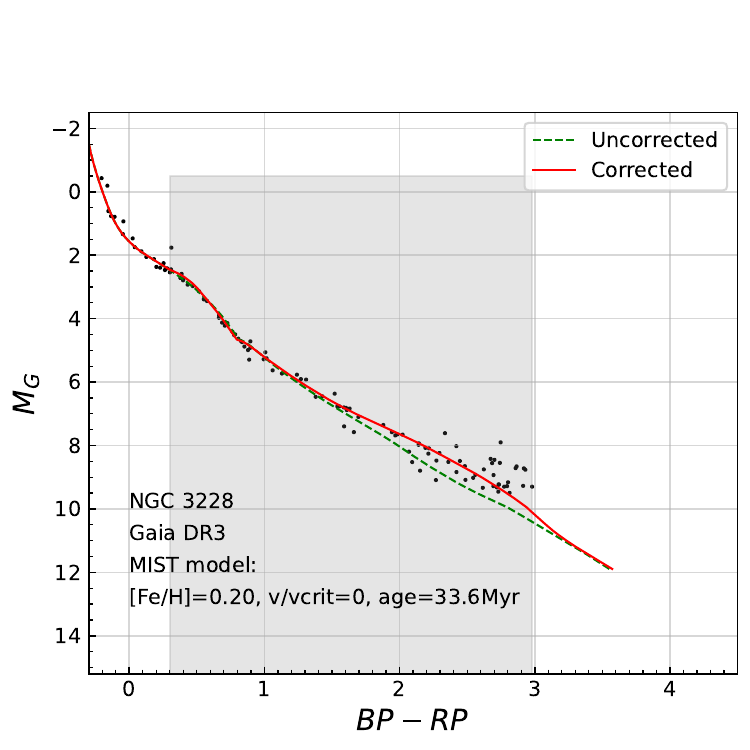}\quad
    \includegraphics[width=0.22\textwidth, trim=0.1cm 0.2cm 0.0cm 1.7cm, clip]{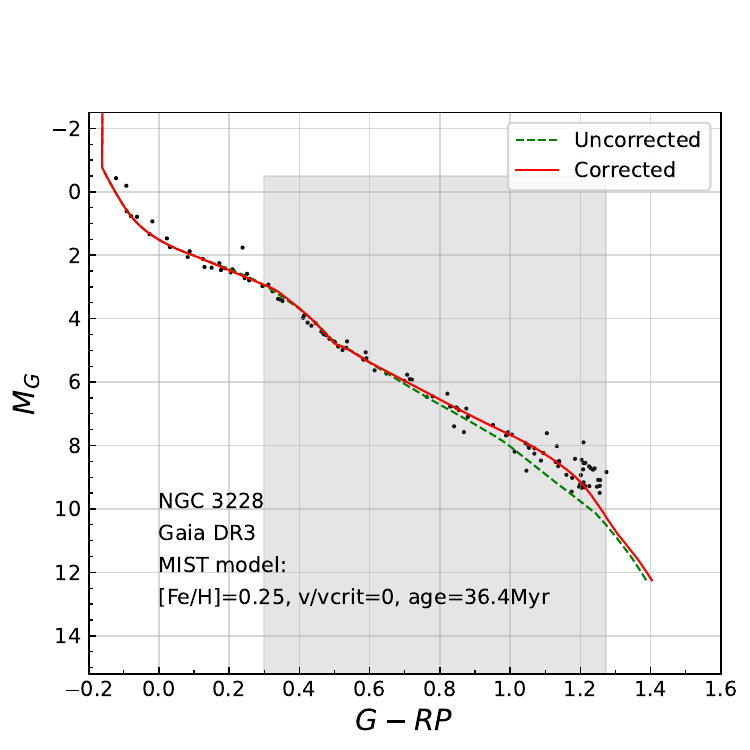}
    }
    {%
    \includegraphics[width=0.22\textwidth, trim=0.1cm 0.2cm 0.0cm 1.7cm, clip]{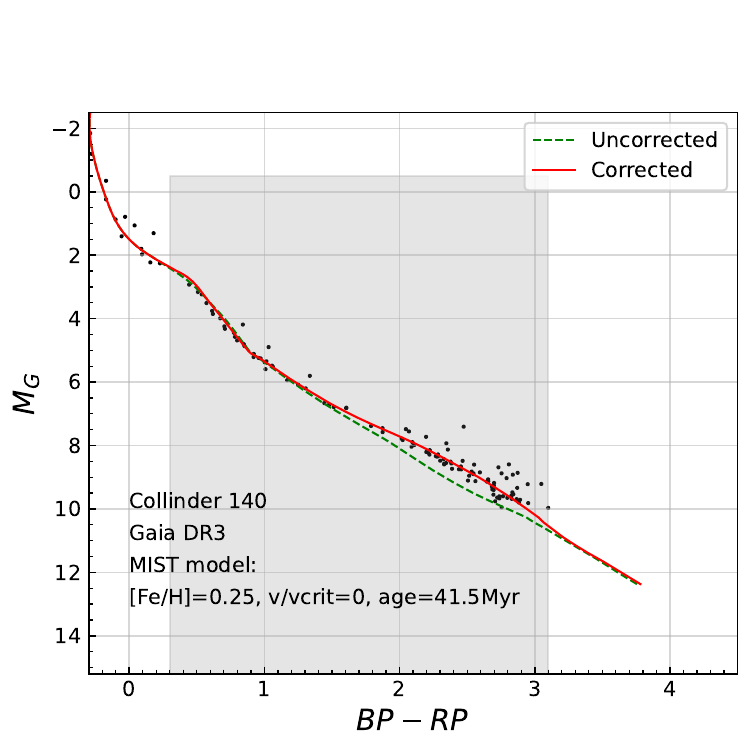}\quad
    \includegraphics[width=0.22\textwidth, trim=0.1cm 0.2cm 0.0cm 1.7cm, clip]{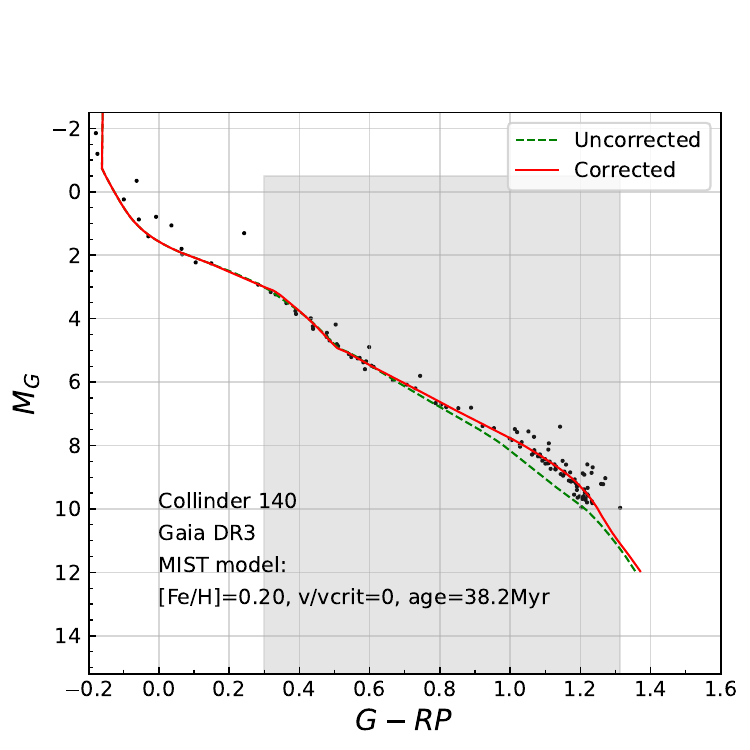}
    }
    {%
    \includegraphics[width=0.22\textwidth, trim=0.1cm 0.2cm 0.0cm 1.7cm, clip]{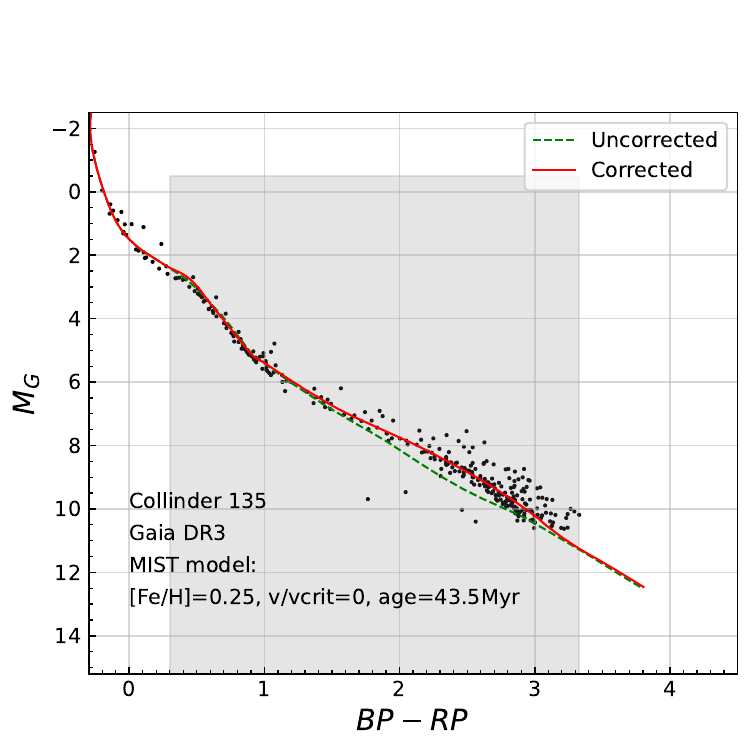}\quad
    \includegraphics[width=0.22\textwidth, trim=0.1cm 0.2cm 0.0cm 1.7cm, clip]{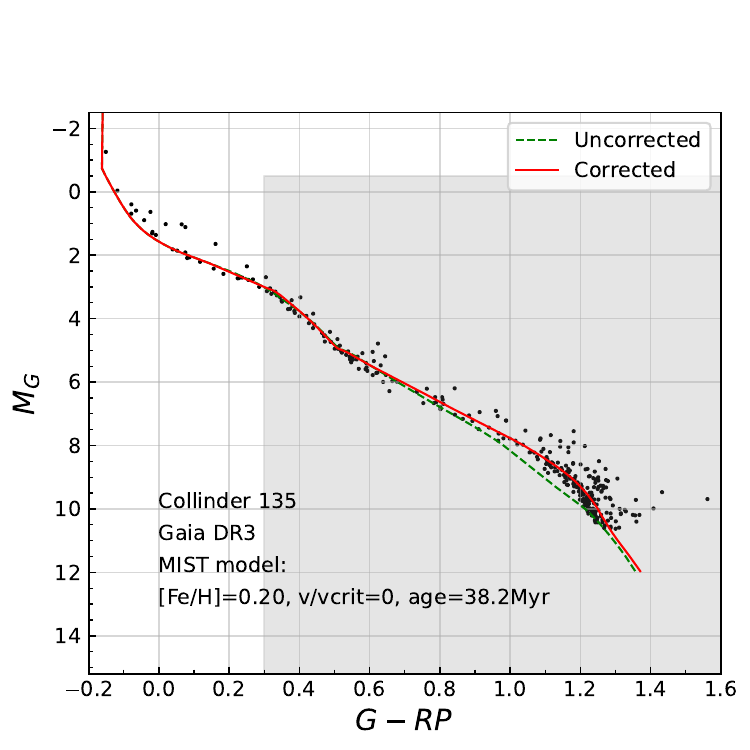}
    }
    {%
    \includegraphics[width=0.22\textwidth, trim=0.1cm 0.2cm 0.0cm 1.7cm, clip]{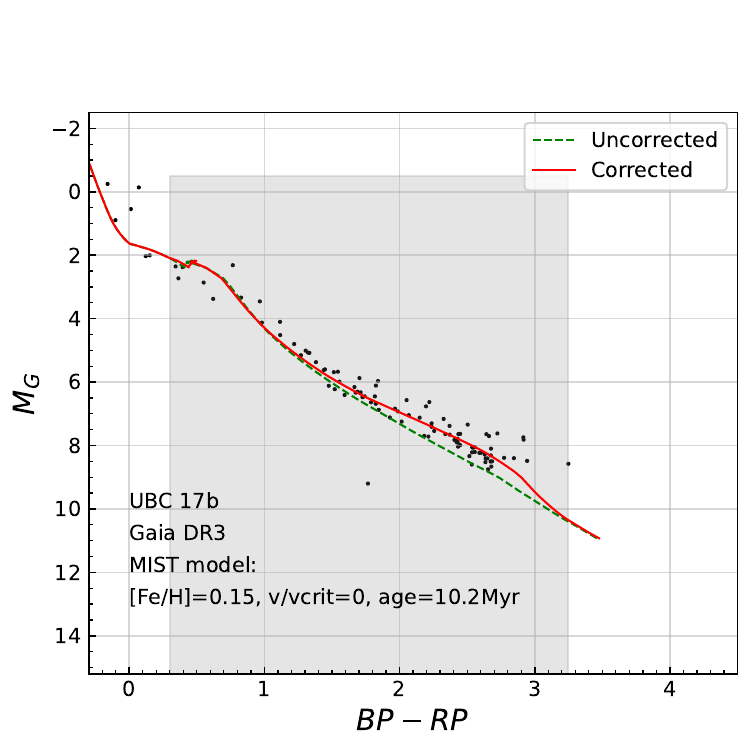}\quad
    \includegraphics[width=0.22\textwidth, trim=0.1cm 0.2cm 0.0cm 1.7cm, clip]{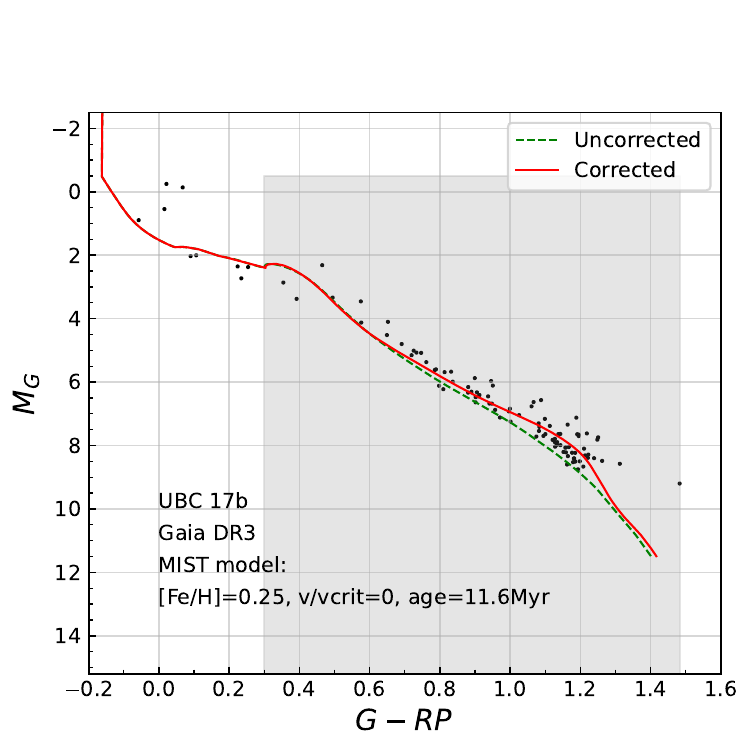}
    }
    {%
    \includegraphics[width=0.22\textwidth, trim=0.1cm 0.9cm 0.0cm 1.6cm, clip]{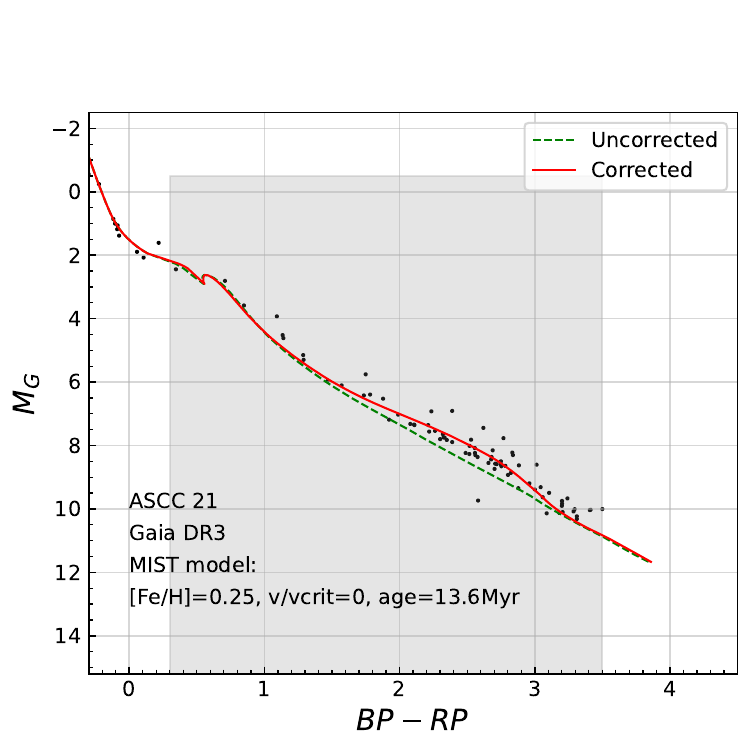}\quad
    \includegraphics[width=0.22\textwidth, trim=0.1cm 0.9cm 0.0cm 1.6cm, clip]{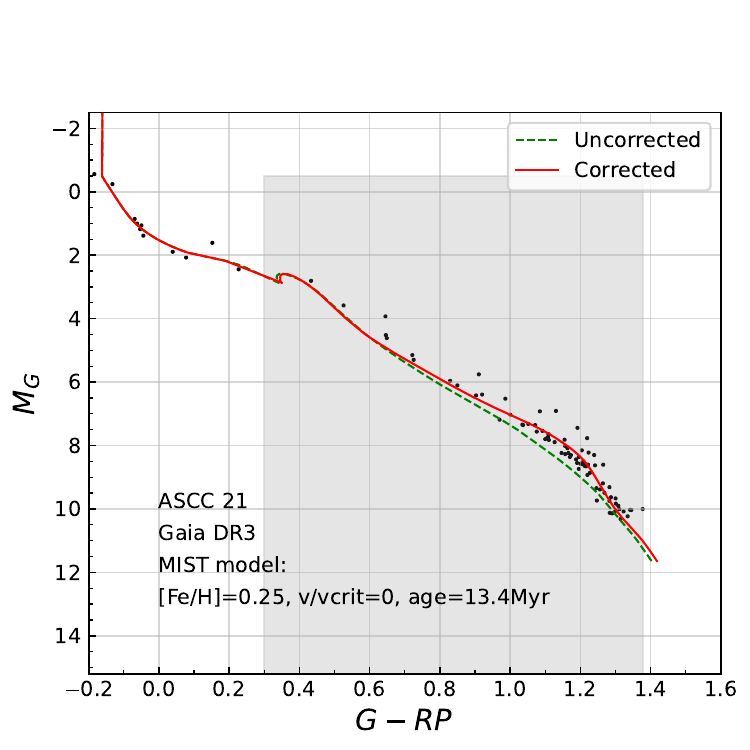}
    }
    {%
    \includegraphics[width=0.22\textwidth, trim=0.1cm 0.9cm 0.0cm 1.6cm, clip]{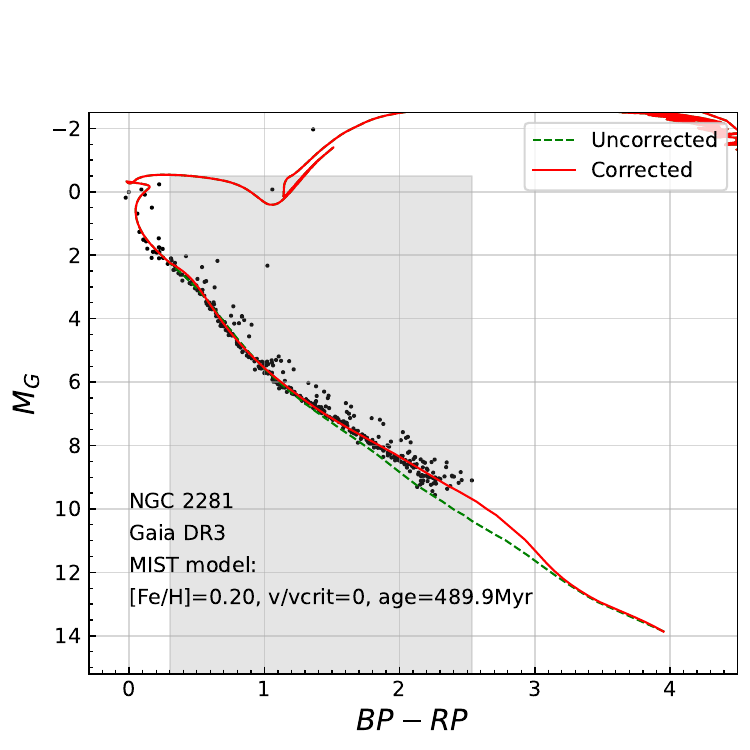}\quad
    \includegraphics[width=0.22\textwidth, trim=0.1cm 0.9cm 0.0cm 1.6cm, clip]{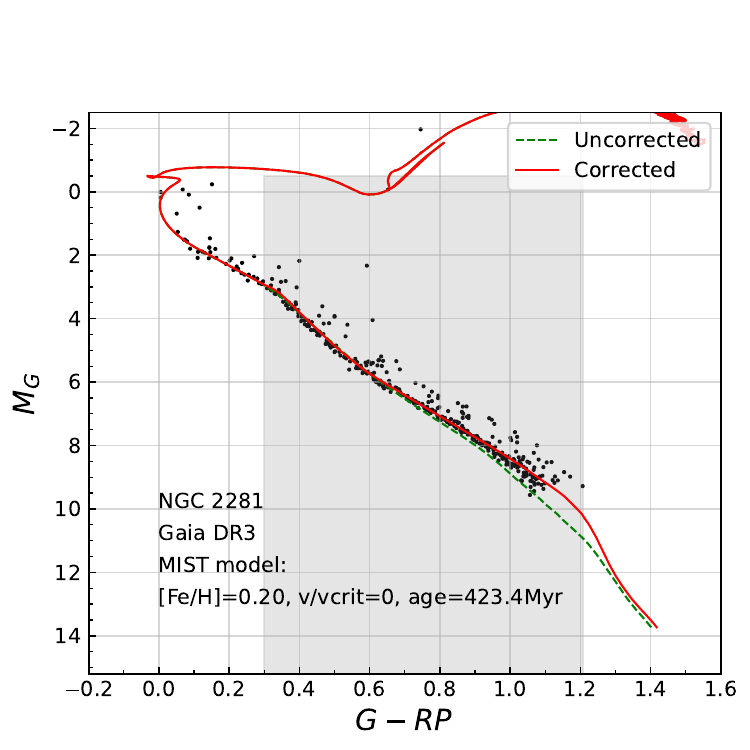}
    }   
\caption{Continued from Fig~\ref{fig:33_bp/g-rp_dr3_mist}}
\end{figure*}

\setcounter{figure}{0}
\begin{figure*}
\centering
    {%
    \includegraphics[width=0.22\textwidth, trim=0.1cm 0.2cm 0.0cm 1.7cm, clip]{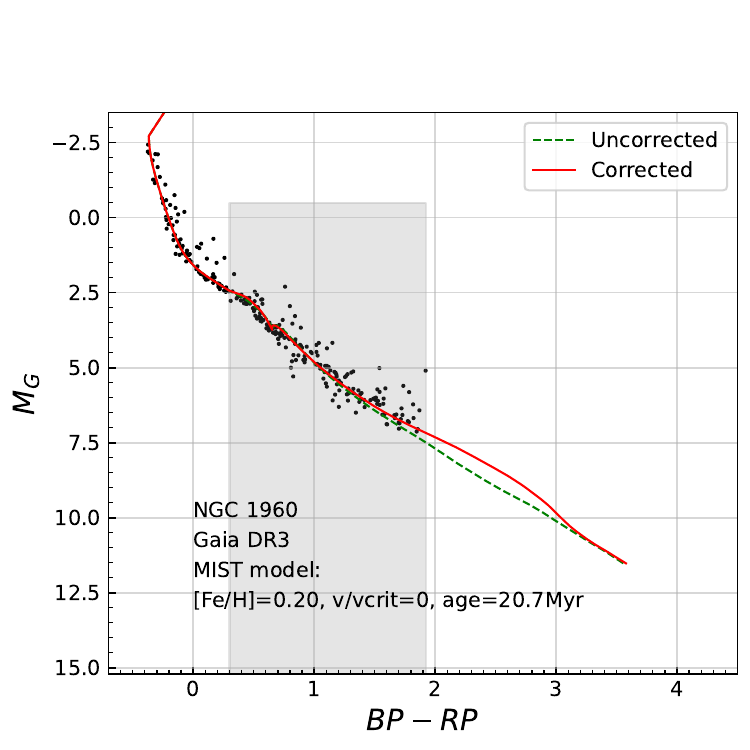}\quad
    \includegraphics[width=0.22\textwidth, trim=0.1cm 0.2cm 0.0cm 1.7cm, clip]{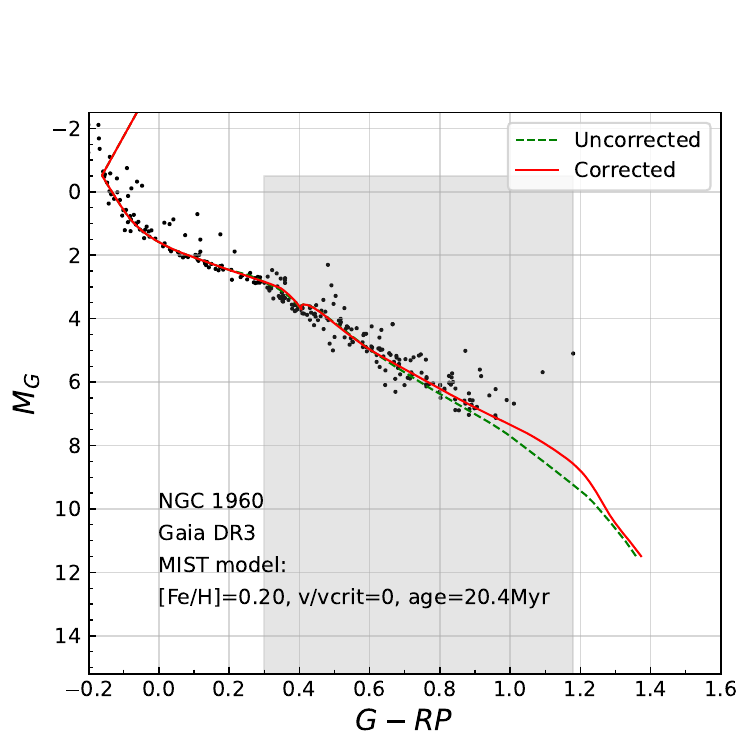}
    } 
    {%
    \includegraphics[width=0.22\textwidth, trim=0.1cm 0.2cm 0.0cm 1.7cm, clip]{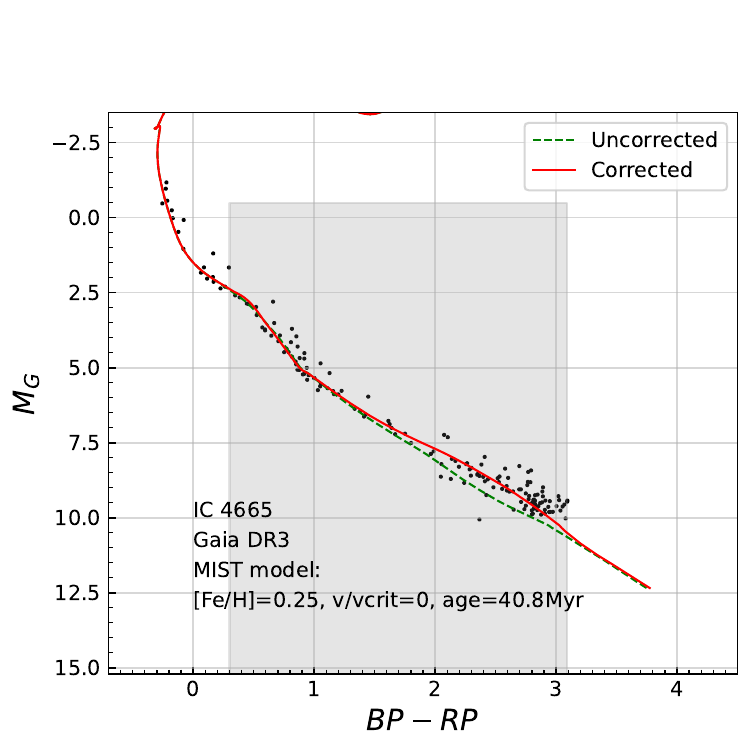}\quad
    \includegraphics[width=0.22\textwidth, trim=0.1cm 0.2cm 0.0cm 1.7cm, clip]{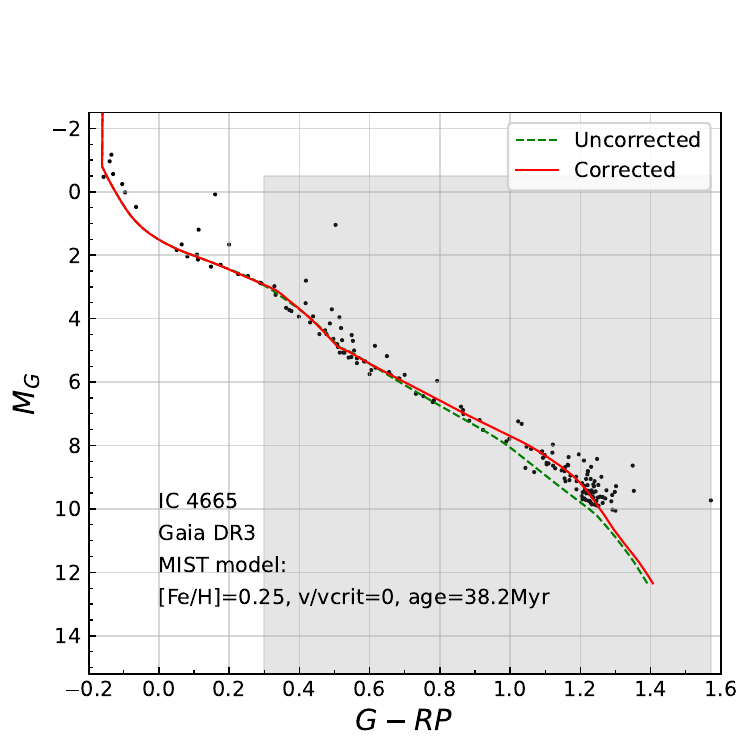}
    }
    {%
    \includegraphics[width=0.22\textwidth, trim=0.1cm 0.2cm 0.0cm 1.7cm, clip]{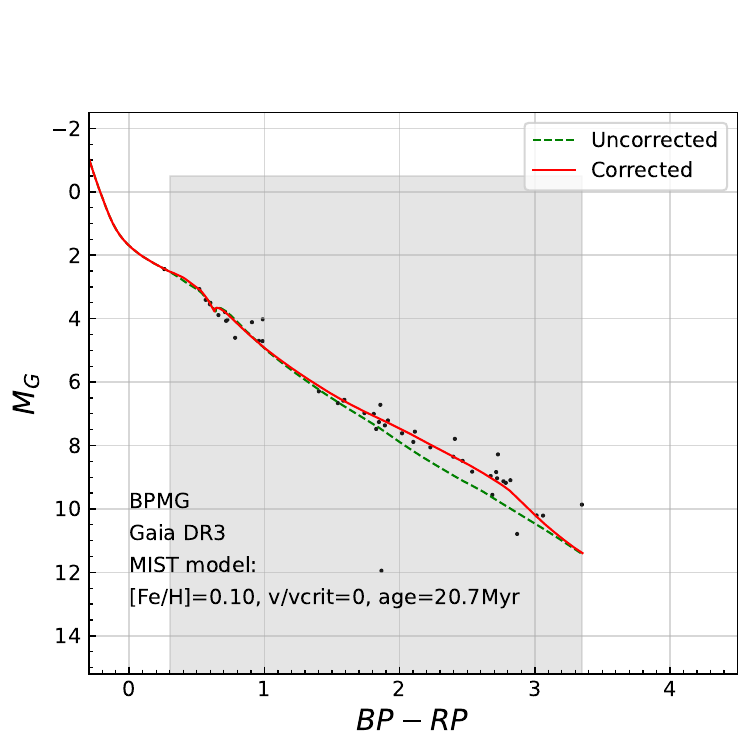}\quad
    \includegraphics[width=0.22\textwidth, trim=0.1cm 0.2cm 0.0cm 1.7cm, clip]{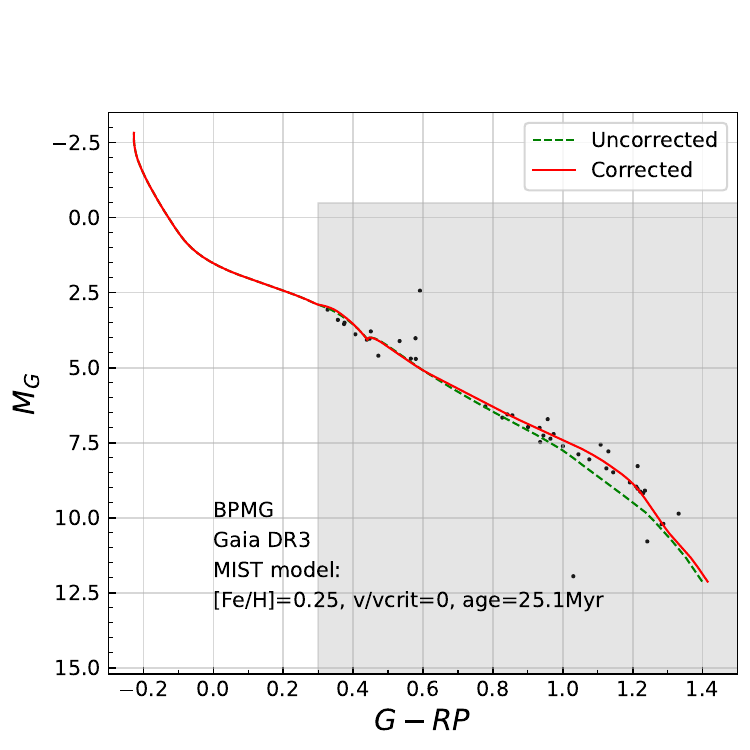}
    }
    {%
    \includegraphics[width=0.22\textwidth, trim=0.1cm 0.2cm 0.0cm 1.7cm, clip]{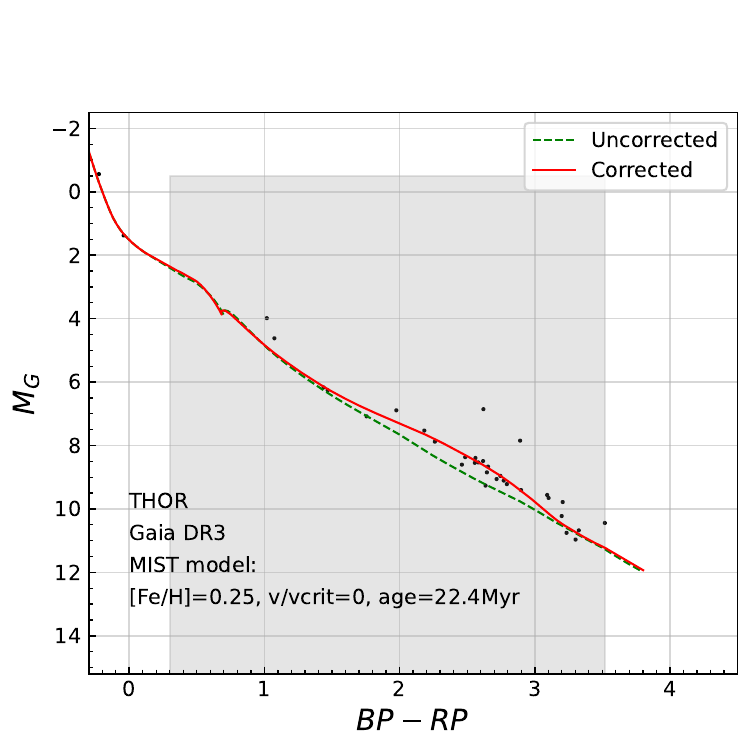}\quad
    \includegraphics[width=0.22\textwidth, trim=0.1cm 0.2cm 0.0cm 1.7cm, clip]{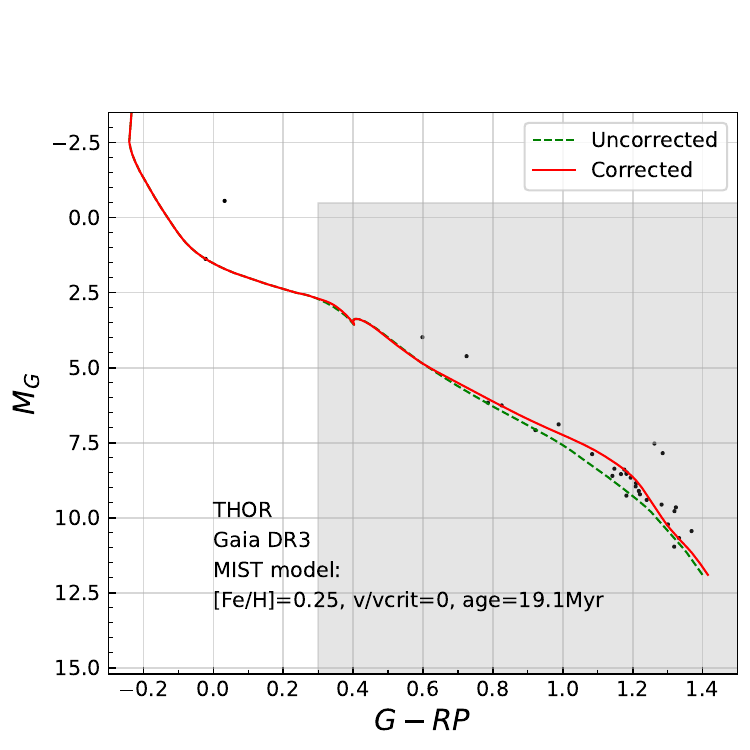}
    }
    {%
    \includegraphics[width=0.22\textwidth, trim=0.1cm 0.2cm 0.0cm 1.7cm, clip]{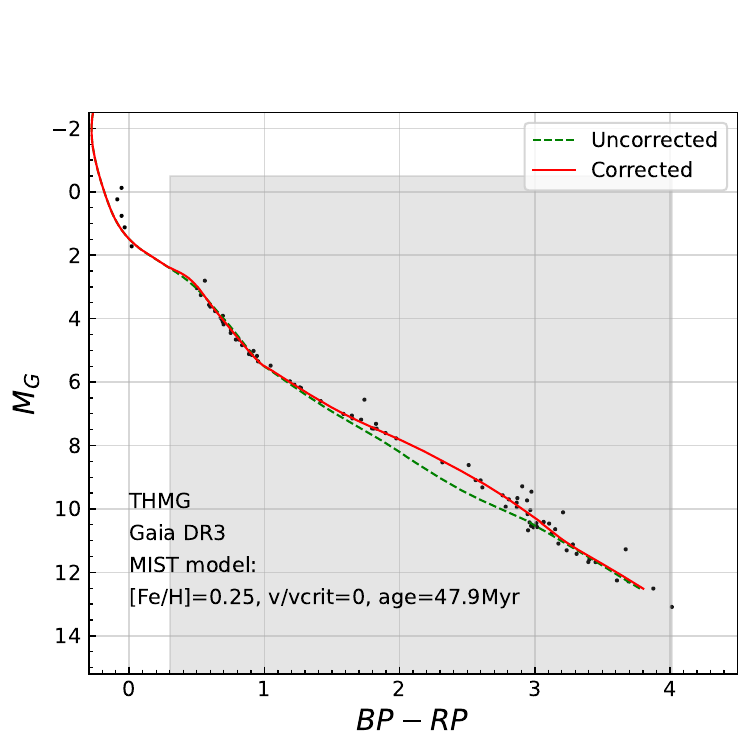}\quad
    \includegraphics[width=0.22\textwidth, trim=0.1cm 0.2cm 0.0cm 1.7cm, clip]{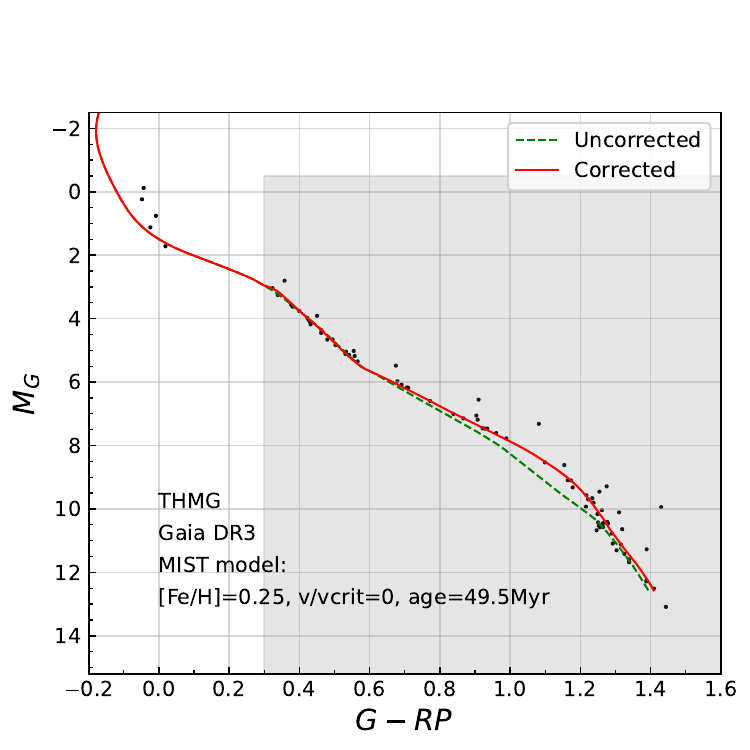}
    }
\caption{Continued from Fig~\ref{fig:33_bp/g-rp_dr3_mist}}
\end{figure*}

\setcounter{figure}{1}
\begin{figure*}
\centering
    {%
    \includegraphics[width=0.22\textwidth, trim=0.1cm 0.2cm 0.0cm 1.7cm, clip]{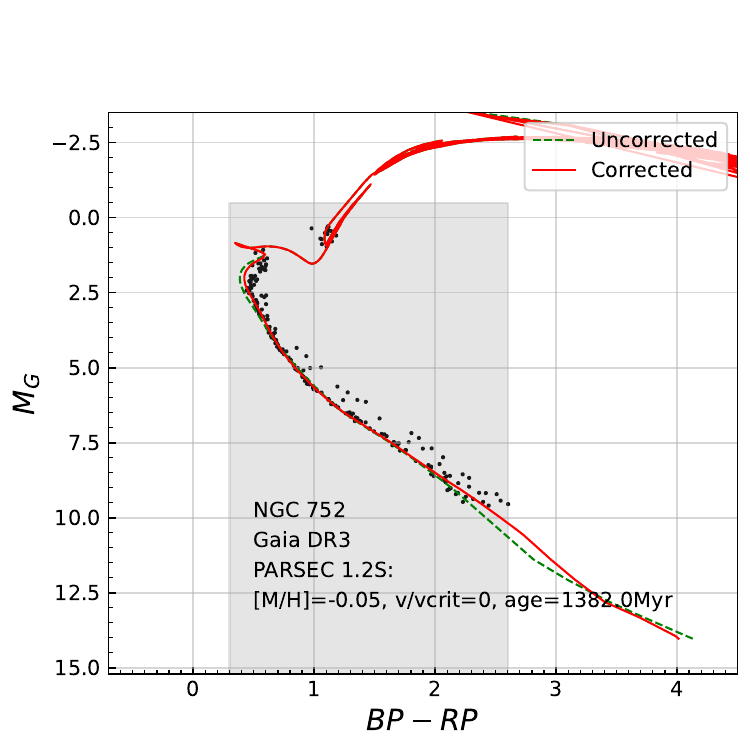}\quad
    \includegraphics[width=0.22\textwidth, trim=0.1cm 0.2cm 0.0cm 1.7cm, clip]{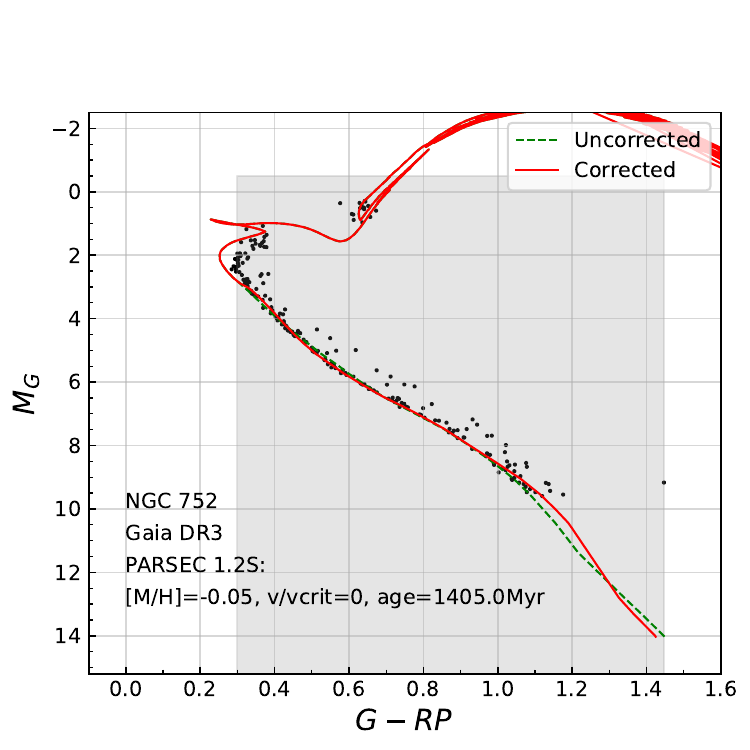}
    }
    {%
    \includegraphics[width=0.22\textwidth, trim=0.1cm 0.2cm 0.0cm 1.7cm, clip]{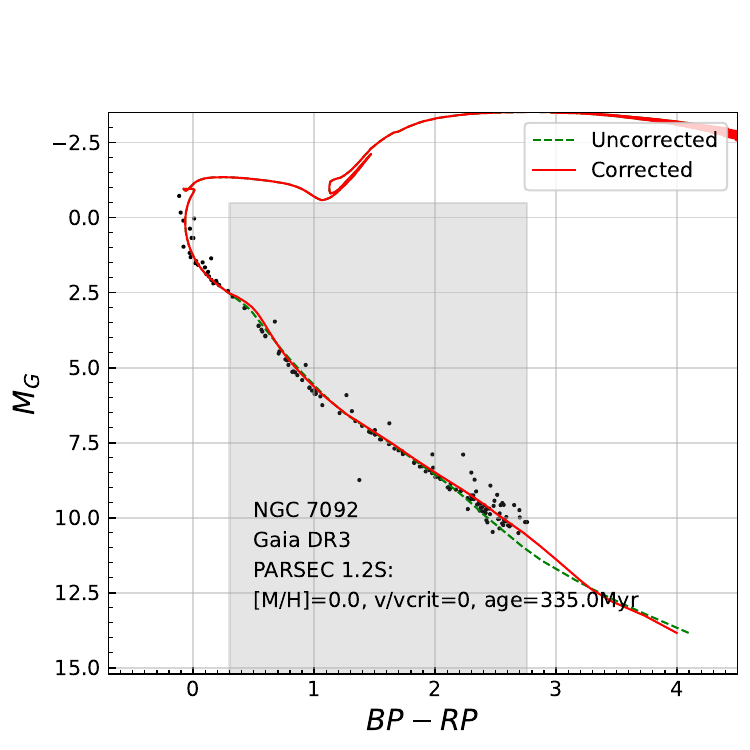}\quad
    \includegraphics[width=0.22\textwidth, trim=0.1cm 0.2cm 0.0cm 1.7cm, clip]{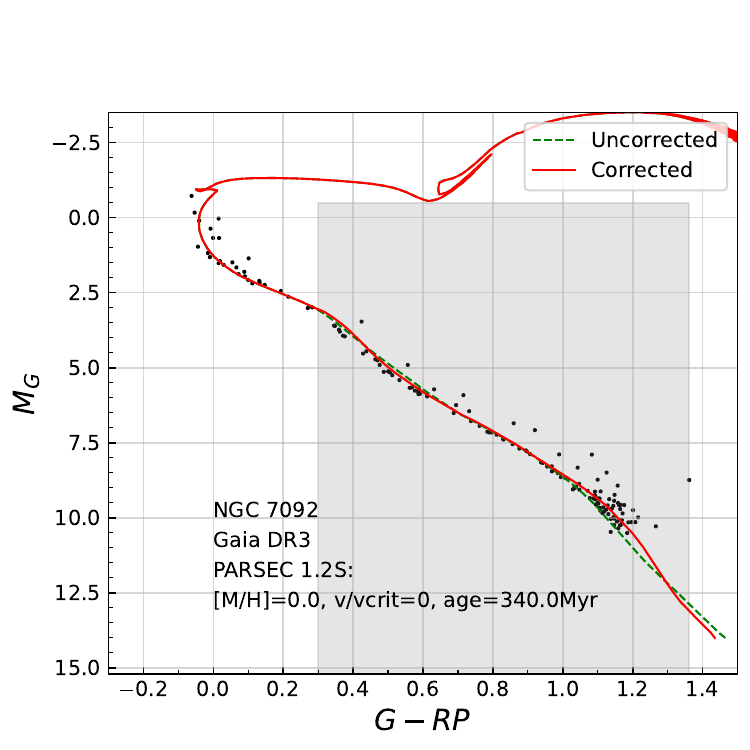}
    }
    {%
    \includegraphics[width=0.22\textwidth, trim=0.1cm 0.2cm 0.0cm 1.7cm, clip]{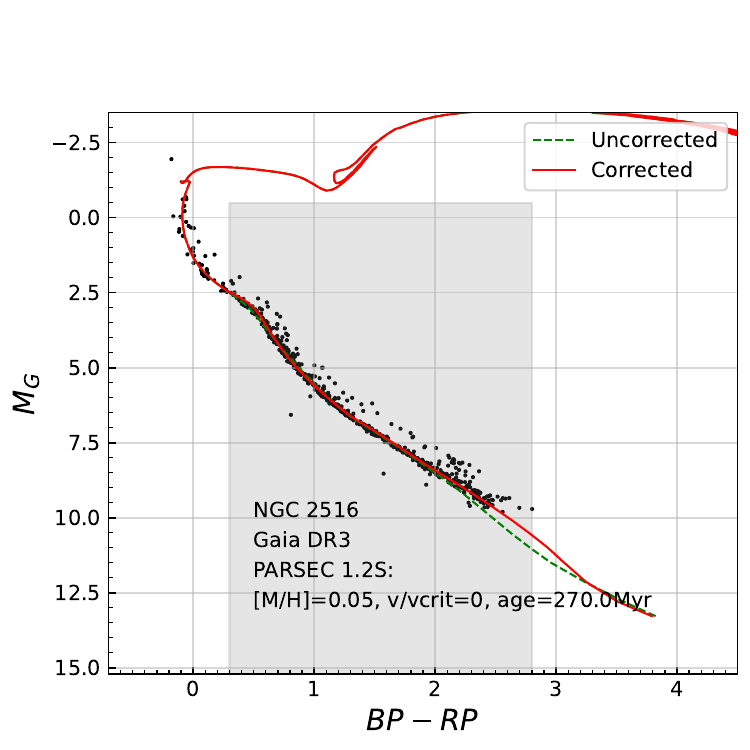}\quad
    \includegraphics[width=0.22\textwidth, trim=0.1cm 0.2cm 0.0cm 1.7cm, clip]{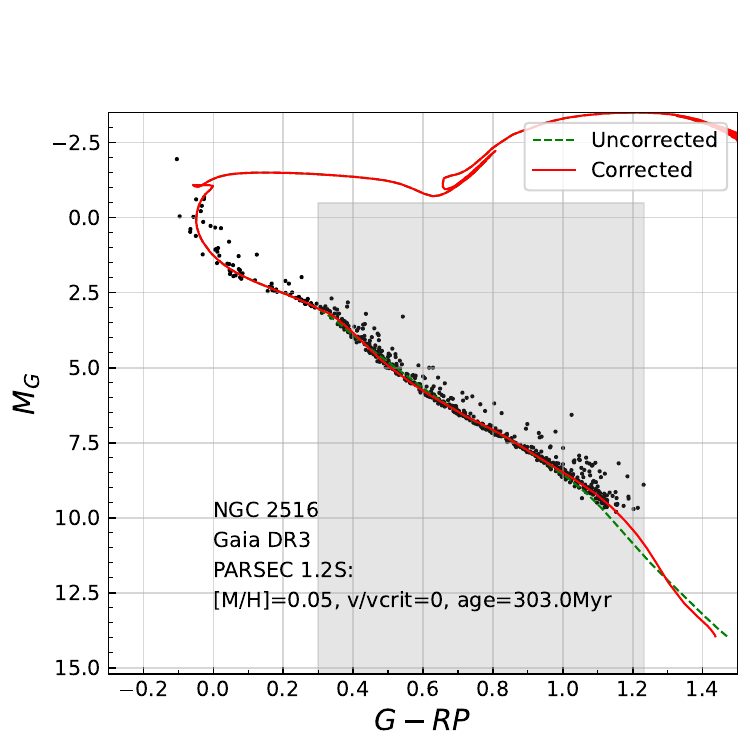}
    }
    {%
    \includegraphics[width=0.22\textwidth, trim=0.1cm 0.2cm 0.0cm 1.7cm, clip]{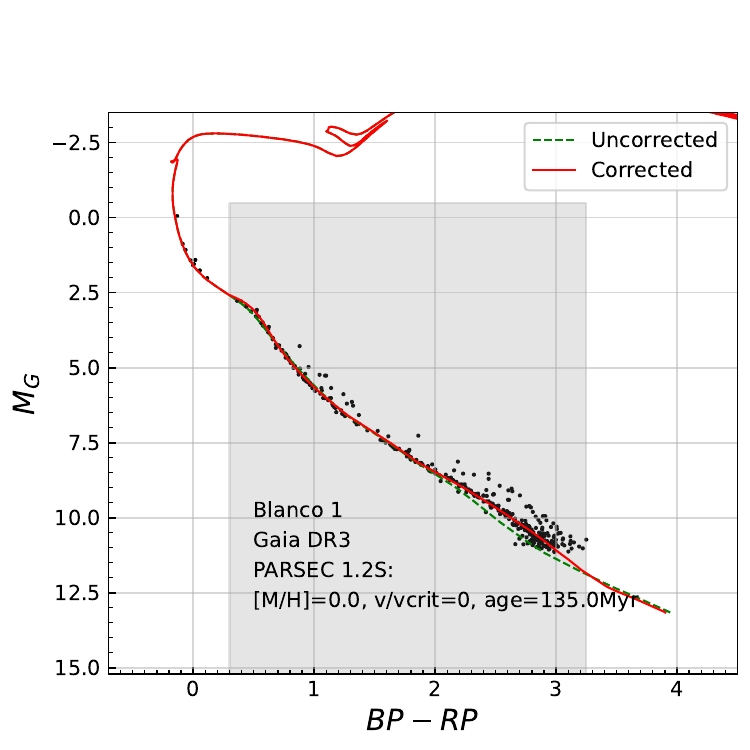}\quad
    \includegraphics[width=0.22\textwidth, trim=0.1cm 0.2cm 0.0cm 1.7cm, clip]{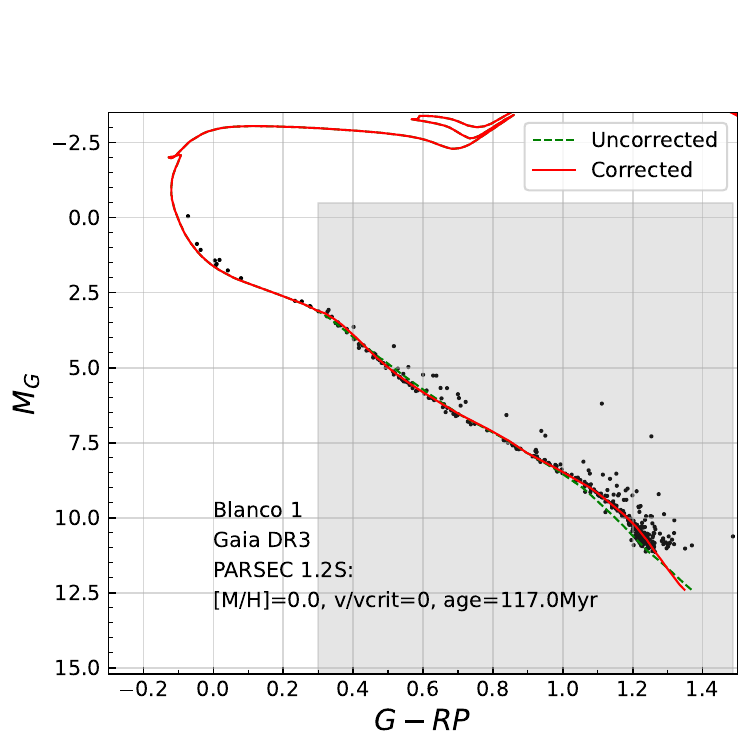}
    }
    {%
    \includegraphics[width=0.22\textwidth, trim=0.1cm 0.2cm 0.0cm 1.7cm, clip]{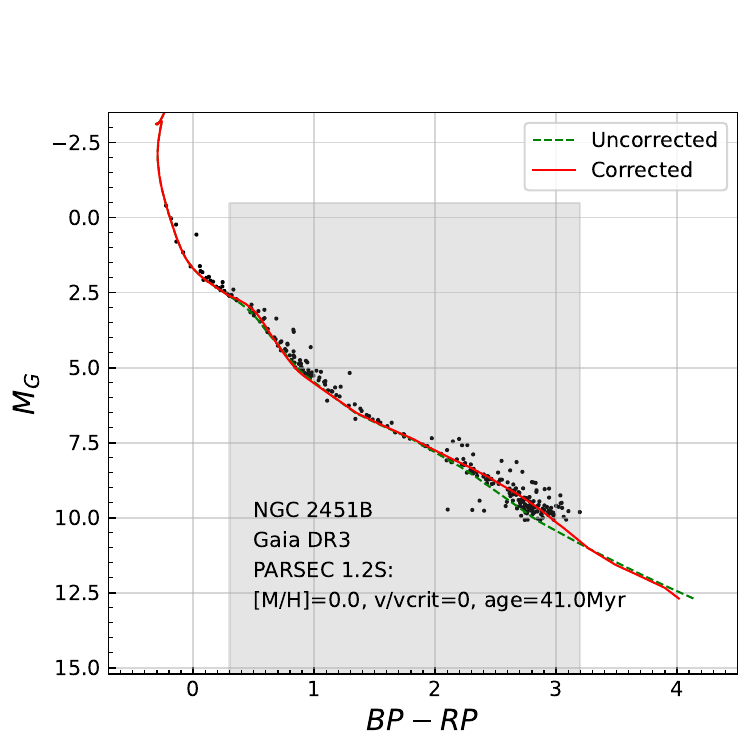}\quad
    \includegraphics[width=0.22\textwidth, trim=0.1cm 0.2cm 0.0cm 1.7cm, clip]{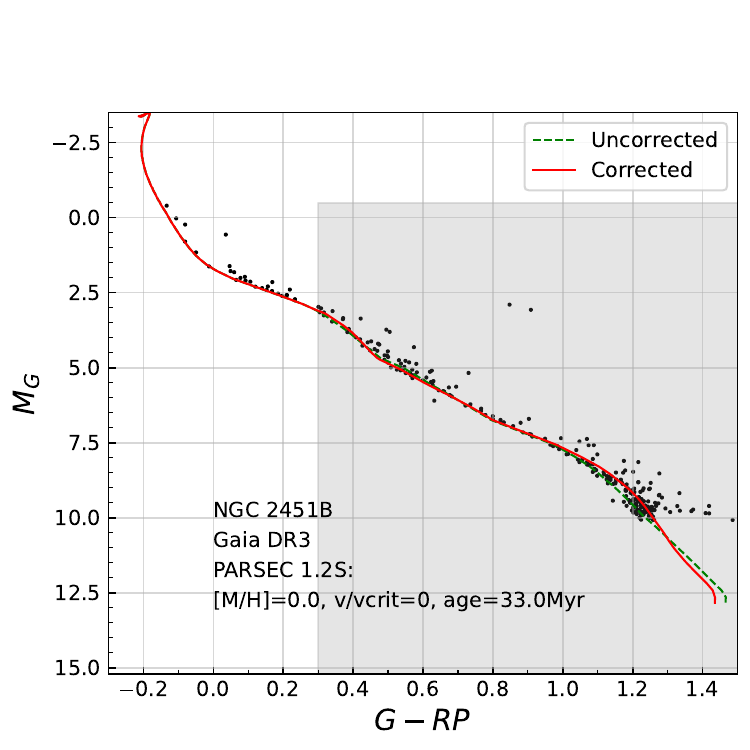}
    }
    {%
    \includegraphics[width=0.22\textwidth, trim=0.1cm 0.2cm 0.0cm 1.7cm, clip]{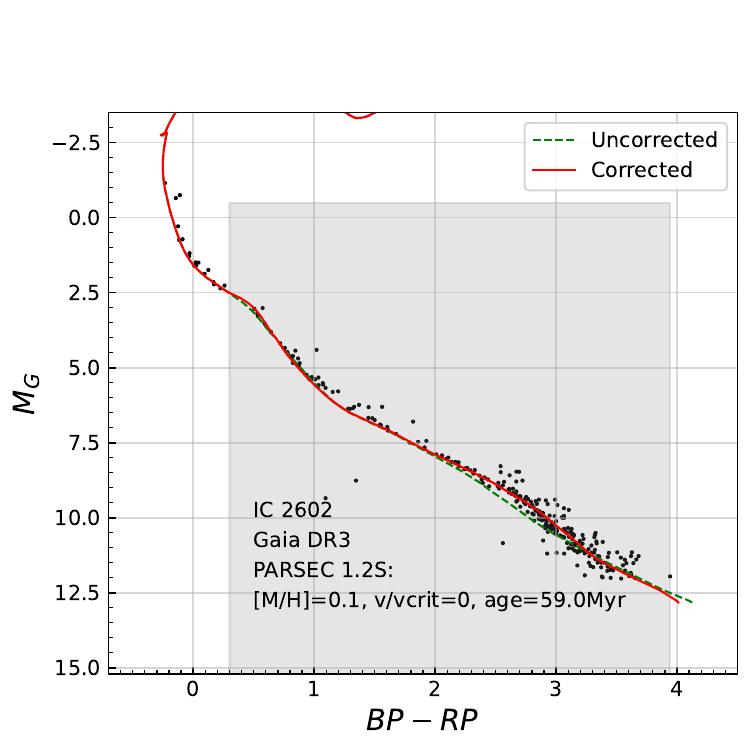}\quad
    \includegraphics[width=0.22\textwidth, trim=0.1cm 0.2cm 0.0cm 1.7cm, clip]{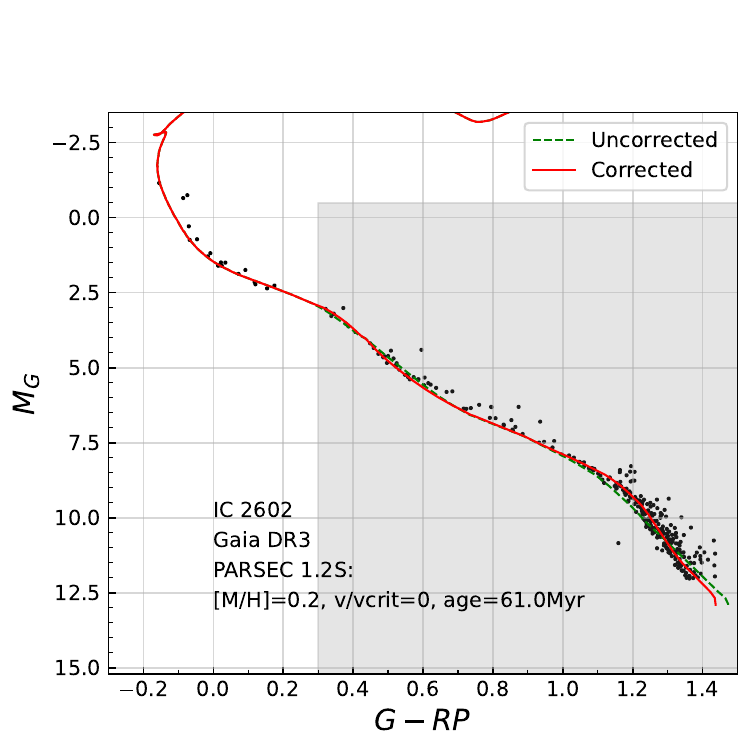}
    }
    {%
    \includegraphics[width=0.22\textwidth, trim=0.1cm 0.2cm 0.0cm 1.7cm, clip]{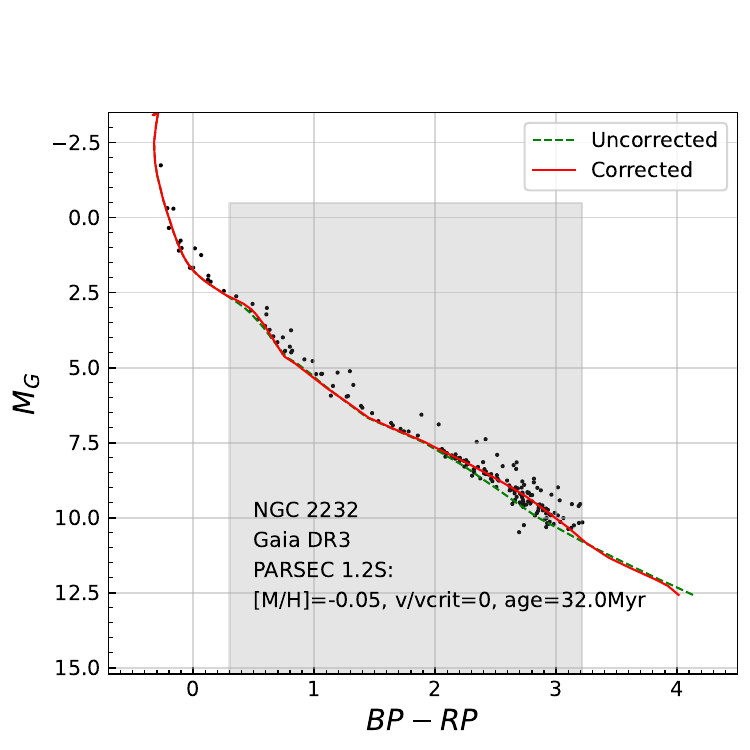}\quad
    \includegraphics[width=0.22\textwidth, trim=0.1cm 0.2cm 0.0cm 1.7cm, clip]{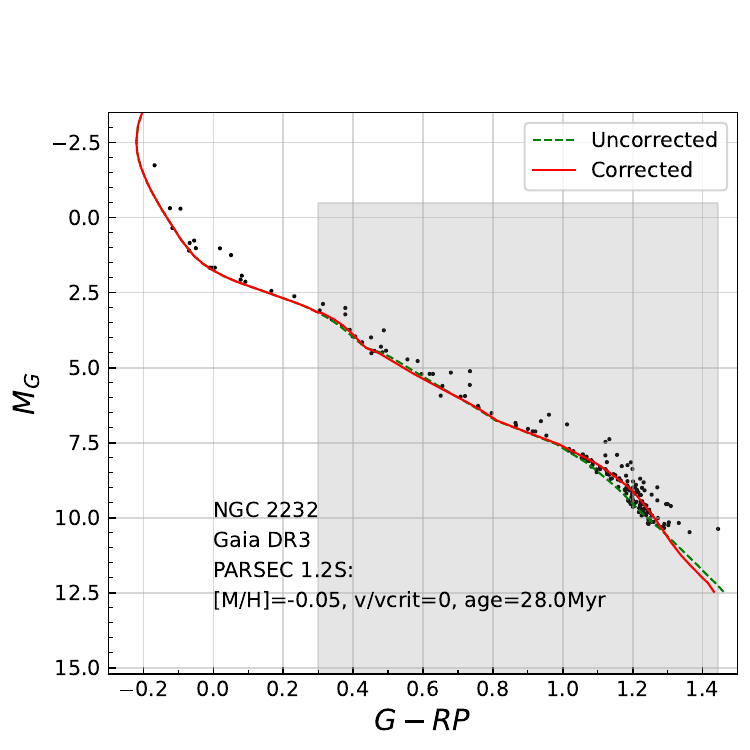}
    }
    {%
    \includegraphics[width=0.22\textwidth, trim=0.1cm 0.2cm 0.0cm 1.7cm, clip]{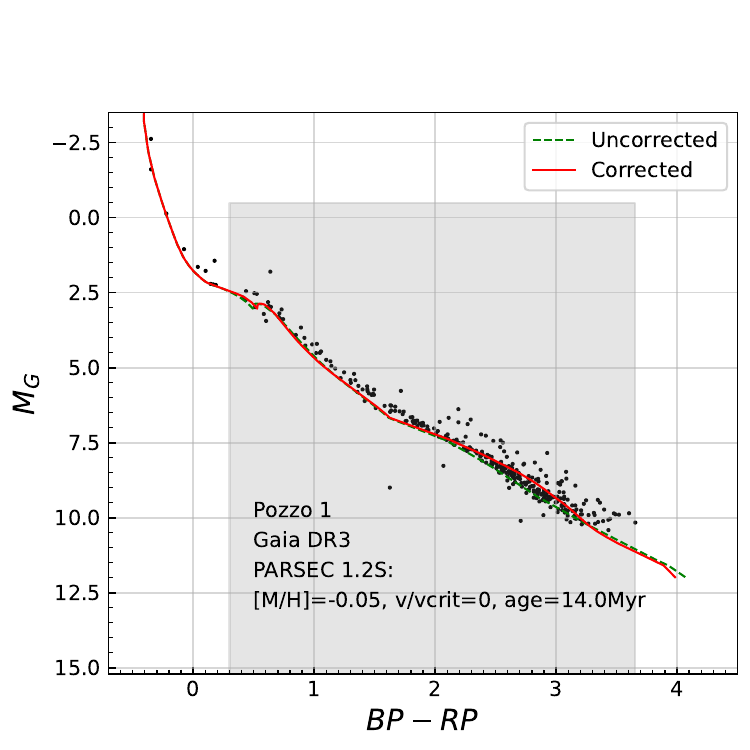}\quad
    \includegraphics[width=0.22\textwidth, trim=0.1cm 0.2cm 0.0cm 1.7cm, clip]{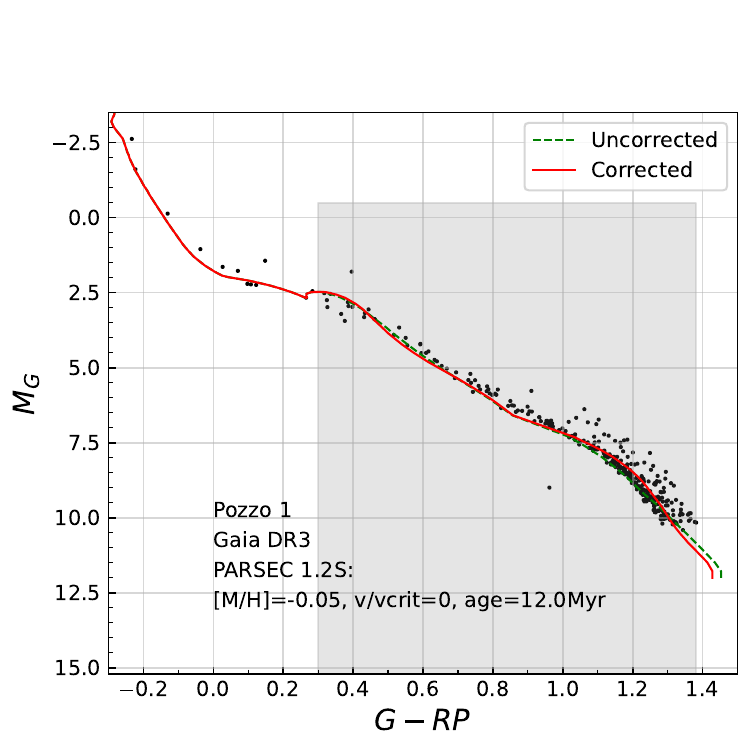}
    }
    {%
    \includegraphics[width=0.22\textwidth, trim=0.1cm 0.2cm 0.0cm 1.7cm, clip]{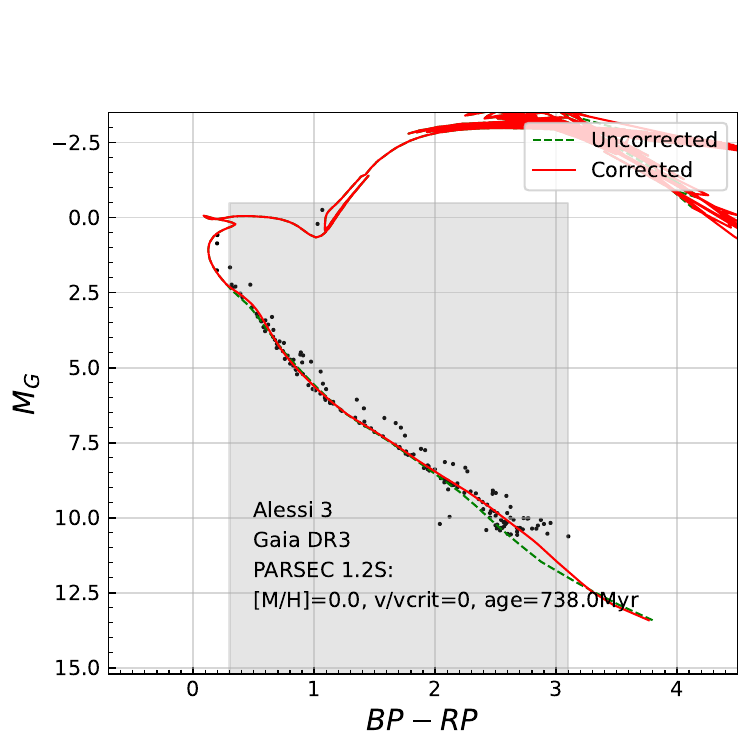}\quad
    \includegraphics[width=0.22\textwidth, trim=0.1cm 0.2cm 0.0cm 1.7cm, clip]{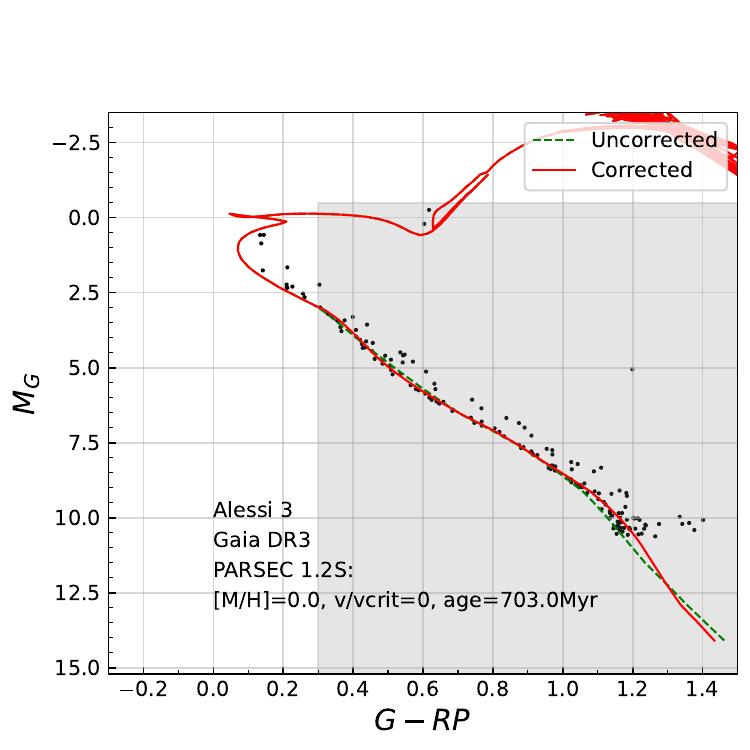}
    }
    {%
    \includegraphics[width=0.22\textwidth, trim=0.1cm 0.2cm 0.0cm 1.7cm, clip]{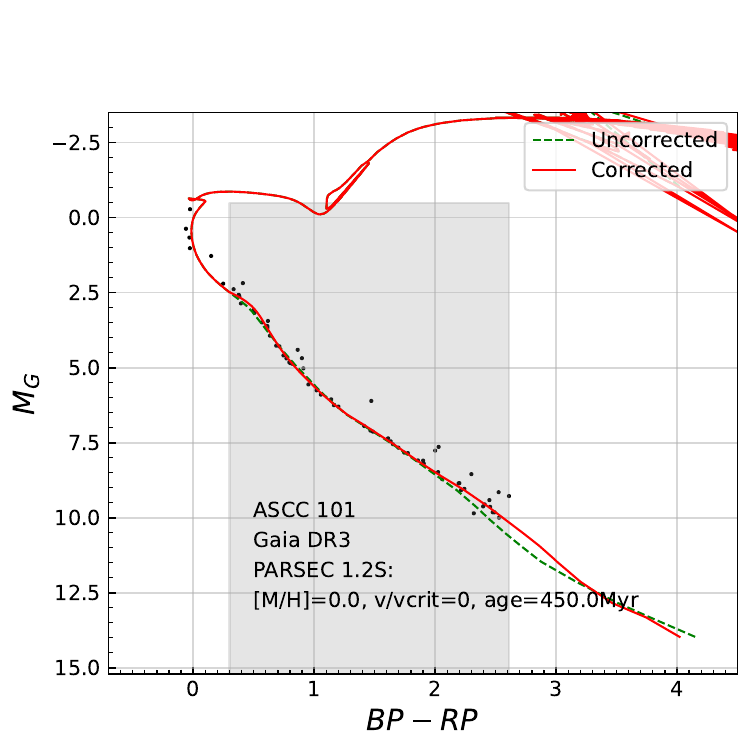}\quad
    \includegraphics[width=0.22\textwidth, trim=0.1cm 0.2cm 0.0cm 1.7cm, clip]{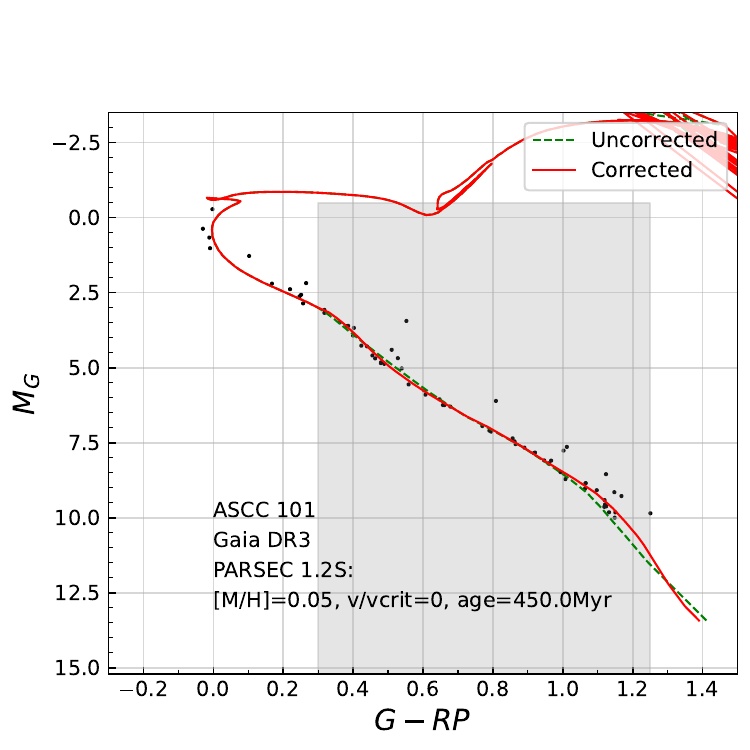}
    }
    {%
    \includegraphics[width=0.22\textwidth, trim=0.1cm 0.2cm 0.0cm 1.7cm, clip]{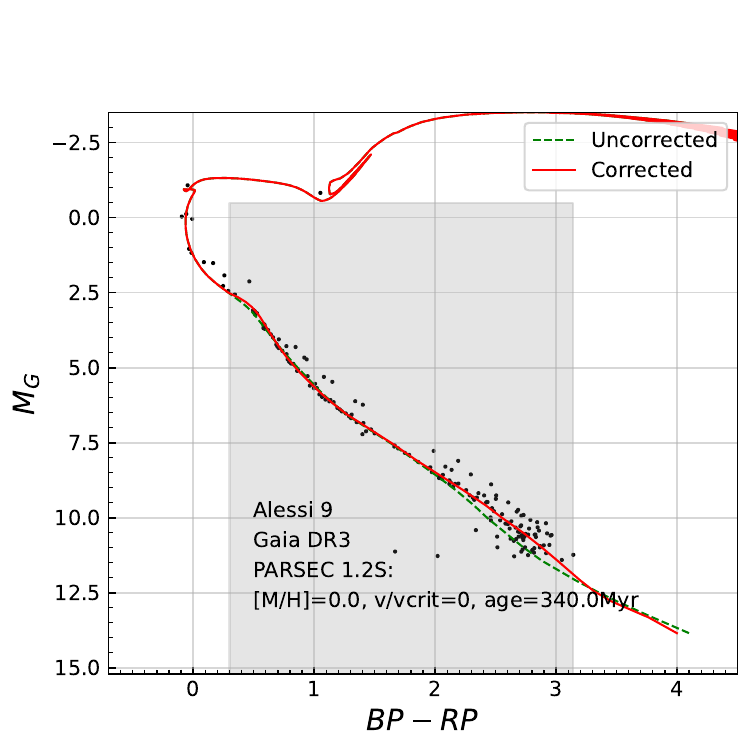}\quad
    \includegraphics[width=0.22\textwidth, trim=0.1cm 0.2cm 0.0cm 1.7cm, clip]{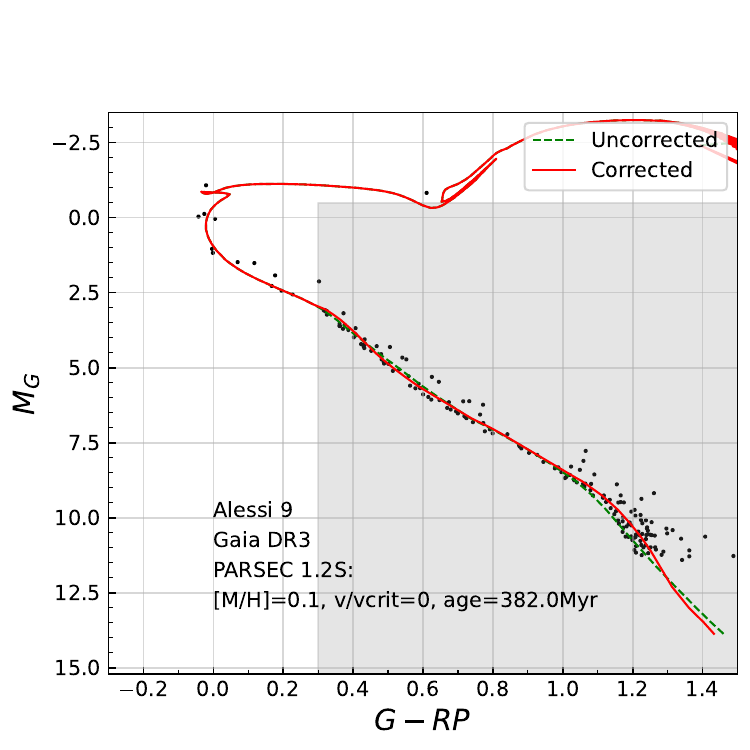}
    }
    {%
    \includegraphics[width=0.22\textwidth, trim=0.1cm 0.2cm 0.0cm 1.7cm, clip]{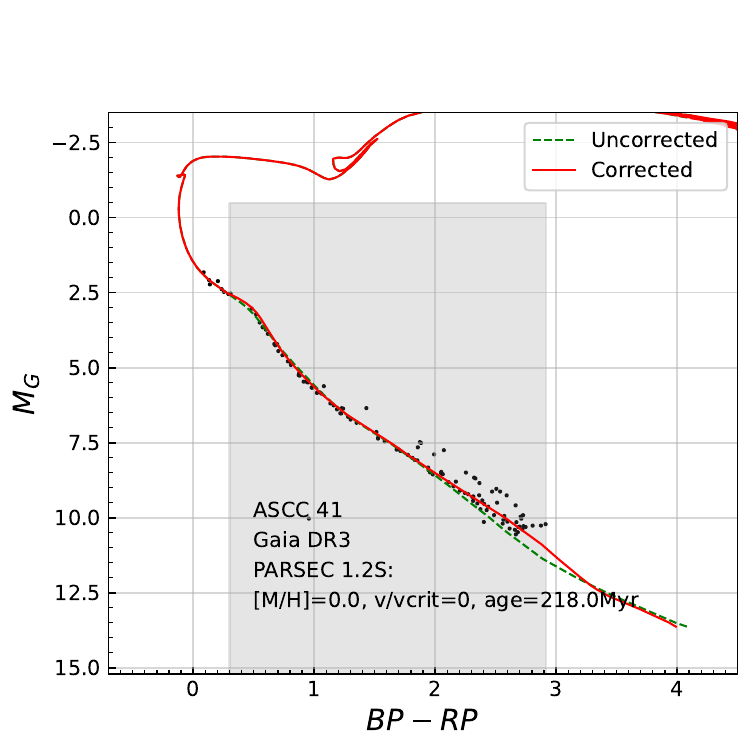}\quad
    \includegraphics[width=0.22\textwidth, trim=0.1cm 0.2cm 0.0cm 1.7cm, clip]{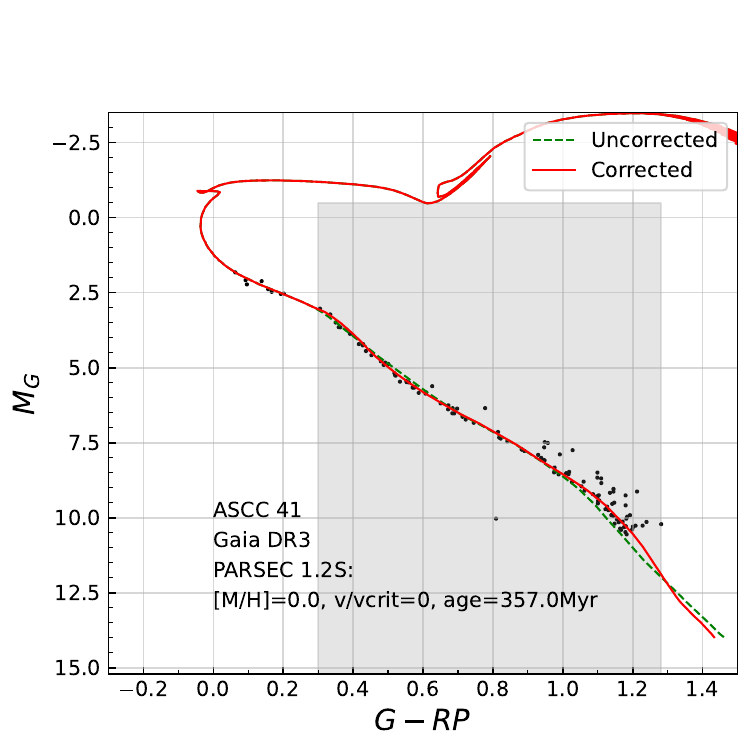}
    }
    {%
    \includegraphics[width=0.22\textwidth, trim=0.1cm 0.2cm 0.0cm 1.7cm, clip]{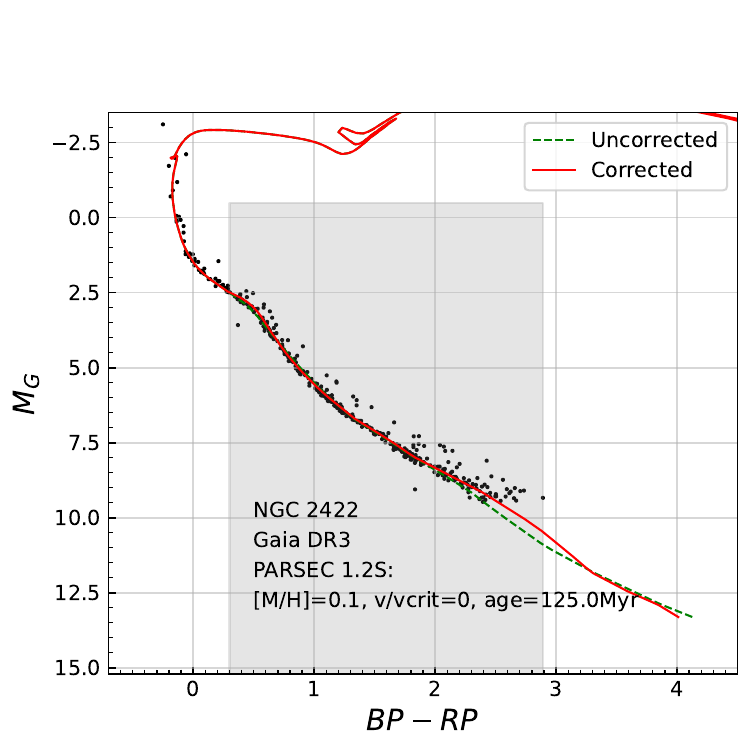}\quad
    \includegraphics[width=0.22\textwidth, trim=0.1cm 0.2cm 0.0cm 1.7cm, clip]{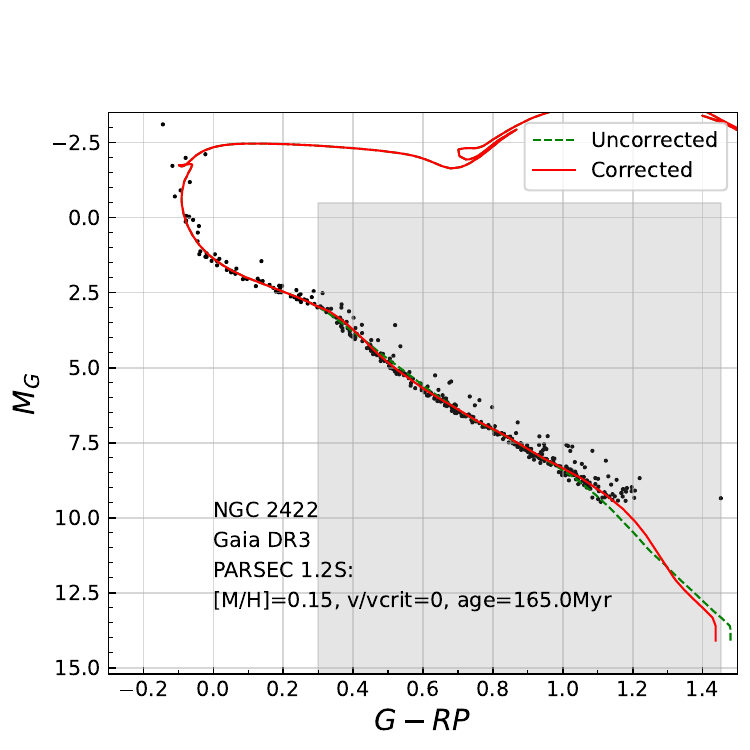}
    }
    {%
    \includegraphics[width=0.22\textwidth, trim=0.1cm 0.2cm 0.0cm 1.7cm, clip]{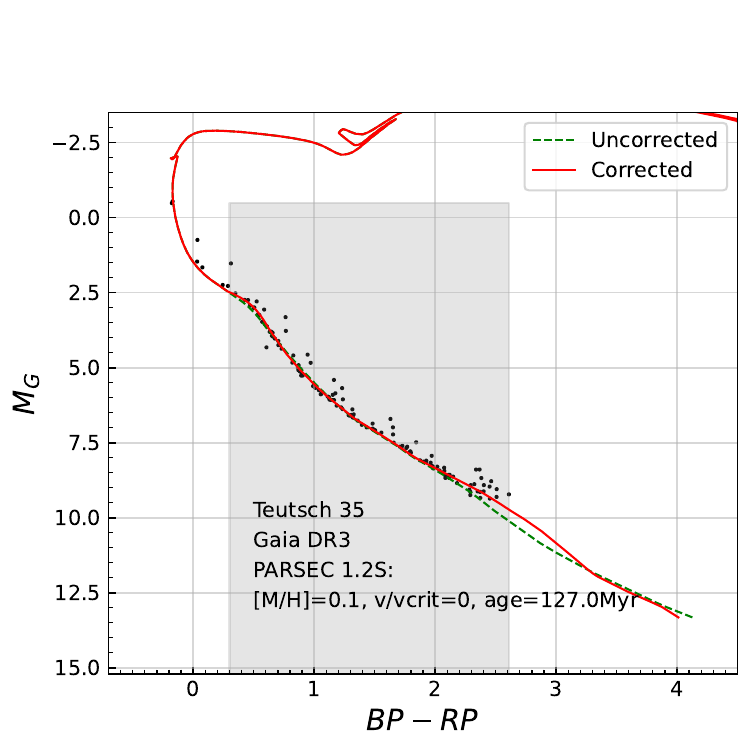}\quad
    \includegraphics[width=0.22\textwidth, trim=0.1cm 0.2cm 0.0cm 1.7cm, clip]{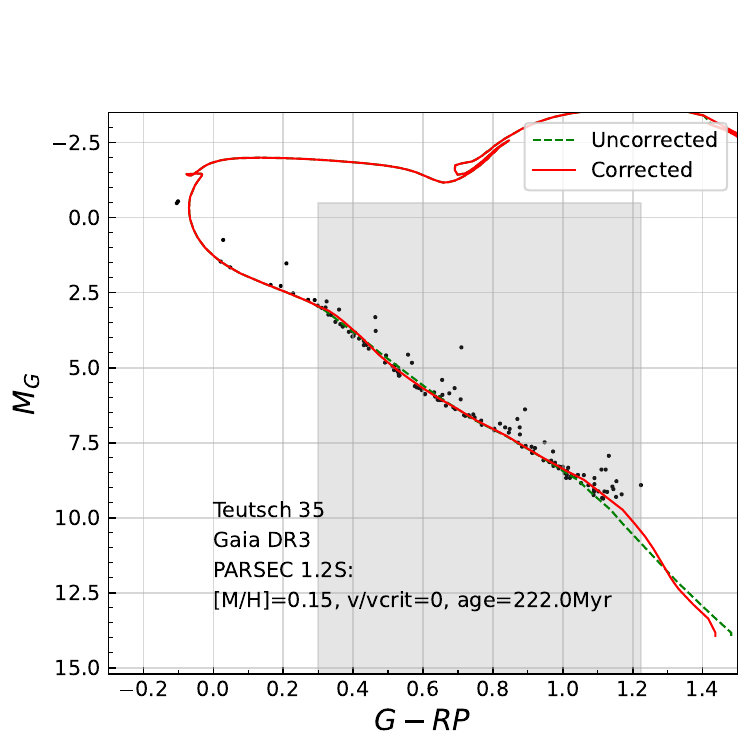}
    }
\caption{CMDs of open clusters used by PARSEC 1.2S model.}
\label{fig:33_bp/g-rp_dr3_parsec}
\end{figure*}

\setcounter{figure}{1}
\begin{figure*}
\centering
{%
    \includegraphics[width=0.22\textwidth, trim=0.1cm 0.2cm 0.0cm 1.7cm, clip]{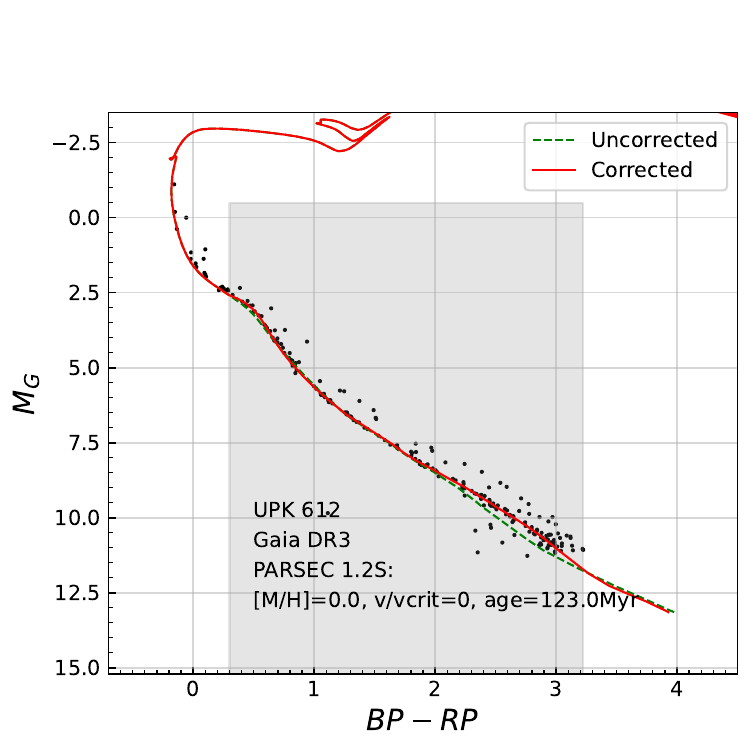}\quad
    \includegraphics[width=0.22\textwidth, trim=0.1cm 0.2cm 0.0cm 1.7cm, clip]{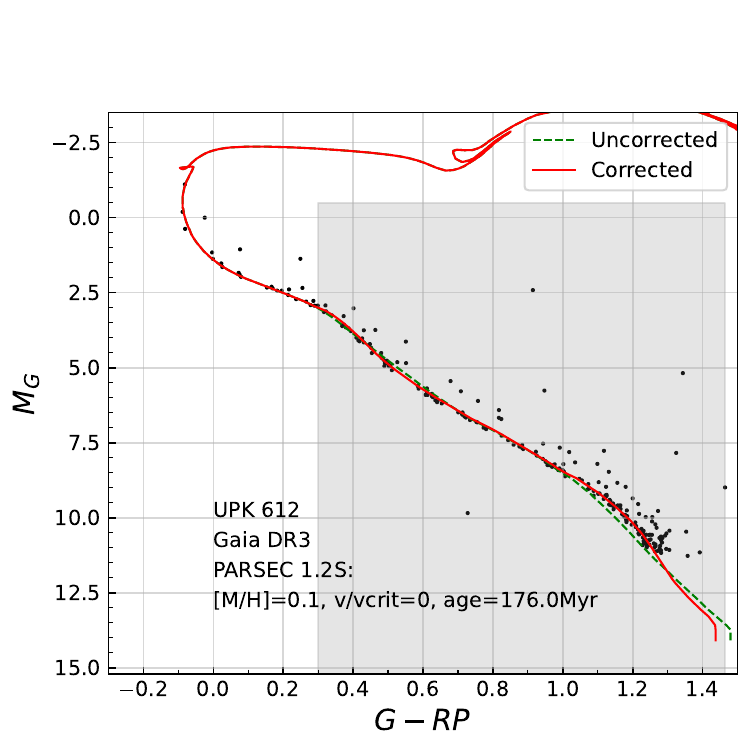}
    }
    {%
    \includegraphics[width=0.22\textwidth, trim=0.1cm 0.2cm 0.0cm 1.7cm, clip]{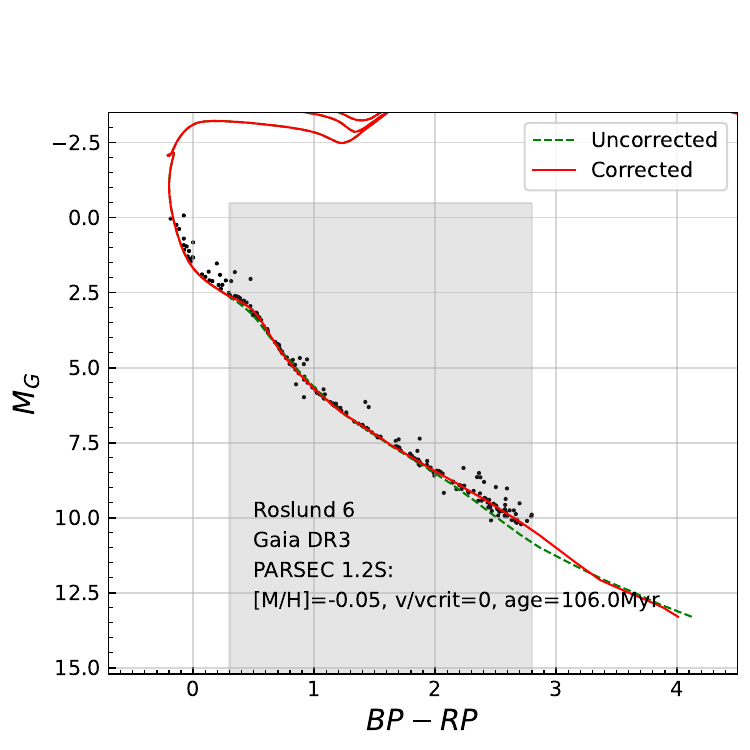}\quad
    \includegraphics[width=0.22\textwidth, trim=0.1cm 0.2cm 0.0cm 1.7cm, clip]{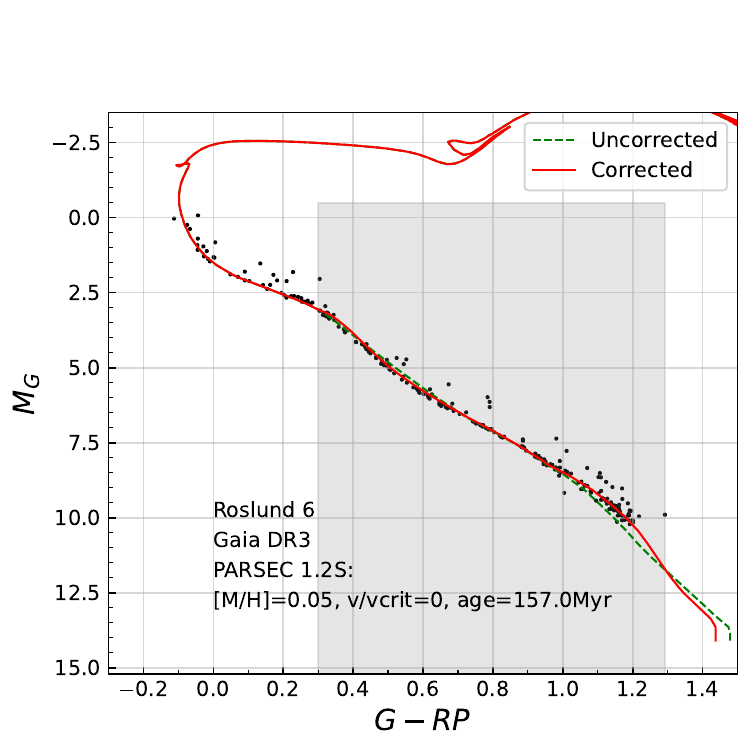}
    }
    {%
    \includegraphics[width=0.22\textwidth, trim=0.1cm 0.2cm 0.0cm 1.7cm, clip]{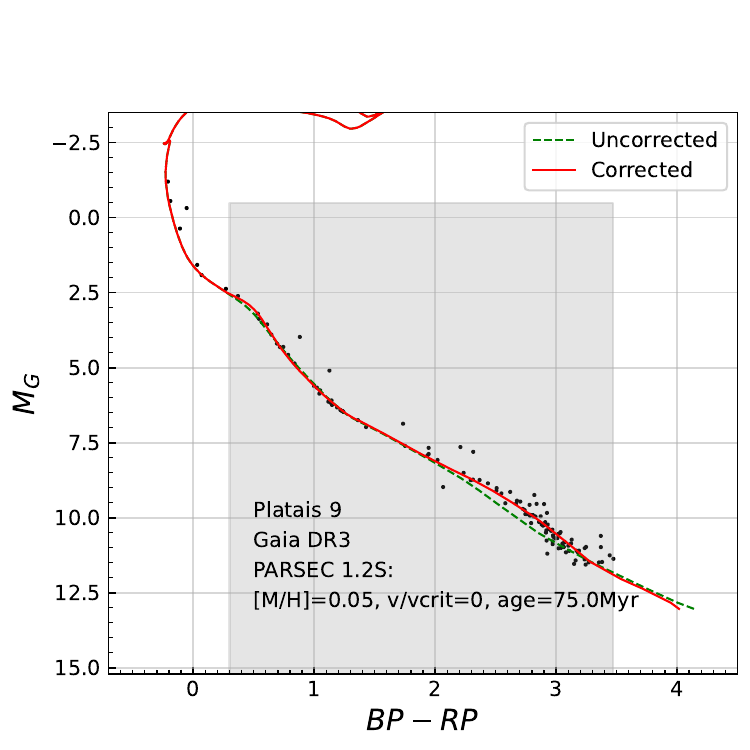}\quad
    \includegraphics[width=0.22\textwidth, trim=0.1cm 0.2cm 0.0cm 1.7cm, clip]{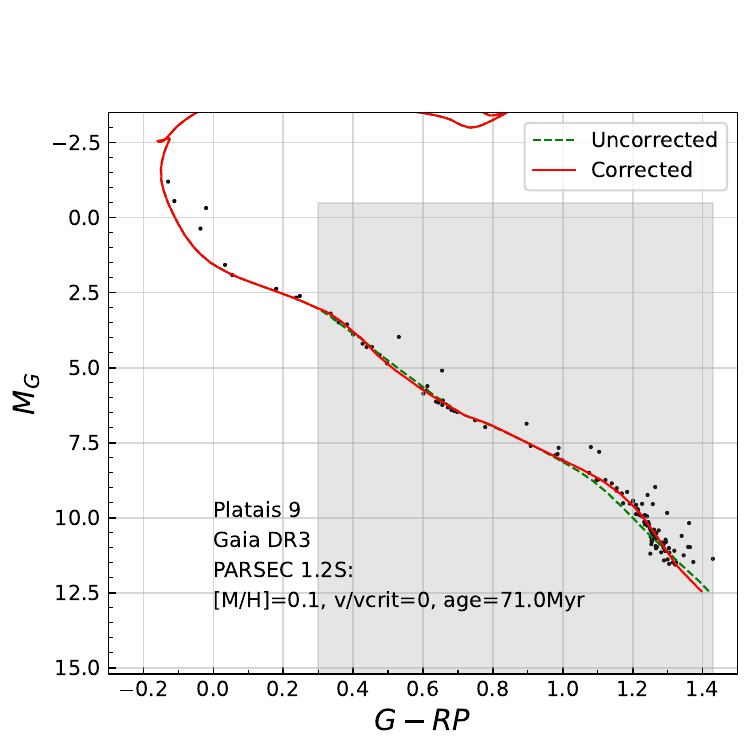}
    }
    {%
    \includegraphics[width=0.22\textwidth, trim=0.1cm 0.2cm 0.0cm 1.7cm, clip]{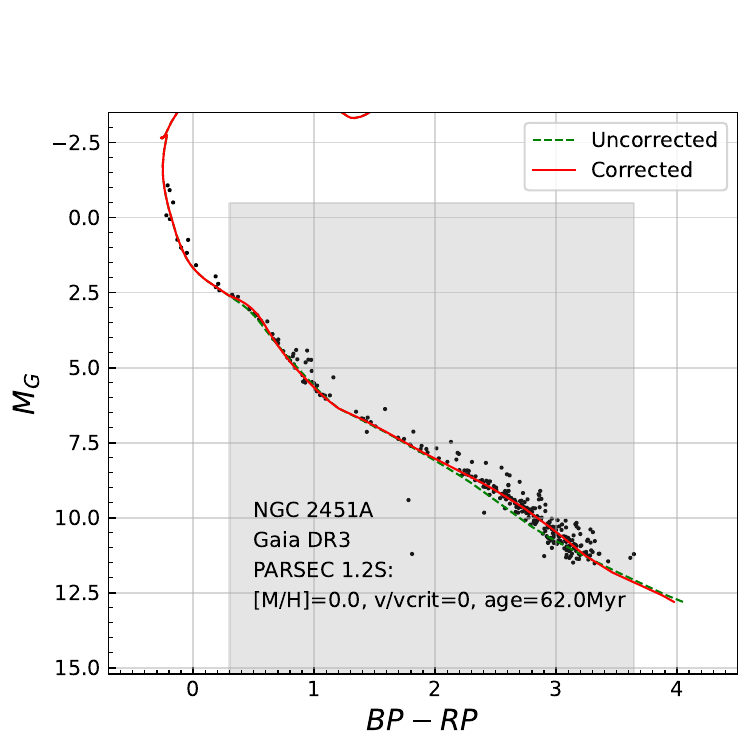}\quad
    \includegraphics[width=0.22\textwidth, trim=0.1cm 0.2cm 0.0cm 1.7cm, clip]{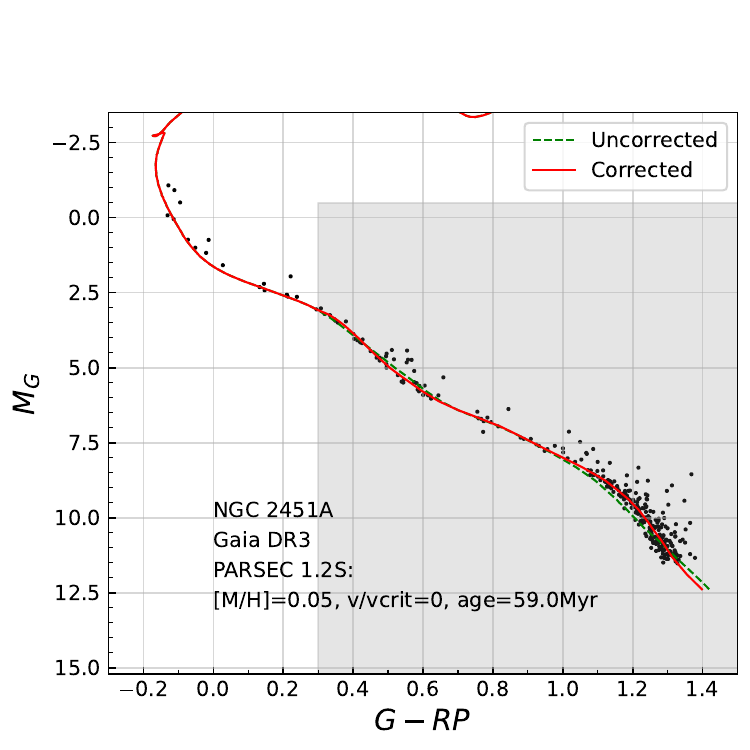}
    }
    {%
    \includegraphics[width=0.22\textwidth, trim=0.1cm 0.2cm 0.0cm 1.7cm, clip]{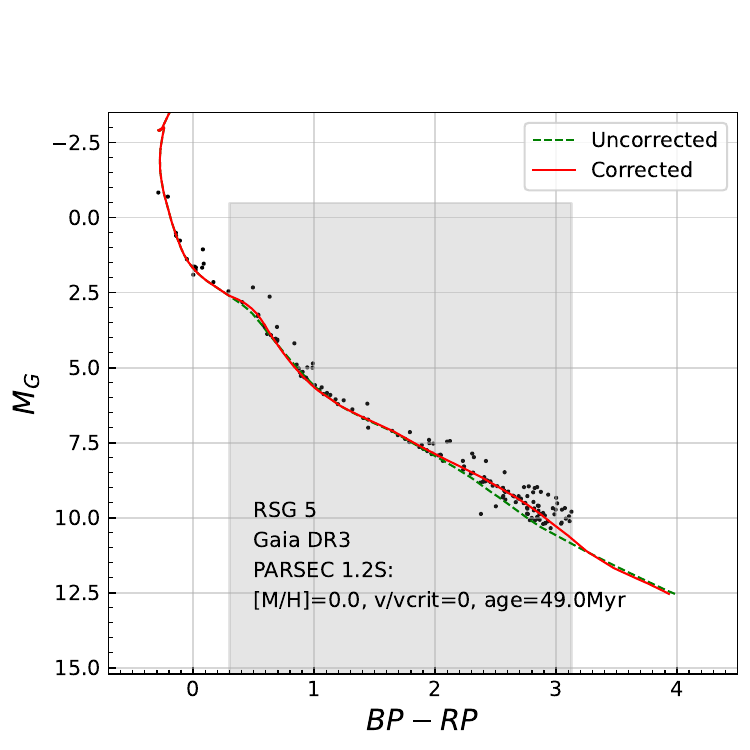}\quad
    \includegraphics[width=0.22\textwidth, trim=0.1cm 0.2cm 0.0cm 1.7cm, clip]{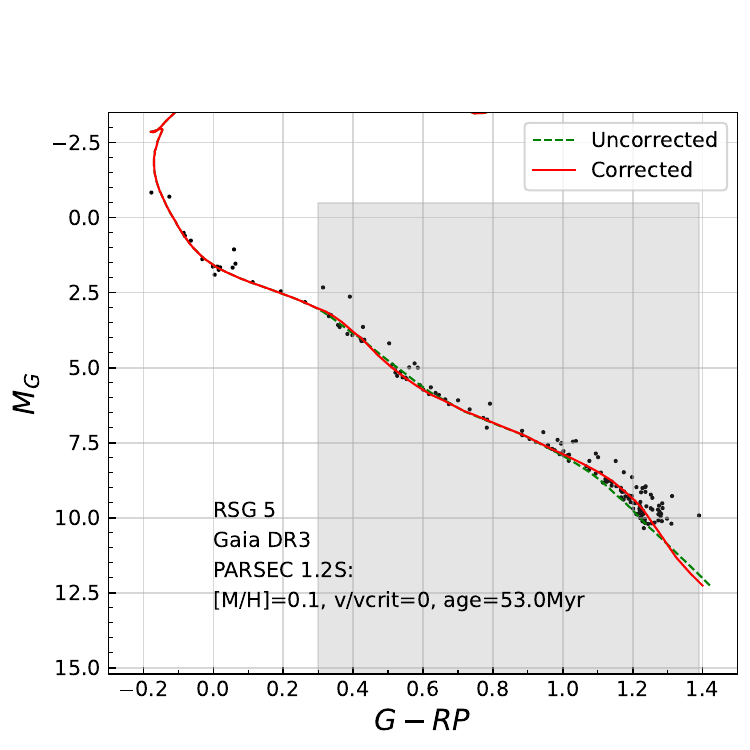}
    }
    {%
    \includegraphics[width=0.22\textwidth, trim=0.1cm 0.2cm 0.0cm 1.7cm, clip]{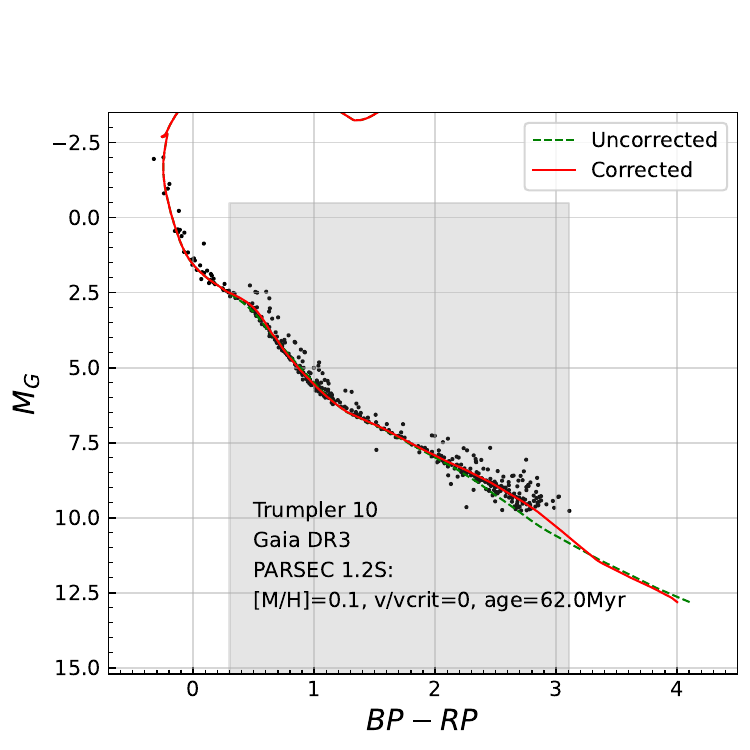}\quad
    \includegraphics[width=0.22\textwidth, trim=0.1cm 0.2cm 0.0cm 1.7cm, clip]{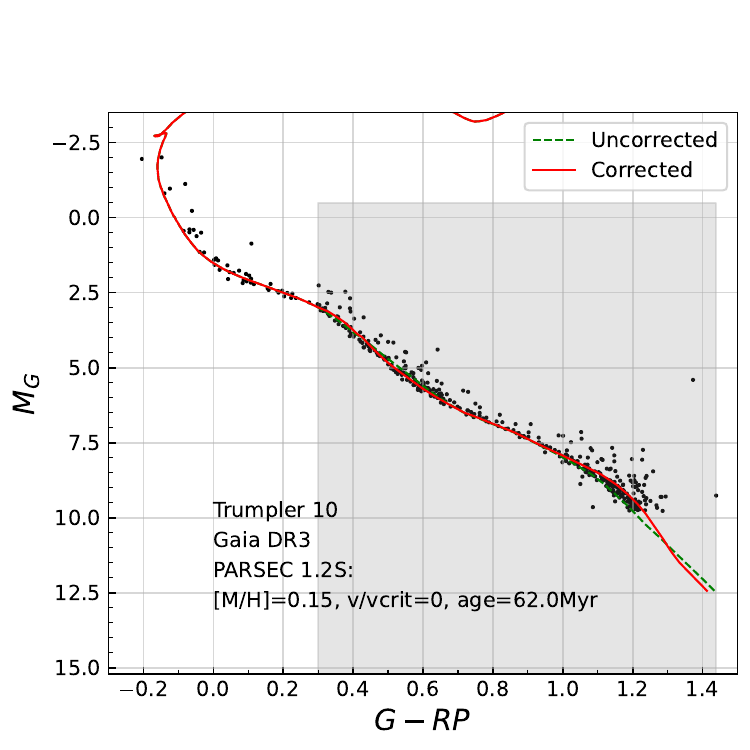}
    }
    {%
    \includegraphics[width=0.22\textwidth, trim=0.1cm 0.2cm 0.0cm 1.7cm, clip]{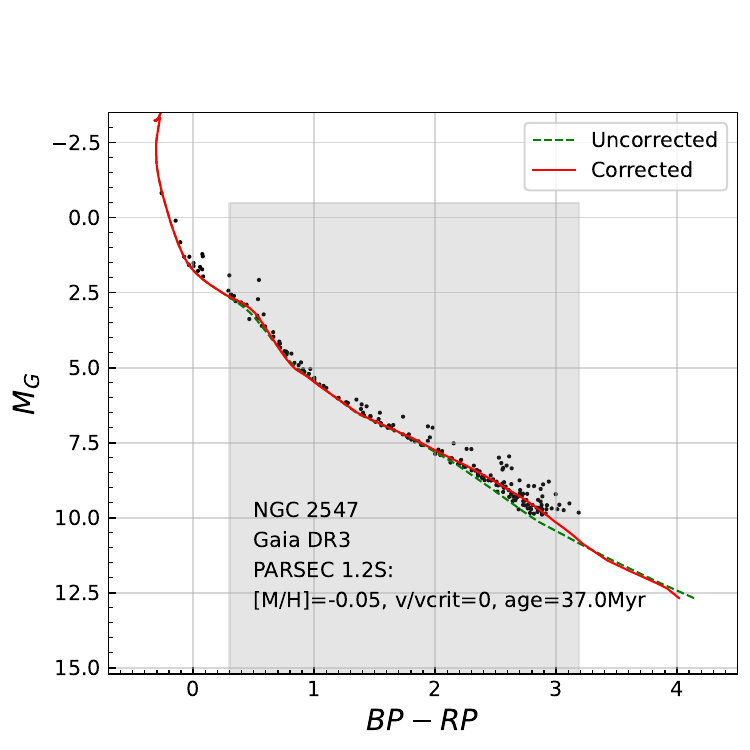}\quad
    \includegraphics[width=0.22\textwidth, trim=0.1cm 0.2cm 0.0cm 1.7cm, clip]{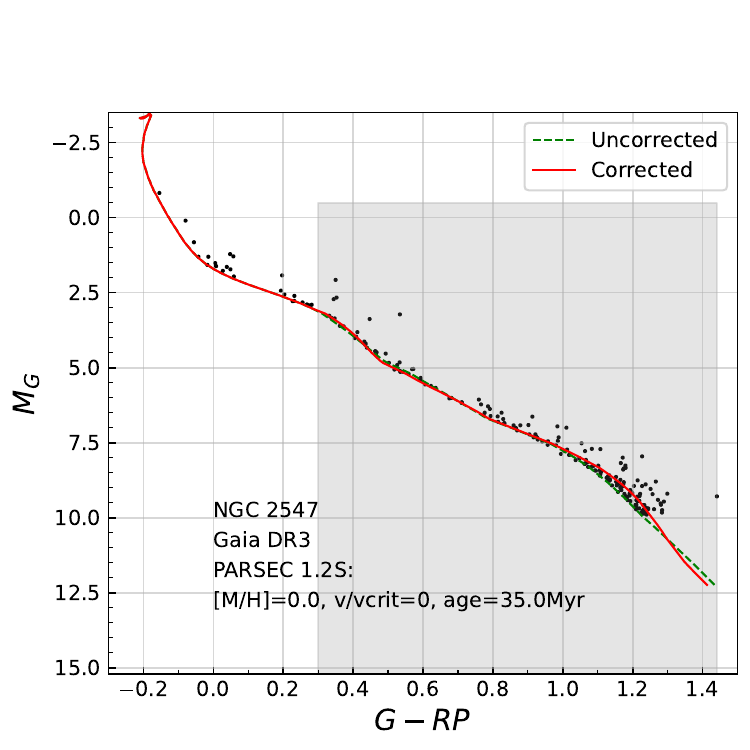}
    }
    {%
    \includegraphics[width=0.22\textwidth, trim=0.1cm 0.2cm 0.0cm 1.7cm, clip]{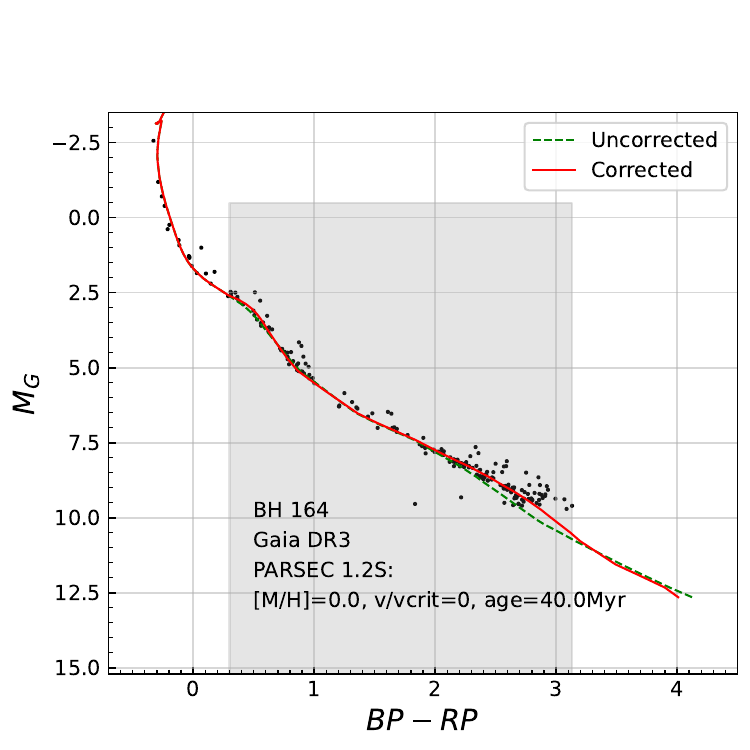}\quad
    \includegraphics[width=0.22\textwidth, trim=0.1cm 0.2cm 0.0cm 1.7cm, clip]{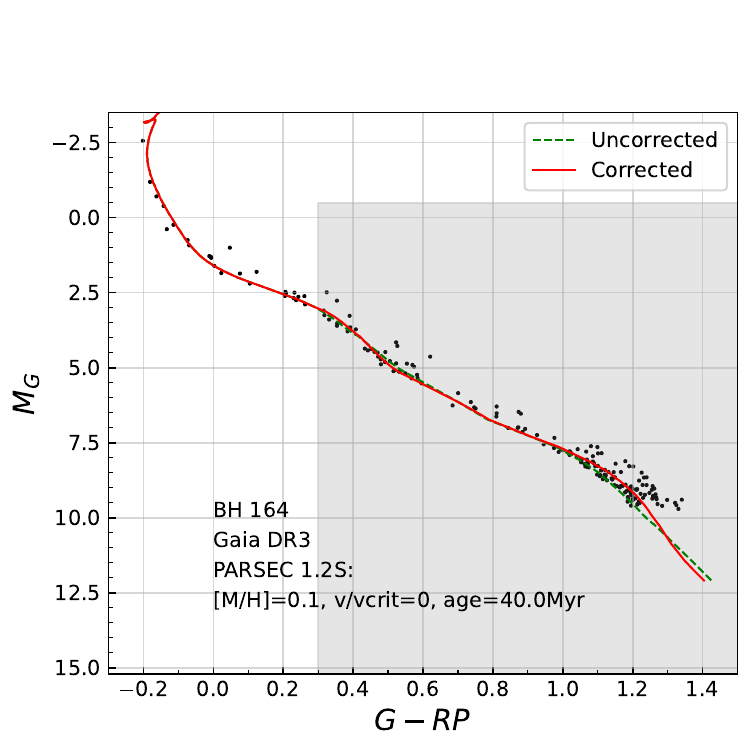}
    }
    {%
    \includegraphics[width=0.22\textwidth, trim=0.1cm 0.2cm 0.0cm 1.7cm, clip]{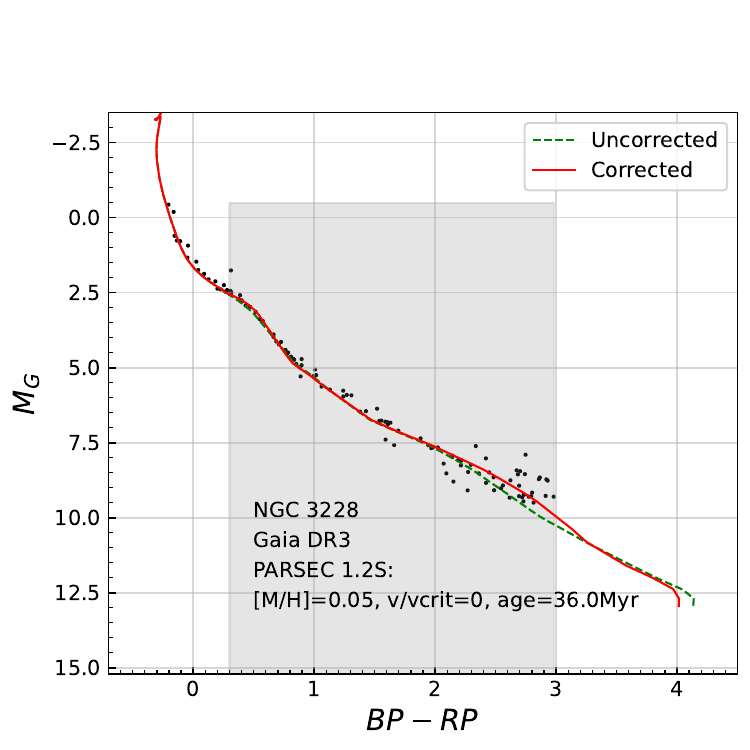}\quad
    \includegraphics[width=0.22\textwidth, trim=0.1cm 0.2cm 0.0cm 1.7cm, clip]{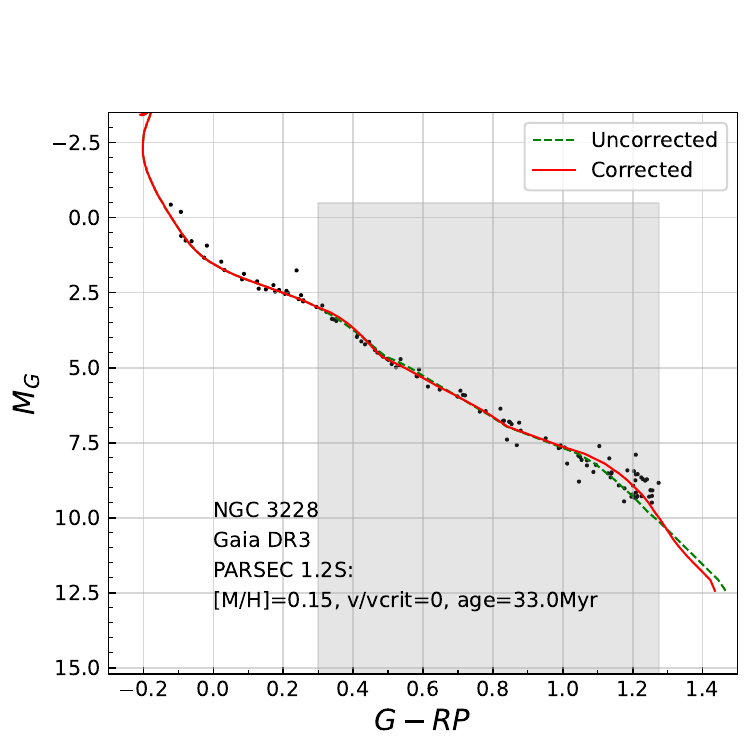}
    }
    {%
    \includegraphics[width=0.22\textwidth, trim=0.1cm 0.2cm 0.0cm 1.7cm, clip]{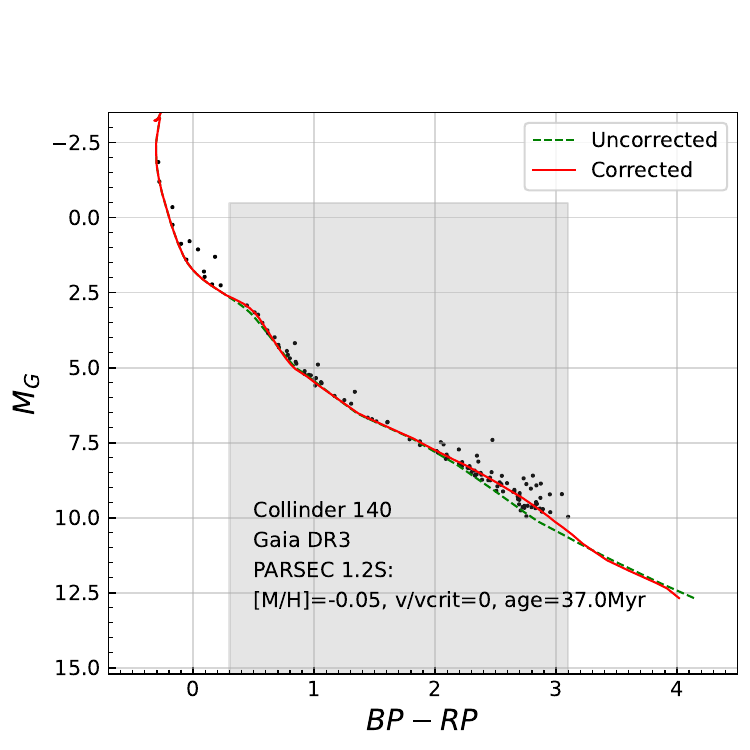}\quad
    \includegraphics[width=0.22\textwidth, trim=0.1cm 0.2cm 0.0cm 1.7cm, clip]{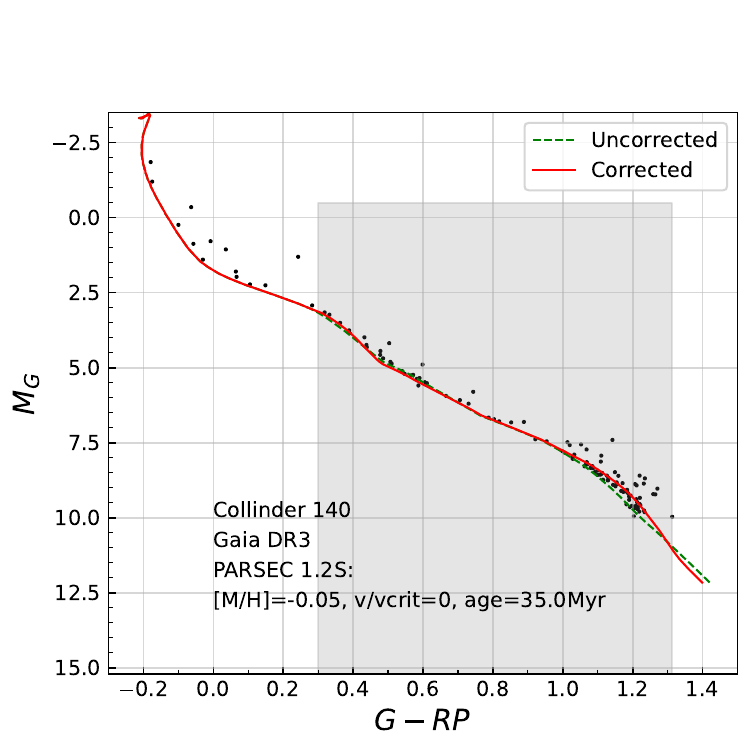}
    }
    {%
    \includegraphics[width=0.22\textwidth, trim=0.1cm 0.2cm 0.0cm 1.7cm, clip]{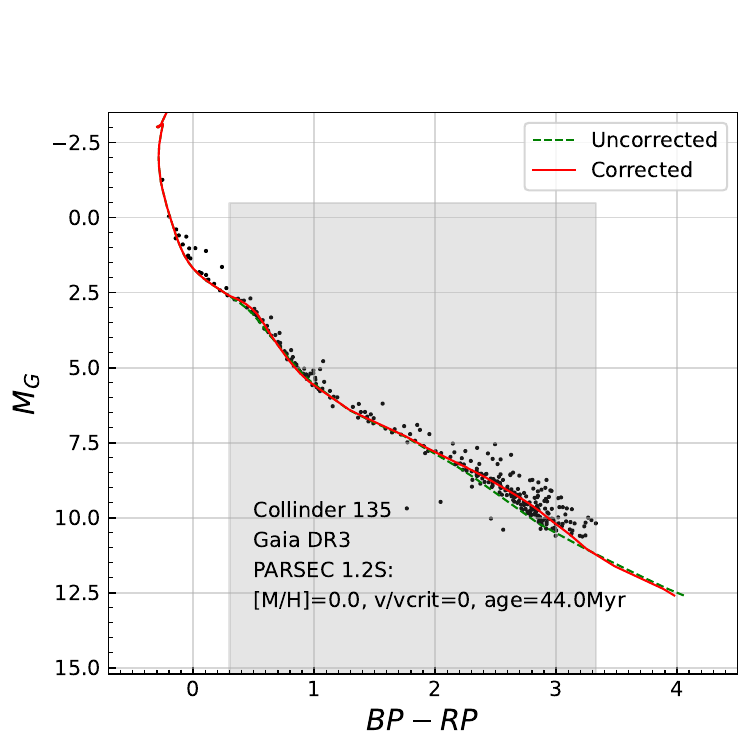}\quad
    \includegraphics[width=0.22\textwidth, trim=0.1cm 0.2cm 0.0cm 1.7cm, clip]{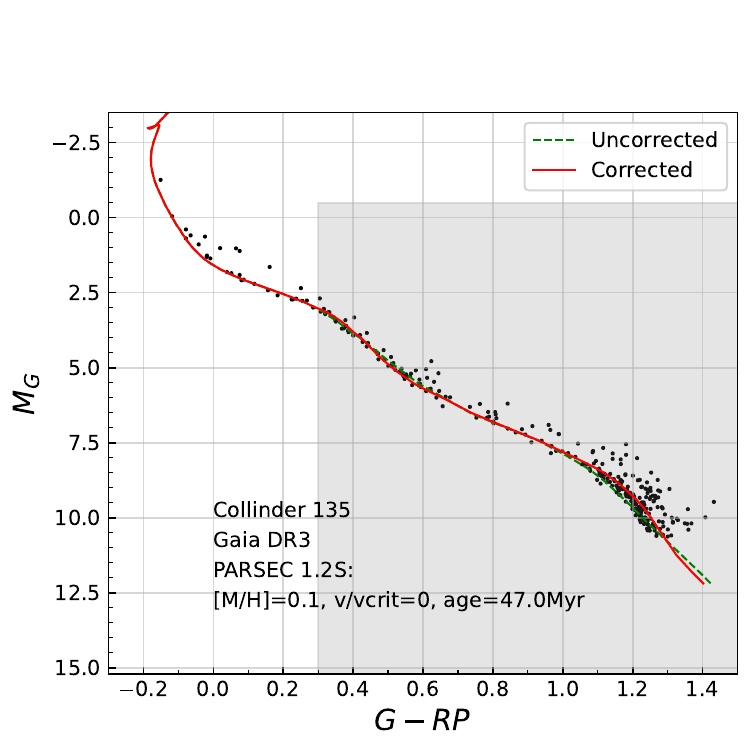}
    }
    {%
    \includegraphics[width=0.22\textwidth, trim=0.1cm 0.2cm 0.0cm 1.7cm, clip]{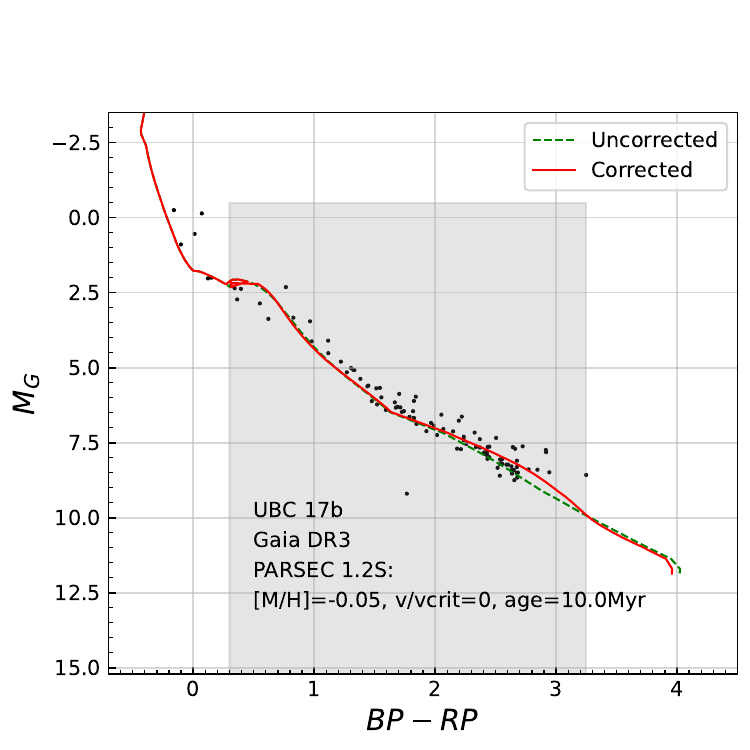}\quad
    \includegraphics[width=0.22\textwidth, trim=0.1cm 0.2cm 0.0cm 1.7cm, clip]{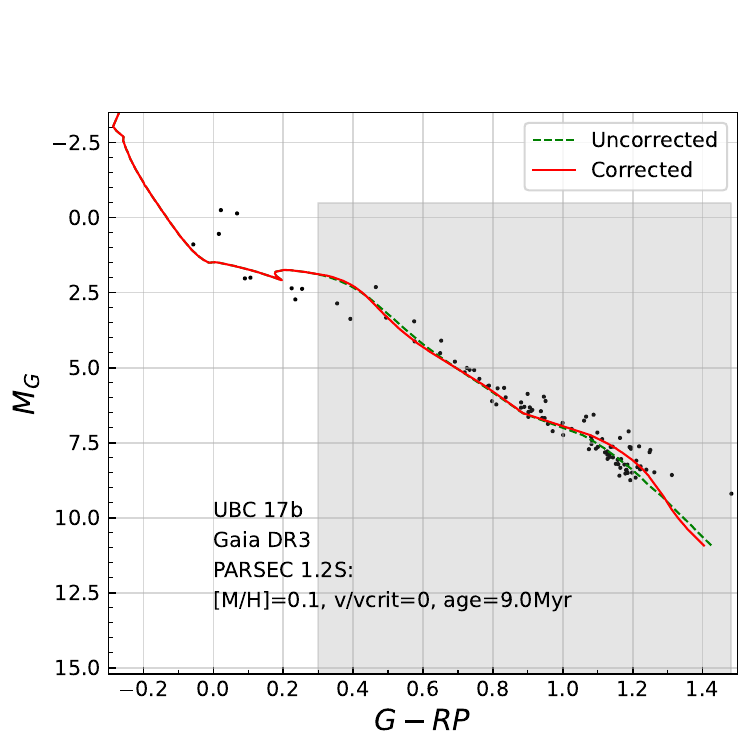}
    }
    {%
    \includegraphics[width=0.22\textwidth, trim=0.1cm 0.2cm 0.0cm 1.7cm, clip]{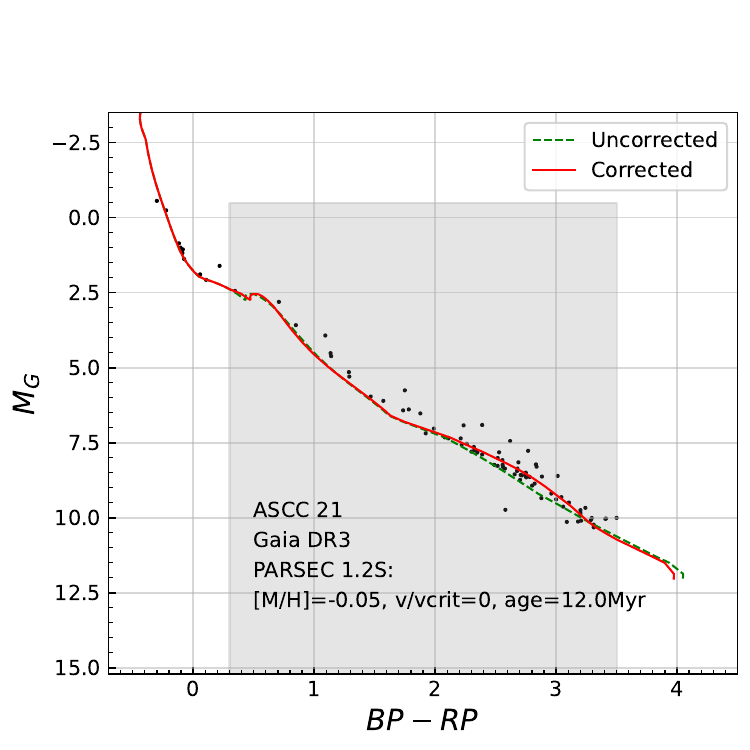}\quad
    \includegraphics[width=0.22\textwidth, trim=0.1cm 0.2cm 0.0cm 1.7cm, clip]{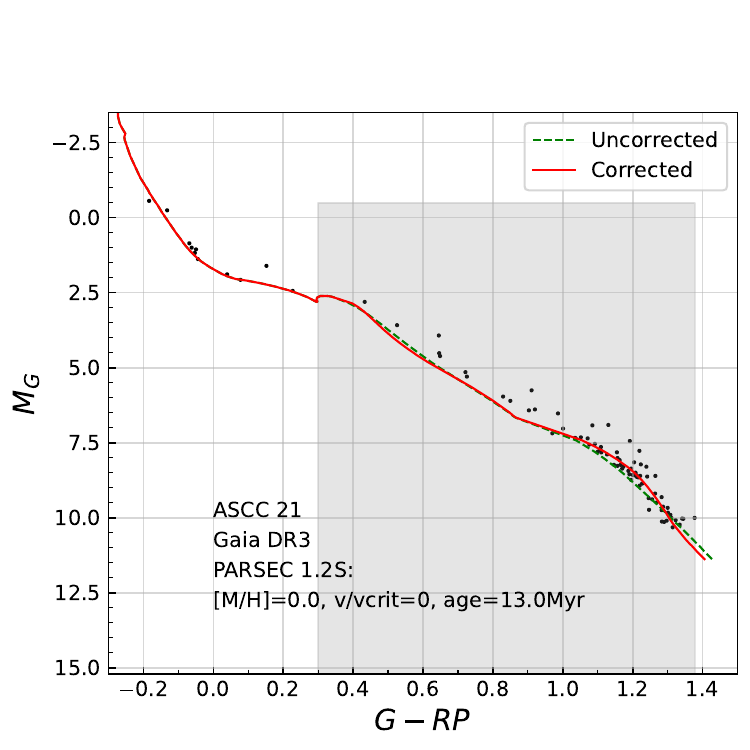}
    }
    {%
    \includegraphics[width=0.22\textwidth, trim=0.1cm 0.2cm 0.0cm 1.7cm, clip]{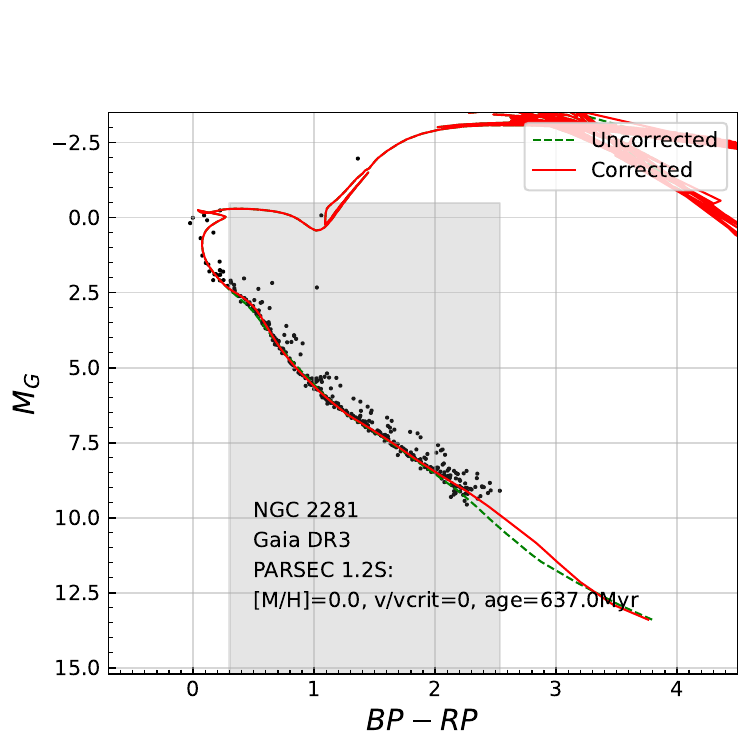}\quad
    \includegraphics[width=0.22\textwidth, trim=0.1cm 0.2cm 0.0cm 1.7cm, clip]{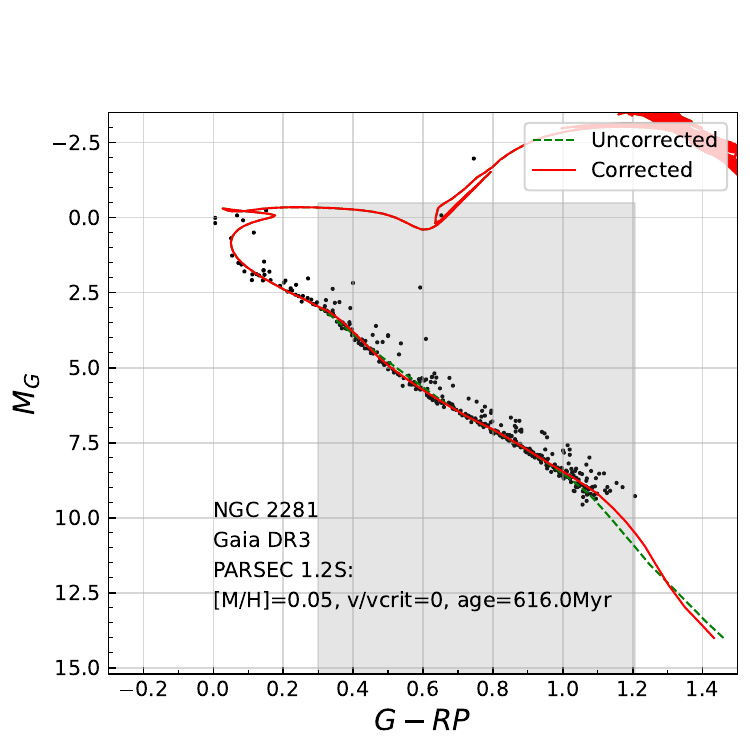}
    }    
\caption{Continued from Fig~\ref{fig:33_bp/g-rp_dr3_parsec}}
\end{figure*}

\setcounter{figure}{1}
\begin{figure*}
\centering
    {%
    \includegraphics[width=0.22\textwidth, trim=0.1cm 0.2cm 0.0cm 1.7cm, clip]{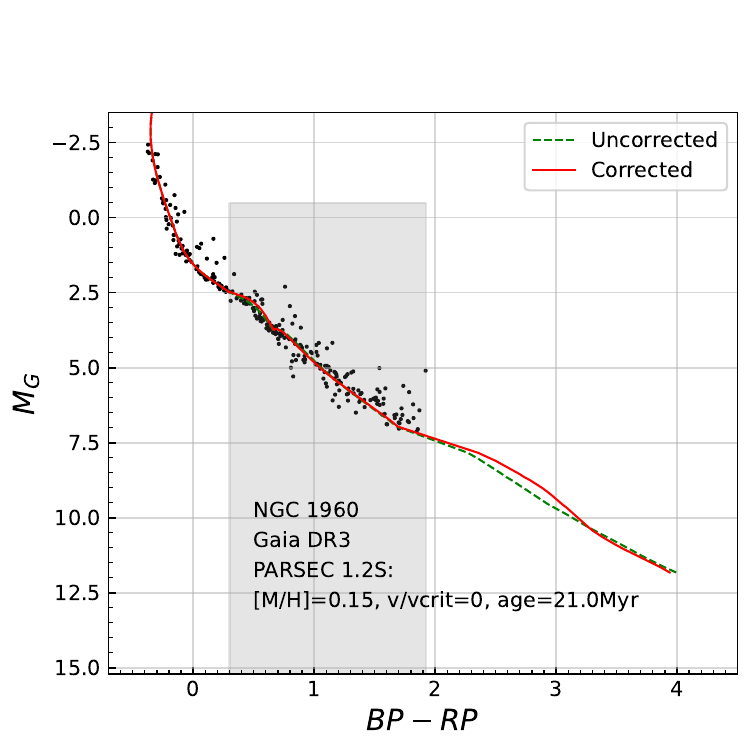}\quad
    \includegraphics[width=0.22\textwidth, trim=0.1cm 0.2cm 0.0cm 1.7cm, clip]{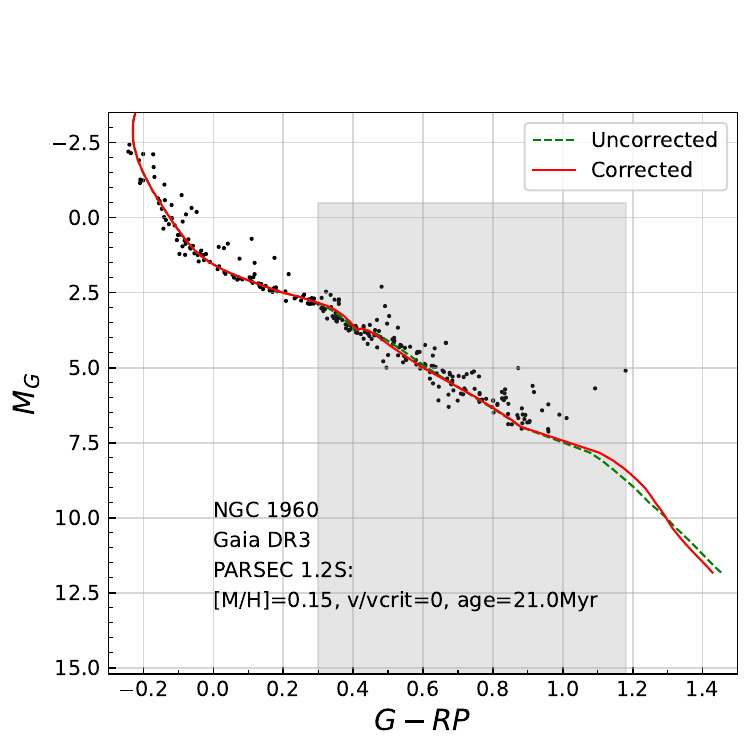}
    } 
    {%
    \includegraphics[width=0.22\textwidth, trim=0.1cm 0.2cm 0.0cm 1.7cm, clip]{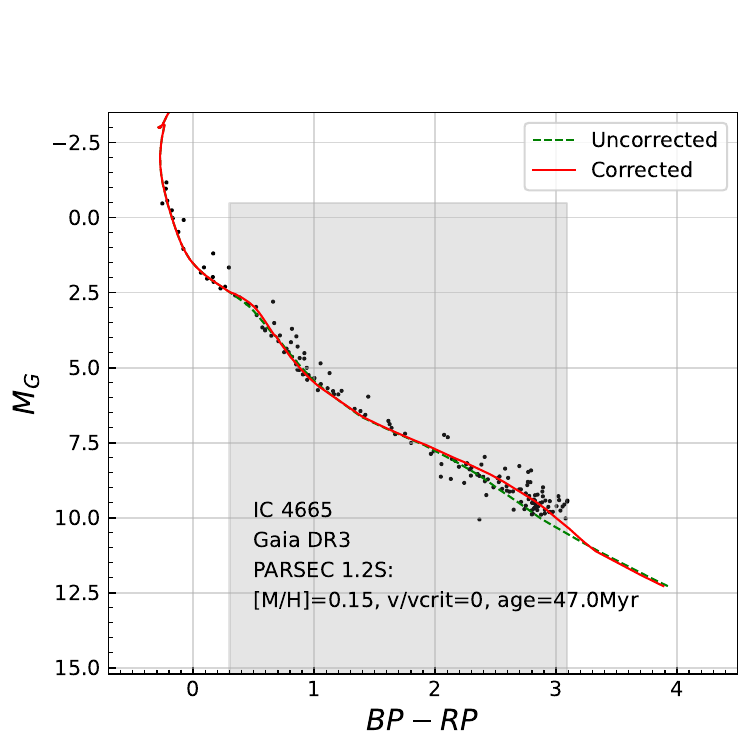}\quad
    \includegraphics[width=0.22\textwidth, trim=0.1cm 0.2cm 0.0cm 1.7cm, clip]{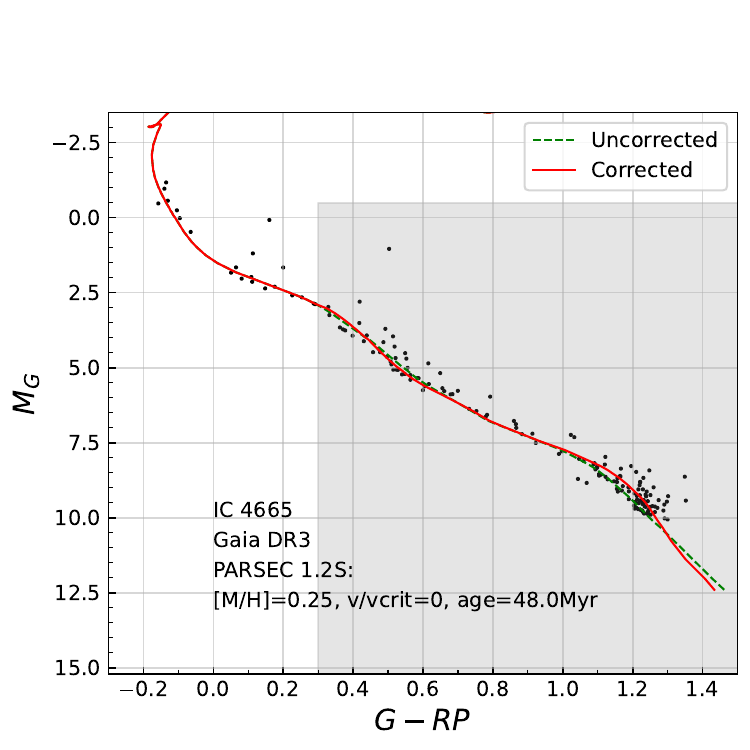}
    }
    {%
    \includegraphics[width=0.22\textwidth, trim=0.1cm 0.2cm 0.0cm 1.7cm, clip]{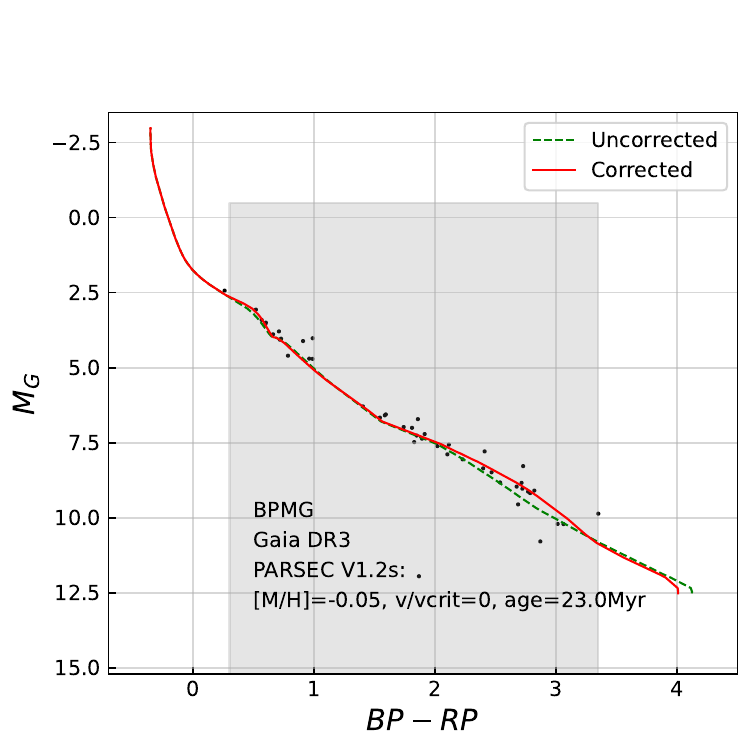}\quad
    \includegraphics[width=0.22\textwidth, trim=0.1cm 0.2cm 0.0cm 1.7cm, clip]{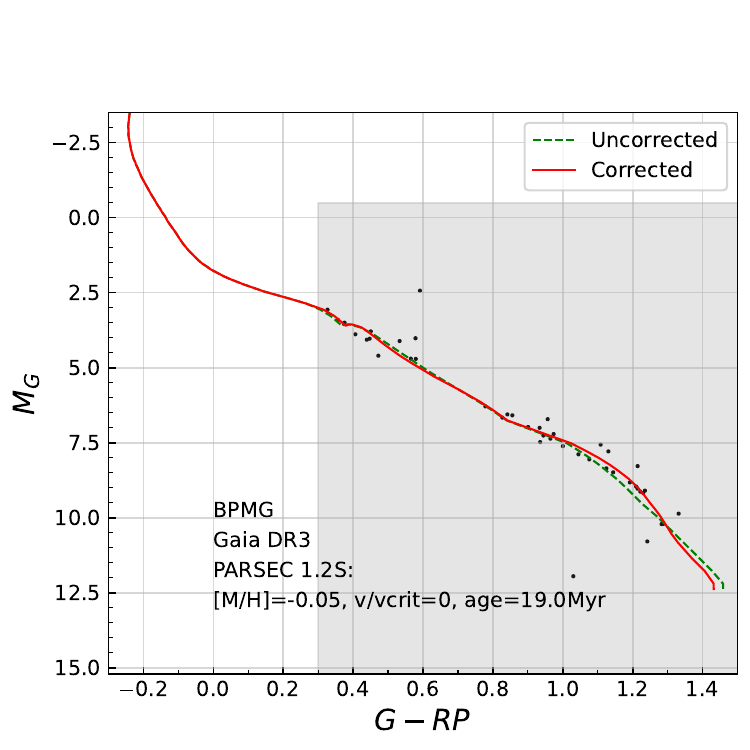}
    }
    {%
    \includegraphics[width=0.22\textwidth, trim=0.1cm 0.2cm 0.0cm 1.7cm, clip]{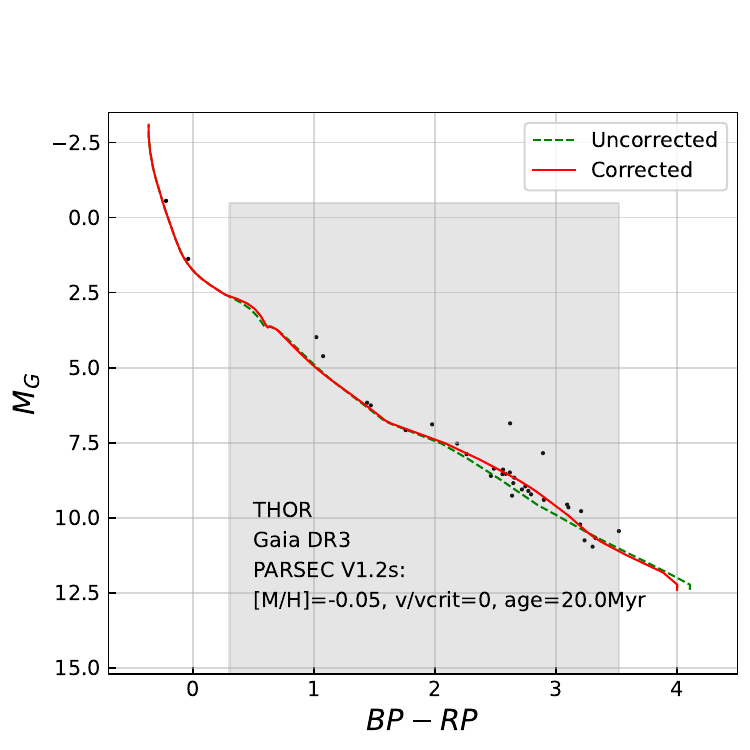}\quad
    \includegraphics[width=0.22\textwidth, trim=0.1cm 0.2cm 0.0cm 1.7cm, clip]{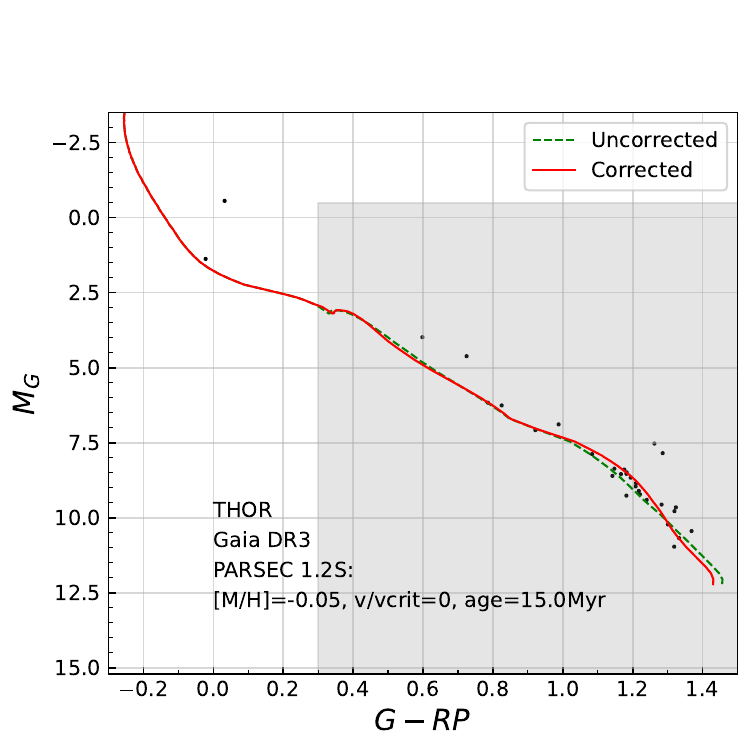}
    }
    {%
    \includegraphics[width=0.22\textwidth, trim=0.1cm 0.2cm 0.0cm 1.7cm, clip]{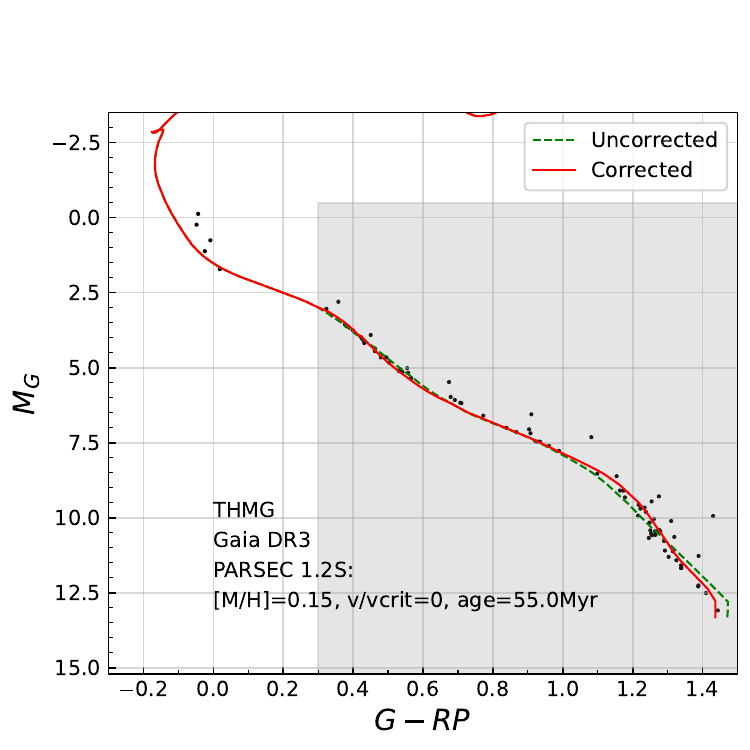}\quad
    \includegraphics[width=0.22\textwidth, trim=0.1cm 0.2cm 0.0cm 1.7cm, clip]{PARSEC_MGs_LDB_THMG_20240128_grp_dr3.pdf}
    }
\caption{Continued from Fig~\ref{fig:33_bp/g-rp_dr3_parsec}}
\end{figure*}

\counterwithin{table}{section}
\setcounter{table}{0}
\setcounter{table}{0}
\begin{center}
\begin{longrotatetable}
\renewcommand{\arraystretch}{0.8}
\begin{deluxetable*}{cccccc}
\tabletypesize{\tiny} 
\movetabledown=1in
\tablecaption{\label{table:summa_litera_10OCs}Literature age results for the 31 open clusters and 3 moving groups with isochronal or LDB method. The corresponding [Fe/H] results are also listed. For results with the LDB method, the spectral resolution R is listed as well.}
\tablehead{\colhead{Clusters} &\colhead{[Fe/H]} & \colhead{ref} & \colhead{Age} & \colhead{method} &\colhead{ref} \\
\colhead{} & \colhead{dex} & \colhead{} & \colhead{Myr} & \colhead{}  & \colhead{} \\} 
\startdata
    NGC\,752 & 0.01$\pm$0.04, 0.04$\pm$0.01, $-$0.063$\pm$0.013, $+$0.08$\pm$0.04 & 1,3, 5, 7 & $\approx$2000,1340$\pm$60, 1450, 160, 1780,$\approx$1413 &L, I, I, I, I, I &2, 4, 6, 7, 9, 10     \\    
    & $-$0.01$\pm$0.06, $-$0.03$\pm$0.06, $-$0.064, $-$0.037$\pm$0.049, $-$0.082$\pm$0.071  & 8, 21, 29, 33, 34 & $\approx$1520,1450$\pm$50, 1175, $1521^{+101}_{-95}$, 1247 &I, I, I, I, I & 11, 27, 28, 33, 41     \\    
    & $-$0.054$\pm$0.005, $-$0.04$\pm$0.01, $-$0.040$\pm$0.007, $-$0.053$\pm$0.006 & 37, 69, 71, 72 &  &  & \\
\hline
    NGC\,7092 &$-$0.047$\pm$0.072, $-$0.29 &33, 34 &$310^{+74}_{-58}$, 350, $457^{+70}_{-61}$, $501^{+74}_{-206}$ & I, I, I, I &17, 30, 33, 39    \\
\hline
    NGC\,2516 & $+$0.05$\pm$0.11, $-$0.008$\pm$0.028, $+$0.05, $−$0.08$\pm$0.01, 0.1 &21, 33, 39, 43, 66 &123, $277^{+60}_{-49}$, 150, 251, 112&I, I, I, I, I&15, 33, 39, 41, 66     \\
\hline
    Blanco\,1 & +0.04$\pm$0.02, +0.03$\pm$0.07, $-$0.03, $-$0.016$\pm$0.023, $-$0.09$\pm$0.13 &13, 21, 32, 33, 65 &$115^{+23}_{-15}$, $94^{+5}_{-7}$, $146^{+13}_{-14}$, 132$\pm$24, $126^{+13}_{-14}$ & I, I, I, L, L &15, 17, 22, 23, 24      \\    
    & $+$0.01, $+$0.04$\pm$0.07, $-$0.056$\pm$0.048  &66, 69, 71 & 105, $103^{+24}_{-19}$, $144^{+417}_{-53}$, 93.33 & I, I, I, I &32, 33 , 40, 66     \\                  
\hline
    NGC\,2451B & 0.0, $-$0.02, 0.05$\pm$0.054, $+$0.06 & 21, 32, 33, 66 & 39, 50, 41, $47^{+7}_{-6}$ & I, I, I, I & 17, 30, 32, 33 \\ 
              &  &  & 170, 45, 50, 39 & I, I, I, I & 35, 36, 38, 66 \\ 
\hline
    IC\,2602 & $-$0.05$\pm$0.05, $-$0.02$\pm$0.02, $+$0.02$\pm$0.02, $-$0.06 & 14, 21, 31, 32 &  $40^{+10}_{-11}$, $44^{+18}_{-16}$, $35.3^{+1.3}_{-1.1}$, $46^{+6}_{-5}$ & I, I, I, L & 15, 16, 17, 18 \\   
            & $-$0.016 $\pm$0.012, $-$0.02$\pm$0.02, $-$0.02$\pm$0.05, $-$0.035$\pm$0.136 & 33, 69, 70, 71 & 44, 36, $47^{+2.6}_{-2.4}$, $100^{+12}_{-45}$, 35, 32 & I, I, I, I, I, I & 30, 32, 33, 40, 41, 66 \\ 
\hline
        IC\,2391 & $-$0.03$\pm$0.07, $-$0.01$\pm$0.03, 0.05$\pm$0.02, $-$0.06 & 14, 21, 31, 32 & 50$\pm$5, $50^{+14}_{-13}$, 36.4$\pm$2.0, 53$\pm$5, 48$\pm$5 & L, I, I, L, L & 12, 15, 17, 19, 20 \\
                & 0.00 $\pm$0.011, $+$0.03, $-$0.03$\pm$0.04 & 33, 66, 69 & 49, 29, $49^{+5.0}_{-4.5}$, $81^{+118}_{-51}$, 46, 36.31 & I, I, I, I, I, I & 30, 32, 33, 40, 41, 66 \\
\hline
        NGC\,2232 & $+$0.11, 0.27$\pm$0.08, $-$0.03, $+$0.016$\pm$0.06, $-$0.094$\pm$0.099 & 21, 26, 32, 33, 34 & 25, 38$\pm$3, 24, 18, $31^{+3.0}_{-2.8}$ & I, L, I, I, I & 15, 25, 30, 32, 33 \\
                 & $-$0.02$\pm$0.12, $+$0.03, $-$0.121$\pm$0.037 & 37, 66, 71 & $105^{+37}_{-50}$, 55, 52.48 & I, I, I & 40, 41, 66 \\      
\hline
        Pozzo\,1 & $-$0.069$\pm$0.064, $-$0.04$\pm$0.05, $-$0.04$\pm$0.05, $+$0.07 & 33, 61, 62, 66 & 15, $12.2^{+2.7}_{-2.2}$, 20.0$\pm$4.6, 14 & I, I, I, I & 17, 33, 63, 66 \\
\hline
        Alessi\,3 & $-$0.057$\pm$0.045 & 33 & 631, $798^{+111}_{-98}$, $794^{+183}_{-163}$, 637 & I, I, I, I & 28, 33, 40, 41 \\        
\hline
        ASCC\,101 & 0.004$\pm$0.064 & 33 & 490, $332^{+150}_{-103}$, 189 & I, I, I & 28, 33, 41 \\        
\hline
        Alessi\,9 & $-$0.002$\pm$0.050 & 33 & 282, $392^{+71}_{-60}$, 291 & I, I, I & 28, 33, 41 \\        
\hline
        ASCC\,41 & $-$0.089$\pm$0.056, $-$0.11$\pm$0.065, $-$0.08$\pm$0.018 & 33, 34, 37 & 214, $281^{+188}_{-113}$, 100 & I, I, I & 28, 33, 41 \\        
\hline
       NGC\,2422 & $+$0.09$\pm$0.03, 0.132$\pm$0.054, $−$0.05$\pm$0.02, $+$0.07 & 21, 33, 43, 66 & 110, $155^{+39}_{-31}$, 117, 72.44 & I, I, I, I & 28, 33, 41, 66 \\       
\hline
     Teutsch\,35&0.078$\pm$0.068 &33      &103, $231^{+103}_{-71}$ & I, I &28, 33 \\
\hline
      UPK\,612 &$-$0.076$\pm$0.097, $+$0.021$\pm$0.027 & 33, 71 & 100, $141^{+78}_{-50}$ & I, I & 28, 33 \\
\hline
      Roslund\,6& $+$0.078$\pm$0.058, $-$0.010$\pm$0.104, $+$0.032$\pm$0.011 &33, 34, 37 & $259^{+49}_{-41}$ & I & 33     \\                
\hline
      Platais\,9&$-$0.025$\pm$0.047 &33     &50, $56^{+7.9}_{-6.9}$, 54 & I, I, I & 28, 33, 41     \\        
\hline
     NGC\,2451A&$-$0.08, $+$0.050$\pm$0.057, 0.0 &21, 33, 66 & 36, $55^{+6.2}_{-5.6}$, $148^{+98}_{-76}$, 50, 43.65  & I, I, I, I, I & 28, 33, 40, 41, 66 \\              
\hline
        RSG\,5 & $+$0.012$\pm$0.076, $+$0.065$\pm$0.094, 0.105$\pm$0.016 & 33, 34, 37 & 35, $52^{+5.4}_{-4.9}$ & I ,I & 28, 33 \\              
\hline
  Trumpler\,10 & $-$0.12$\pm$0.06, $+$0.043$\pm$0.050, $+$0.03 & 21, 33, 66 & 33, $57^{+3.5}_{-3.3}$, 56, 35 & I, I, I, I & 28, 33, 41, 66 \\                
\hline
     NGC\,2547 & $−$0.14$\pm$0.10, $-$0.014$\pm$0.05, $−$0.03$\pm$0.06, $+$0.04, $−$0.01$\pm$0.01 & 21, 33, 64, 66, 68 & 33, $39^{+2.2}_{-2.1}$, 35, 31, 34.7$\pm$4.0, 27, 57$\pm$5.7 & I, I, I, I, L, I, I & 28, 33, 39, 41, 42, 66, 68 \\     
\hline
       BH\,164 & $-$0.002$\pm$0.050 & 33     &33, $46^{+6.5}_{-5.7}$ & I, I & 28, 33     \\       
\hline
     NGC\,3228 & $+$0.01, $+$0.143$\pm$0.08 & 21, 33 & 31, $46^{+7.7}_{-6.6}$, 32 & I, I, I & 28, 33, 41    \\     
\hline
      Col\,140 & $+$0.01$\pm$0.0, $+$0.011$\pm$0.04 & 21, 33 & 27, $40^{+3.5}_{-3.3}$ & I, I & 28, 33     \\       
\hline
      Col\,135 & $+$0.111$\pm$0.05, $+$0.03, $-$0.011$\pm$0.044 & 33, 66, 71 & 26, $47^{+5.5}_{-4.9}$, $47^{+5.5}_{-4.9}$ & I, I  & 28, 66, 33 \\            
\hline
      UBC\,17b & $+$0.013$\pm$0.012, $-$0.070$\pm$0.032 & 37, 71 & 12 & I &28     \\      
\hline
     ASCC\,21 & 0.0, $-$0.008$\pm$0.029, $-$0.108$\pm$0.318 & 17, 33, 34 & $11.0^{+0.0}_{-0.3}$, 9, $13^{+1.2}_{-1.1}$ & I, I, I & 17, 28, 33 \\     
\hline
    NGC\,2281 & $-$0.037$\pm$0.061, $-$0.085$\pm$0.11, $-$0.015$\pm$0.007, $+$0.06 & 33, 34, 37, 66 & 617, $590^{+89}_{-77}$, 590, 354.82 & I, I, I, I & 28, 33, 41, 66 \\    
\hline
    NGC\,1960 & $-$0.030$\pm$0.085, $-$0.20$\pm$0.097, $-$0.090$\pm$0.025, $+$0.21 & 33, 34, 37, 66 & 25, $30.2^{+4.4}_{-3.8}$, $22\pm4$, $26.3^{+3.2}_{-5.2}$, 26 & I, I, L, I, I & 17, 33, 44, 49, 66 \\    
\hline
    IC\,4665 & $-$0.03$\pm$0.04, $-$0.005$\pm$0.032, $-$0.029$\pm$0.011, $+$0.08$\pm$0.015, $+$0.12 & 21, 33, 34, 37, 66 & $38^{+5}_{-2}$, $53^{11}_{-9}$, $28\pm4$, $23.2^{+3.5}_{-3.1}$, $52^{+8}_{-6}$ & I, I, L, L, L & 17, 33, 45, 46, 47 \\
             & $-$0.03$\pm$0.08, $-$0.01$\pm$0.02, $-$0.034$\pm$0.078 & 67, 69, 71 & $32^{+4}_{-5}$, 38 & L, I & 48, 66 \\      
\hline
        BPMG & $-$0.01$\pm$0.08 & 17 & $21\pm4$, $26\pm3$, $25\pm3$, $22\pm6$, $12-22$, $20\pm10$, $24\pm3$ & L, L, L, L, I, I, I & 50, 51, 52, 53, 54, 55, 57 \\              
\hline
        THOR & & & $23\pm4$, $22^{+4}_{-3}$, $25\pm2.5$ & L, I, I & 56, 57, 58 \\            
\hline
        THMG & $+$0.03$\pm$0.0 & 60 & $10-40$, $45\pm4$, $40\pm3$, 30 & I, I, L, I & 54, 57, 59, 60 \\  
\hline
\hline
\enddata
\tablerefs{
1. \citet{Sestito2004AA809S}, 2. \citet{Sestito2004AA809S}, 3. \citet{Blanco-Cuaresma2014AA111B}, 4. \citet{Agueros2018ApJ33A}, 5. \citet{Maderak2013AJ143M}, 6. \citet{Anthony-Twarog2009AJ1171A}, 7. \citet{Carrera2011AA30C}, 8. \citet{Lum2019ApJ99L}, 9. \citet{Daniel1994PASP281D}, 10. \citet{Siegel2019AJ35S}, 11. \citet{Bocek-Topcu2020MN544B}, 12. \citet{Barrado2004ApJ386B}, 13. \citet{Ford2005MN272F}, 14. \citet{Randich2001AA862R}, 15. \citet{GaiaCollaboration2018AA10G}, 16. \citet{Naylor2009MN432N}, 17. \citet{Bossin2019AA108B}, 18. \citet{Dobbie2010MN1002D}, 19. \citet{Barrado1999ApJ53B}, 20. \citet{Burke2004ApJ272B}, 21. \citet{Netopil2016AA150N}, 22. \citet{Cargile2014ApJ29C}, 23. \citet{Cargile2010ApJ111C}, 24. \citet{Juarez2014ApJ143J}, 25. \citet{Binks2021MN1280B}, 26. \citet{Monroe2010AJ2109M}, 27. \citet{Twarog2023AJ105T}, 28. \citet{CantatGaudin2020AA}, 29. \citet{Spoo2022AJ229S}, 30. \citet{Pang2023AJ110P}, 31. \citet{Nisak2022AJ}, 32. \citet{Bragaglia2022AA200B}, 33. \citet{Dias2021MN356D}, 34. \citet{Zhong2020AA127Z}, 35. \citet{Kharchenko2013AA53K}, 36. \citet{Dias2002AA871D}, 37. \citet{Fu2022AA668A}, 38. \citet{Hunsch2003cs787H}, 39. \citet{Godoy-Rivera2021Ap46G}, 40. \citet{Yen2018yCat12Y}, 41. \citet{Kounkel2019AJ122K}, 42. \citet{Jeffries2005MNR13J}, 43. \citet{Bailey2018MN1609B}, 44. \citet{Jeffries2013MN}, 45. \citet{Manzi2008AA}, 46. \citet{Randich2018AA}, 47. \citet{Jeffries2023MN}, 48. \citet{Jeffries2023MN1260J}, 49. \citet{Bell2013MN}, 50. \citet{Binks2014MN}, 51. \citet{Malo2014ApJ}, 52. \citet{Messina2016AA}, 53. \citet{Shkolnik2017AJ}, 54. \citet{Malo2013ApJ}, 55. \citet{Barrado1999ApJ}, 56. \citet{Bell2017MN}, 57. \citet{Bell2015MN}, 58. \citet{Mamajek2007IAUS}, 59. \citet{Kraus2014AJ}, 60. \citet{Torres2008hsf2}, 61. \citet{Jeffries2014AA}, 62. \citet{Spina2014AA}, 63. \citet{Jeffries2009MNRAS}, 64. \citet{Magrini2015AA}, 65. \citet{Buder2021MN}, 66. \citet{Rain2021AA}, 67. \citet{Mermilliod1981AA235M}, 68. \citet{Claria1982AAS}, 69. \citet{Netopil2022MN}, 70. \citet{Randich2001AA}, 71. \citet{Spina2021MN}, 72. \citet{Casamiquela2021AA}
}
\tablecomments{Lithium depletion boundary (LDB, L), isochrone fitting (iso, I).}
\end{deluxetable*}
\end{longrotatetable}
\end{center}

\end{document}